\documentstyle[]{article}
\begin{document}
\overfullrule 0 mm
\language 0
\centerline { \bf{ CLASSICAL TUNNELING }}
\centerline { \bf{OF SOMMERFELD SPHERE }}
\centerline { \bf {IN CLASSICAL ELECTRODYNAMICS}}
\vskip 0.5 cm

\centerline {\bf{ Alexander A.  Vlasov}}
\vskip 0.3 cm
\centerline {{  High Energy and Quantum Theory}}
\centerline {{Department of Physics}}
\centerline {{ Moscow State University}}
\centerline {{  Moscow, 119899}}
\centerline {{ Russia}}
\vskip 0.3cm
{\it   One-dimensional motion of Sommerfeld sphere in the case of
potential barrier  is numerically investigated. The effect of
classical tunneling is found out - Sommerfeld
sphere overcomes the barrier and finds itself in the forbidden, from
classical point of view, area}

03.50.De
\vskip 0.3 cm
Here numerical investigations of one-dimensional motion of Sommerfeld
sphere with total charge $Q$, mechanical mass $m$ and radius $a$ are
continued (for prev. results - see [1,2]).

We consider  one-dimensional motion for the case of
potential barrier, produced by homogeneous static electric field
$E$, stretched in $z$ - direction for $0<z<L$ (like in plane
condenser):
$$ E= \left\{\matrix{
0,& z<0; \cr
E,& 0<z<L; \cr
0,& L<z; \cr
}\right.$$

For dimensionless variables $y= R/L,\ \ x=ct/L,\ \ a^{*}=2a/L$
 the  equation of motion of Sommerfeld sphere for our problem is

$${d^2 y \over dx^2} =\left(1-({dy\over dx})^2\right)^{3/2}$$
$$\left\{ k \cdot \left[ - \int\limits_{x^{-}}^{x^{+}} dz {z-a^{*}
\over L^2} + \ln {{L^{+}\over L^{-}}} + ({1\over \beta^2}-1)\ln {
{1+\beta \over 1-\beta}} \right] +\lambda \cdot F_Z \right\}
\eqno(1)$$ here     $$x^{\pm}=a^{*} \pm L^{\pm},\ \
L^{\pm}=y(x)-y(x-x^{\pm}),\ \ L=y(x)-y(x-z),$$
$$ \beta=dy/dx, \
\ \ \ k= {Q^2 \over 2 m c^2 a},\ \ \lambda={L Q E \over m c^2}$$
$$F_Z=\left\{\matrix{
0,& y<-a^{*}/2; \cr
(2y/a^{*} +1)/2,& -a^{*}/2<y<a^{*}/2; \cr
1,& a^{*}/2<y<1-a^{*}/2; \cr
(-2y/a^{*} +1+2/a^{*})/2,& 1-a^{*}/2<y<1+a^{*}/2;\cr
0,& 1+a^{*}/2<y; \cr
}\right.$$

Later on we take $k=1$.

It is useful to compare solutions of (1) with  classical point
charge motion in the same field, governed by the following
relativistic equation without radiation force:
$${d^2 y \over dx^2} =\left(1-({dy\over dx})^2\right)^{3/2}
  \lambda \cdot F_E \eqno(2)$$
here
$$F_E=\lambda \left\{\matrix{
0,& y<0; \cr
1,& 0<y<1; \cr
0,& 1<y; \cr
}\right.$$

\vskip 0.5 cm {\bf A.}

Numerical calculations of eqs. (1-2) show that there is the
effect of classical tunneling for Sommerfeld sphere.

Indeed, classical point particle motion, governed by eq. (2), is
simple:
$$\gamma = \lambda +\gamma_{0} \eqno(3)$$
here $\gamma=1/\sqrt{1-(v/c)^2}$ - Lorentz factor, and $\gamma_{0}$ -
Lorentz factor for initial velocity.

Thus for given initial velocity for $\lambda<1-\gamma_{0}$ there is
the turning point - i.e. classical particle cannot overcome the
potential barrier (for ex., for $(v/c)_{0}=0.3$ the crucial value of
$\lambda$ is approximately $ -0.0482848...$) and the electric field
turns the particle back.

For numerical calculations of eq.(1) we took  $(v/c)_{0}=0.3$ and
$\lambda= -0.04829$.

Numerical results for different values of
$a^{*}$ are shown on fig. (A.1-A.3) (vertical axis is velocity
$dy/dx$, horizontal axis is coordinate $y$; trajectories of
Sommerfeld sphere are compared with trajectories of classical
particle):

we see that for $a^{*}=0.2,\ \  (Fig. A.1)$ and for $a^{*}=0.001,\ \
(Fig. A.2)$ there is the effect of classical tunneling - Sommerfeld
sphere overcomes the barrier and finds itself in the forbidden, from
classical point of view, area with resulting velocities $(v/c)_r
=0.139630$ for $a^{*}=0.2$ and $(v/c)_r
=0.037190$ for $a^{*}=0.001$;

while for $a^{*}=0.0002,\ \ (Fig.
A.3)$ - this effect is absent (the barrier is too high... ).

These results confirm the results of the works [3,4] where
the effects of "pumping"  of extended object in the potential hole
[4] (on the example of some model in classical field theory) and
classical tunneling [3] for Lorentz-Dirac eq. were found.

We see that these unusual (from point of view of classical mechanics)
results are not the consequences of "point-like" description of
charged particles but reflect the general features of equations with
retardation (difference-differential equations).

 \centerline {\bf{REFERENCES}}

  \begin{enumerate}
\item Alexander A.Vlasov, physics/9901051.
\item Alexander A.Vlasov, physics/9902018.
\item W.Troost et al.,  preprint hep-th/9602066.
 \item Alexander A.Vlasov, Theoretical and Mathematical
Physics, 109, n.3, 1608(1996).

\end{enumerate}
\newcount\numpoint
\newcount\numpointo
\numpoint=1 \numpointo=1
\def\emmoveto#1#2{\offinterlineskip
\hbox to 0 true cm{\vbox to 0
true cm{\vskip - #2 true mm
\hskip #1 true mm \special{em:point
\the\numpoint}\vss}\hss}
\numpointo=\numpoint
\global\advance \numpoint by 1}
\def\emlineto#1#2{\offinterlineskip
\hbox to 0 true cm{\vbox to 0
true cm{\vskip - #2 true mm
\hskip #1 true mm \special{em:point
\the\numpoint}\vss}\hss}
\special{em:line
\the\numpointo,\the\numpoint}
\numpointo=\numpoint
\global\advance \numpoint by 1}
\def\emshow#1#2#3{\offinterlineskip
\hbox to 0 true cm{\vbox to 0
true cm{\vskip - #2 true mm
\hskip #1 true mm \vbox to 0
true cm{\vss\hbox{#3\hss
}}\vss}\hss}}
\special{em:linewidth 0.8pt}

\vrule width 0 mm height                0 mm depth 90.000 true mm

\special{em:linewidth 0.8pt}
\emmoveto{130.000}{10.000}
\emlineto{12.000}{10.000}
\emlineto{12.000}{80.000}
\emmoveto{71.000}{10.000}
\emlineto{71.000}{80.000}
\emmoveto{12.000}{45.000}
\emlineto{130.000}{45.000}
\emmoveto{130.000}{10.000}
\emlineto{130.000}{80.000}
\emlineto{12.000}{80.000}
\emlineto{12.000}{10.000}
\emlineto{130.000}{10.000}
\special{em:linewidth 0.4pt}
\emmoveto{12.000}{17.000}
\emlineto{130.000}{17.000}
\emmoveto{12.000}{24.000}
\emlineto{130.000}{24.000}
\emmoveto{12.000}{31.000}
\emlineto{130.000}{31.000}
\emmoveto{12.000}{38.000}
\emlineto{130.000}{38.000}
\emmoveto{12.000}{45.000}
\emlineto{130.000}{45.000}
\emmoveto{12.000}{52.000}
\emlineto{130.000}{52.000}
\emmoveto{12.000}{59.000}
\emlineto{130.000}{59.000}
\emmoveto{12.000}{66.000}
\emlineto{130.000}{66.000}
\emmoveto{12.000}{73.000}
\emlineto{130.000}{73.000}
\emmoveto{23.800}{10.000}
\emlineto{23.800}{80.000}
\emmoveto{35.600}{10.000}
\emlineto{35.600}{80.000}
\emmoveto{47.400}{10.000}
\emlineto{47.400}{80.000}
\emmoveto{59.200}{10.000}
\emlineto{59.200}{80.000}
\emmoveto{71.000}{10.000}
\emlineto{71.000}{80.000}
\emmoveto{82.800}{10.000}
\emlineto{82.800}{80.000}
\emmoveto{94.600}{10.000}
\emlineto{94.600}{80.000}
\emmoveto{106.400}{10.000}
\emlineto{106.400}{80.000}
\emmoveto{118.200}{10.000}
\emlineto{118.200}{80.000}
\special{em:linewidth 0.8pt}
\emmoveto{21.077}{80.000}
\emlineto{21.349}{79.937}
\emmoveto{21.349}{79.927}
\emlineto{21.621}{79.863}
\emmoveto{21.621}{79.853}
\emlineto{21.892}{79.790}
\emmoveto{21.892}{79.780}
\emlineto{22.163}{79.716}
\emmoveto{22.163}{79.706}
\emlineto{22.434}{79.643}
\emmoveto{22.434}{79.633}
\emlineto{22.704}{79.569}
\emmoveto{22.704}{79.559}
\emlineto{22.974}{79.496}
\emmoveto{22.974}{79.486}
\emlineto{23.243}{79.422}
\emmoveto{23.243}{79.412}
\emlineto{23.512}{79.349}
\emmoveto{23.512}{79.339}
\emlineto{23.781}{79.275}
\emmoveto{23.781}{79.265}
\emlineto{24.049}{79.201}
\emmoveto{24.049}{79.191}
\emlineto{24.317}{79.127}
\emmoveto{24.317}{79.117}
\emlineto{24.585}{79.054}
\emmoveto{24.585}{79.044}
\emlineto{24.852}{78.980}
\emmoveto{24.852}{78.970}
\emlineto{25.119}{78.906}
\emmoveto{25.119}{78.896}
\emlineto{25.385}{78.832}
\emmoveto{25.385}{78.822}
\emlineto{25.651}{78.759}
\emmoveto{25.651}{78.749}
\emlineto{25.917}{78.685}
\emmoveto{25.917}{78.675}
\emlineto{26.182}{78.611}
\emmoveto{26.182}{78.601}
\emlineto{26.447}{78.537}
\emmoveto{26.447}{78.527}
\emlineto{26.711}{78.463}
\emmoveto{26.711}{78.453}
\emlineto{26.975}{78.389}
\emmoveto{26.975}{78.379}
\emlineto{27.239}{78.315}
\emmoveto{27.239}{78.305}
\emlineto{27.502}{78.241}
\emmoveto{27.502}{78.231}
\emlineto{27.765}{78.167}
\emmoveto{27.765}{78.157}
\emlineto{28.028}{78.093}
\emmoveto{28.028}{78.083}
\emlineto{28.290}{78.018}
\emmoveto{28.290}{78.008}
\emlineto{28.552}{77.944}
\emmoveto{28.552}{77.934}
\emlineto{28.813}{77.870}
\emmoveto{28.813}{77.860}
\emlineto{29.074}{77.796}
\emmoveto{29.074}{77.786}
\emlineto{29.335}{77.721}
\emmoveto{29.335}{77.711}
\emlineto{29.595}{77.647}
\emmoveto{29.595}{77.637}
\emlineto{29.855}{77.573}
\emmoveto{29.855}{77.563}
\emlineto{30.114}{77.498}
\emmoveto{30.114}{77.488}
\emlineto{30.374}{77.424}
\emmoveto{30.374}{77.414}
\emlineto{30.632}{77.349}
\emmoveto{30.632}{77.339}
\emlineto{30.891}{77.275}
\emmoveto{30.891}{77.265}
\emlineto{31.148}{77.201}
\emmoveto{31.148}{77.191}
\emlineto{31.406}{77.126}
\emmoveto{31.406}{77.116}
\emlineto{31.663}{77.052}
\emmoveto{31.663}{77.042}
\emlineto{31.920}{76.977}
\emmoveto{31.920}{76.967}
\emlineto{32.176}{76.902}
\emmoveto{32.176}{76.892}
\emlineto{32.432}{76.828}
\emmoveto{32.432}{76.818}
\emlineto{32.688}{76.753}
\emmoveto{32.688}{76.743}
\emlineto{32.943}{76.678}
\emmoveto{32.943}{76.668}
\emlineto{33.198}{76.604}
\emmoveto{33.198}{76.594}
\emlineto{33.452}{76.529}
\emmoveto{33.452}{76.519}
\emlineto{33.706}{76.454}
\emmoveto{33.706}{76.444}
\emlineto{33.960}{76.379}
\emmoveto{33.960}{76.369}
\emlineto{34.213}{76.305}
\emmoveto{34.213}{76.295}
\emlineto{34.466}{76.230}
\emmoveto{34.466}{76.220}
\emlineto{34.719}{76.155}
\emmoveto{34.719}{76.145}
\emlineto{34.971}{76.080}
\emmoveto{34.971}{76.070}
\emlineto{35.223}{76.005}
\emmoveto{35.223}{75.995}
\emlineto{35.474}{75.930}
\emmoveto{35.474}{75.920}
\emlineto{35.725}{75.855}
\emmoveto{35.725}{75.845}
\emlineto{35.976}{75.780}
\emmoveto{35.976}{75.770}
\emlineto{36.226}{75.705}
\emmoveto{36.226}{75.695}
\emlineto{36.475}{75.630}
\emmoveto{36.475}{75.620}
\emlineto{36.725}{75.555}
\emmoveto{36.725}{75.545}
\emlineto{36.974}{75.479}
\emmoveto{36.974}{75.469}
\emlineto{37.223}{75.404}
\emmoveto{37.223}{75.394}
\emlineto{37.471}{75.329}
\emmoveto{37.471}{75.319}
\emlineto{37.719}{75.254}
\emmoveto{37.719}{75.244}
\emlineto{37.966}{75.179}
\emmoveto{37.966}{75.169}
\emlineto{38.213}{75.103}
\emmoveto{38.213}{75.093}
\emlineto{38.460}{75.028}
\emmoveto{38.460}{75.018}
\emlineto{38.706}{74.953}
\emmoveto{38.706}{74.943}
\emlineto{38.952}{74.877}
\emmoveto{38.952}{74.867}
\emlineto{39.197}{74.802}
\emmoveto{39.197}{74.792}
\emlineto{39.442}{74.726}
\emmoveto{39.442}{74.716}
\emlineto{39.687}{74.651}
\emmoveto{39.687}{74.641}
\emlineto{39.932}{74.575}
\emmoveto{39.932}{74.565}
\emlineto{40.175}{74.500}
\emmoveto{40.175}{74.490}
\emlineto{40.419}{74.424}
\emmoveto{40.419}{74.414}
\emlineto{40.662}{74.349}
\emmoveto{40.662}{74.339}
\emlineto{40.905}{74.273}
\emmoveto{40.905}{74.263}
\emlineto{41.147}{74.198}
\emmoveto{41.147}{74.188}
\emlineto{41.389}{74.122}
\emmoveto{41.389}{74.112}
\emlineto{41.631}{74.046}
\emmoveto{41.631}{74.036}
\emlineto{41.872}{73.971}
\emmoveto{41.872}{73.961}
\emlineto{42.113}{73.895}
\emmoveto{42.113}{73.885}
\emlineto{42.353}{73.819}
\emmoveto{42.353}{73.809}
\emlineto{42.593}{73.743}
\emmoveto{42.593}{73.733}
\emlineto{42.833}{73.667}
\emmoveto{42.833}{73.657}
\emlineto{43.072}{73.592}
\emmoveto{43.072}{73.582}
\emlineto{43.311}{73.516}
\emmoveto{43.311}{73.506}
\emlineto{43.549}{73.440}
\emmoveto{43.549}{73.430}
\emlineto{43.787}{73.364}
\emmoveto{43.787}{73.354}
\emlineto{44.025}{73.288}
\emmoveto{44.025}{73.278}
\emlineto{44.262}{73.212}
\emmoveto{44.262}{73.202}
\emlineto{44.499}{73.136}
\emmoveto{44.499}{73.126}
\emlineto{44.735}{73.060}
\emmoveto{44.735}{73.050}
\emlineto{44.971}{72.984}
\emmoveto{44.971}{72.974}
\emlineto{45.207}{72.907}
\emmoveto{45.207}{72.897}
\emlineto{45.442}{72.831}
\emmoveto{45.442}{72.821}
\emlineto{45.677}{72.755}
\emmoveto{45.677}{72.745}
\emlineto{45.912}{72.679}
\emmoveto{45.912}{72.669}
\emlineto{46.146}{72.603}
\emmoveto{46.146}{72.593}
\emlineto{46.379}{72.526}
\emmoveto{46.379}{72.516}
\emlineto{46.613}{72.450}
\emmoveto{46.613}{72.440}
\emlineto{46.846}{72.374}
\emmoveto{46.846}{72.364}
\emlineto{47.078}{72.298}
\emmoveto{47.078}{72.288}
\emlineto{47.310}{72.221}
\emmoveto{47.310}{72.211}
\emlineto{47.542}{72.145}
\emmoveto{47.542}{72.135}
\emlineto{47.773}{72.069}
\emmoveto{47.773}{72.059}
\emlineto{48.004}{71.992}
\emmoveto{48.004}{71.982}
\emlineto{48.235}{71.916}
\emmoveto{48.235}{71.906}
\emlineto{48.465}{71.839}
\emmoveto{48.465}{71.829}
\emlineto{48.695}{71.763}
\emmoveto{48.695}{71.753}
\emlineto{48.924}{71.686}
\emmoveto{48.924}{71.676}
\emlineto{49.153}{71.609}
\emmoveto{49.153}{71.599}
\emlineto{49.381}{71.533}
\emmoveto{49.381}{71.523}
\emlineto{49.610}{71.456}
\emmoveto{49.610}{71.446}
\emlineto{49.837}{71.380}
\emmoveto{49.837}{71.370}
\emlineto{50.065}{71.303}
\emmoveto{50.065}{71.293}
\emlineto{50.292}{71.226}
\emmoveto{50.292}{71.216}
\emlineto{50.518}{71.149}
\emmoveto{50.518}{71.139}
\emlineto{50.744}{71.073}
\emmoveto{50.744}{71.063}
\emlineto{50.970}{70.996}
\emmoveto{50.970}{70.986}
\emlineto{51.195}{70.919}
\emmoveto{51.195}{70.909}
\emlineto{51.420}{70.842}
\emmoveto{51.420}{70.832}
\emlineto{51.645}{70.765}
\emmoveto{51.645}{70.755}
\emlineto{51.869}{70.688}
\emmoveto{51.869}{70.678}
\emlineto{52.093}{70.612}
\emmoveto{52.093}{70.602}
\emlineto{52.316}{70.535}
\emmoveto{52.316}{70.525}
\emlineto{52.539}{70.458}
\emmoveto{52.539}{70.448}
\emlineto{52.762}{70.381}
\emmoveto{52.762}{70.371}
\emlineto{52.984}{70.304}
\emmoveto{52.984}{70.294}
\emlineto{53.206}{70.227}
\emmoveto{53.206}{70.217}
\emlineto{53.427}{70.149}
\emmoveto{53.427}{70.139}
\emlineto{53.648}{70.072}
\emmoveto{53.648}{70.062}
\emlineto{53.868}{69.995}
\emmoveto{53.868}{69.985}
\emlineto{54.089}{69.918}
\emmoveto{54.089}{69.908}
\emlineto{54.308}{69.841}
\emmoveto{54.308}{69.831}
\emlineto{54.528}{69.764}
\emmoveto{54.528}{69.754}
\emlineto{54.747}{69.686}
\emmoveto{54.747}{69.676}
\emlineto{54.965}{69.609}
\emmoveto{54.965}{69.599}
\emlineto{55.183}{69.532}
\emmoveto{55.183}{69.522}
\emlineto{55.401}{69.455}
\emmoveto{55.401}{69.445}
\emlineto{55.619}{69.377}
\emmoveto{55.619}{69.367}
\emlineto{55.836}{69.300}
\emmoveto{55.836}{69.290}
\emlineto{56.052}{69.222}
\emmoveto{56.052}{69.212}
\emlineto{56.268}{69.145}
\emmoveto{56.268}{69.135}
\emlineto{56.484}{69.068}
\emmoveto{56.484}{69.058}
\emlineto{56.699}{68.990}
\emmoveto{56.699}{68.980}
\emlineto{56.914}{68.913}
\emmoveto{56.914}{68.903}
\emlineto{57.129}{68.835}
\emmoveto{57.129}{68.825}
\emlineto{57.343}{68.758}
\emmoveto{57.343}{68.748}
\emlineto{57.557}{68.680}
\emmoveto{57.557}{68.670}
\emlineto{57.770}{68.602}
\emmoveto{57.770}{68.592}
\emlineto{57.983}{68.525}
\emmoveto{57.983}{68.515}
\emlineto{58.196}{68.447}
\emmoveto{58.196}{68.437}
\emlineto{58.408}{68.369}
\emmoveto{58.408}{68.359}
\emlineto{58.619}{68.292}
\emmoveto{58.619}{68.282}
\emlineto{58.831}{68.214}
\emmoveto{58.831}{68.204}
\emlineto{59.042}{68.136}
\emmoveto{59.042}{68.126}
\emlineto{59.252}{68.058}
\emmoveto{59.252}{68.048}
\emlineto{59.462}{67.981}
\emmoveto{59.462}{67.971}
\emlineto{59.672}{67.903}
\emmoveto{59.672}{67.893}
\emlineto{59.881}{67.825}
\emmoveto{59.881}{67.815}
\emlineto{60.090}{67.747}
\emmoveto{60.090}{67.737}
\emlineto{60.299}{67.669}
\emmoveto{60.299}{67.659}
\emlineto{60.507}{67.591}
\emmoveto{60.507}{67.581}
\emlineto{60.714}{67.513}
\emmoveto{60.714}{67.503}
\emlineto{60.922}{67.435}
\emmoveto{60.922}{67.425}
\emlineto{61.129}{67.357}
\emmoveto{61.129}{67.347}
\emlineto{61.335}{67.279}
\emmoveto{61.335}{67.269}
\emlineto{61.541}{67.201}
\emmoveto{61.541}{67.191}
\emlineto{61.747}{67.123}
\emmoveto{61.747}{67.113}
\emlineto{61.952}{67.045}
\emmoveto{61.952}{67.035}
\emlineto{62.157}{66.967}
\emmoveto{62.157}{66.957}
\emlineto{62.361}{66.889}
\emmoveto{62.361}{66.879}
\emlineto{62.566}{66.810}
\emmoveto{62.566}{66.800}
\emlineto{62.769}{66.732}
\emmoveto{62.769}{66.722}
\emlineto{62.972}{66.654}
\emmoveto{62.972}{66.644}
\emlineto{63.175}{66.576}
\emmoveto{63.175}{66.566}
\emlineto{63.378}{66.498}
\emmoveto{63.378}{66.488}
\emlineto{63.580}{66.419}
\emmoveto{63.580}{66.409}
\emlineto{63.781}{66.341}
\emmoveto{63.781}{66.331}
\emlineto{63.982}{66.263}
\emmoveto{63.982}{66.253}
\emlineto{64.183}{66.184}
\emmoveto{64.183}{66.174}
\emlineto{64.384}{66.106}
\emmoveto{64.384}{66.096}
\emlineto{64.584}{66.027}
\emmoveto{64.584}{66.017}
\emlineto{64.783}{65.949}
\emmoveto{64.783}{65.939}
\emlineto{64.982}{65.870}
\emmoveto{64.982}{65.860}
\emlineto{65.181}{65.792}
\emmoveto{65.181}{65.782}
\emlineto{65.380}{65.713}
\emmoveto{65.380}{65.703}
\emlineto{65.577}{65.635}
\emmoveto{65.577}{65.625}
\emlineto{65.775}{65.556}
\emmoveto{65.775}{65.546}
\emlineto{65.972}{65.478}
\emmoveto{65.972}{65.468}
\emlineto{66.169}{65.399}
\emmoveto{66.169}{65.389}
\emlineto{66.365}{65.320}
\emmoveto{66.365}{65.310}
\emlineto{66.561}{65.242}
\emmoveto{66.561}{65.232}
\emlineto{66.757}{65.163}
\emmoveto{66.757}{65.153}
\emlineto{66.952}{65.084}
\emmoveto{66.952}{65.074}
\emlineto{67.146}{65.006}
\emmoveto{67.146}{64.996}
\emlineto{67.341}{64.927}
\emmoveto{67.341}{64.917}
\emlineto{67.535}{64.848}
\emmoveto{67.535}{64.838}
\emlineto{67.728}{64.769}
\emmoveto{67.728}{64.759}
\emlineto{67.921}{64.690}
\emmoveto{67.921}{64.680}
\emlineto{68.114}{64.611}
\emmoveto{68.114}{64.601}
\emlineto{68.306}{64.532}
\emmoveto{68.306}{64.522}
\emlineto{68.498}{64.454}
\emmoveto{68.498}{64.444}
\emlineto{68.689}{64.375}
\emmoveto{68.689}{64.365}
\emlineto{68.880}{64.296}
\emmoveto{68.880}{64.286}
\emlineto{69.071}{64.217}
\emmoveto{69.071}{64.207}
\emlineto{69.261}{64.138}
\emmoveto{69.261}{64.128}
\emlineto{69.451}{64.059}
\emmoveto{69.451}{64.049}
\emlineto{69.640}{63.980}
\emmoveto{69.640}{63.970}
\emlineto{69.829}{63.901}
\emmoveto{69.829}{63.891}
\emlineto{70.018}{63.821}
\emmoveto{70.018}{63.811}
\emlineto{70.206}{63.742}
\emmoveto{70.206}{63.732}
\emlineto{70.393}{63.663}
\emmoveto{70.393}{63.653}
\emlineto{70.581}{63.584}
\emmoveto{70.581}{63.574}
\emlineto{70.768}{63.505}
\emmoveto{70.768}{63.495}
\emlineto{70.954}{63.426}
\emmoveto{70.954}{63.416}
\emlineto{71.140}{63.347}
\emmoveto{71.140}{63.337}
\emlineto{71.326}{63.267}
\emmoveto{71.326}{63.257}
\emlineto{71.511}{63.188}
\emmoveto{71.511}{63.178}
\emlineto{71.696}{63.109}
\emmoveto{71.696}{63.099}
\emlineto{71.880}{63.029}
\emmoveto{71.880}{63.019}
\emlineto{72.065}{62.950}
\emmoveto{72.065}{62.940}
\emlineto{72.248}{62.871}
\emmoveto{72.248}{62.861}
\emlineto{72.431}{62.791}
\emmoveto{72.431}{62.781}
\emlineto{72.614}{62.712}
\emmoveto{72.614}{62.702}
\emlineto{72.797}{62.632}
\emmoveto{72.797}{62.622}
\emlineto{72.978}{62.553}
\emmoveto{72.978}{62.543}
\emlineto{73.160}{62.473}
\emmoveto{73.160}{62.463}
\emlineto{73.341}{62.394}
\emmoveto{73.341}{62.384}
\emlineto{73.522}{62.314}
\emmoveto{73.522}{62.304}
\emlineto{73.702}{62.235}
\emmoveto{73.702}{62.225}
\emlineto{73.882}{62.155}
\emmoveto{73.882}{62.145}
\emlineto{74.062}{62.075}
\emmoveto{74.062}{62.065}
\emlineto{74.241}{61.996}
\emmoveto{74.241}{61.986}
\emlineto{74.419}{61.916}
\emmoveto{74.419}{61.906}
\emlineto{74.598}{61.837}
\emmoveto{74.598}{61.827}
\emlineto{74.775}{61.757}
\emmoveto{74.775}{61.747}
\emlineto{74.953}{61.677}
\emmoveto{74.953}{61.667}
\emlineto{75.130}{61.598}
\emmoveto{75.130}{61.588}
\emlineto{75.306}{61.518}
\emmoveto{75.306}{61.508}
\emlineto{75.483}{61.438}
\emmoveto{75.483}{61.428}
\emlineto{75.658}{61.358}
\emmoveto{75.658}{61.348}
\emlineto{75.834}{61.278}
\emmoveto{75.834}{61.268}
\emlineto{76.009}{61.199}
\emmoveto{76.009}{61.189}
\emlineto{76.183}{61.119}
\emmoveto{76.183}{61.109}
\emlineto{76.357}{61.039}
\emmoveto{76.357}{61.029}
\emlineto{76.531}{60.959}
\emmoveto{76.531}{60.949}
\emlineto{76.704}{60.879}
\emmoveto{76.704}{60.869}
\emlineto{76.877}{60.799}
\emmoveto{76.877}{60.789}
\emlineto{77.050}{60.719}
\emmoveto{77.050}{60.709}
\emlineto{77.222}{60.639}
\emmoveto{77.222}{60.629}
\emlineto{77.393}{60.559}
\emmoveto{77.393}{60.549}
\emlineto{77.565}{60.479}
\emmoveto{77.565}{60.469}
\emlineto{77.735}{60.399}
\emmoveto{77.735}{60.389}
\emlineto{77.906}{60.319}
\emmoveto{77.906}{60.309}
\emlineto{78.076}{60.239}
\emmoveto{78.076}{60.229}
\emlineto{78.245}{60.159}
\emmoveto{78.245}{60.149}
\emlineto{78.414}{60.078}
\emmoveto{78.414}{60.068}
\emlineto{78.583}{59.998}
\emmoveto{78.583}{59.988}
\emlineto{78.751}{59.918}
\emmoveto{78.751}{59.908}
\emlineto{78.919}{59.838}
\emmoveto{78.919}{59.828}
\emlineto{79.087}{59.758}
\emmoveto{79.087}{59.748}
\emlineto{79.254}{59.677}
\emmoveto{79.254}{59.667}
\emlineto{79.420}{59.597}
\emmoveto{79.420}{59.587}
\emlineto{79.587}{59.517}
\emmoveto{79.587}{59.507}
\emlineto{79.753}{59.436}
\emmoveto{79.753}{59.426}
\emlineto{79.918}{59.356}
\emmoveto{79.918}{59.346}
\emlineto{80.083}{59.276}
\emmoveto{80.083}{59.266}
\emlineto{80.247}{59.196}
\emmoveto{80.247}{59.186}
\emlineto{80.411}{59.115}
\emmoveto{80.411}{59.105}
\emlineto{80.575}{59.035}
\emmoveto{80.575}{59.025}
\emlineto{80.739}{58.954}
\emmoveto{80.739}{58.944}
\emlineto{80.901}{58.874}
\emmoveto{80.901}{58.864}
\emlineto{81.064}{58.793}
\emmoveto{81.064}{58.783}
\emlineto{81.226}{58.713}
\emmoveto{81.226}{58.703}
\emlineto{81.388}{58.632}
\emmoveto{81.388}{58.622}
\emlineto{81.549}{58.552}
\emmoveto{81.549}{58.542}
\emlineto{81.710}{58.471}
\emmoveto{81.710}{58.461}
\emlineto{81.870}{58.391}
\emmoveto{81.870}{58.381}
\emlineto{82.030}{58.310}
\emmoveto{82.030}{58.300}
\emlineto{82.189}{58.229}
\emmoveto{82.189}{58.219}
\emlineto{82.349}{58.149}
\emmoveto{82.349}{58.139}
\emlineto{82.507}{58.068}
\emmoveto{82.507}{58.058}
\emlineto{82.666}{57.987}
\emmoveto{82.666}{57.977}
\emlineto{82.824}{57.907}
\emmoveto{82.824}{57.897}
\emlineto{82.981}{57.826}
\emmoveto{82.981}{57.816}
\emlineto{83.138}{57.745}
\emmoveto{83.138}{57.735}
\emlineto{83.295}{57.664}
\emmoveto{83.295}{57.654}
\emlineto{83.451}{57.584}
\emmoveto{83.451}{57.574}
\emlineto{83.607}{57.503}
\emmoveto{83.607}{57.493}
\emlineto{83.762}{57.422}
\emmoveto{83.762}{57.412}
\emlineto{83.917}{57.341}
\emmoveto{83.917}{57.331}
\emlineto{84.071}{57.260}
\emmoveto{84.071}{57.250}
\emlineto{84.226}{57.180}
\emmoveto{84.226}{57.170}
\emlineto{84.379}{57.098}
\emmoveto{84.379}{57.088}
\emlineto{84.532}{57.018}
\emmoveto{84.532}{57.008}
\emlineto{84.685}{56.937}
\emmoveto{84.685}{56.927}
\emlineto{84.838}{56.856}
\emmoveto{84.838}{56.846}
\emlineto{84.990}{56.775}
\emmoveto{84.990}{56.765}
\emlineto{85.141}{56.694}
\emmoveto{85.141}{56.684}
\emlineto{85.292}{56.613}
\emmoveto{85.292}{56.603}
\emlineto{85.443}{56.532}
\emmoveto{85.443}{56.522}
\emlineto{85.594}{56.451}
\emmoveto{85.594}{56.441}
\emlineto{85.743}{56.370}
\emmoveto{85.743}{56.360}
\emlineto{85.893}{56.288}
\emmoveto{85.893}{56.278}
\emlineto{86.042}{56.207}
\emmoveto{86.042}{56.197}
\emlineto{86.191}{56.126}
\emmoveto{86.191}{56.116}
\emlineto{86.339}{56.045}
\emmoveto{86.339}{56.035}
\emlineto{86.487}{55.964}
\emmoveto{86.487}{55.954}
\emlineto{86.634}{55.883}
\emmoveto{86.634}{55.873}
\emlineto{86.781}{55.802}
\emmoveto{86.781}{55.792}
\emlineto{86.927}{55.720}
\emmoveto{86.927}{55.710}
\emlineto{87.074}{55.639}
\emmoveto{87.074}{55.629}
\emlineto{87.219}{55.558}
\emmoveto{87.219}{55.548}
\emlineto{87.365}{55.476}
\emmoveto{87.365}{55.466}
\emlineto{87.509}{55.395}
\emmoveto{87.509}{55.385}
\emlineto{87.654}{55.314}
\emmoveto{87.654}{55.304}
\emlineto{87.798}{55.232}
\emmoveto{87.798}{55.222}
\emlineto{87.941}{55.151}
\emmoveto{87.941}{55.141}
\emlineto{88.084}{55.070}
\emmoveto{88.084}{55.060}
\emlineto{88.227}{54.988}
\emmoveto{88.227}{54.978}
\emlineto{88.370}{54.907}
\emmoveto{88.370}{54.897}
\emlineto{88.511}{54.826}
\emmoveto{88.511}{54.816}
\emlineto{88.653}{54.744}
\emmoveto{88.653}{54.734}
\emlineto{88.794}{54.663}
\emmoveto{88.794}{54.653}
\emlineto{88.935}{54.581}
\emmoveto{88.935}{54.571}
\emlineto{89.075}{54.500}
\emmoveto{89.075}{54.490}
\emlineto{89.215}{54.418}
\emmoveto{89.215}{54.408}
\emlineto{89.354}{54.337}
\emmoveto{89.354}{54.327}
\emlineto{89.493}{54.255}
\emmoveto{89.493}{54.245}
\emlineto{89.631}{54.174}
\emmoveto{89.631}{54.164}
\emlineto{89.769}{54.092}
\emmoveto{89.769}{54.082}
\emlineto{89.907}{54.010}
\emmoveto{89.907}{54.000}
\emlineto{90.044}{53.929}
\emmoveto{90.044}{53.919}
\emlineto{90.181}{53.847}
\emmoveto{90.181}{53.837}
\emlineto{90.318}{53.765}
\emmoveto{90.318}{53.755}
\emlineto{90.454}{53.684}
\emmoveto{90.454}{53.674}
\emlineto{90.589}{53.602}
\emmoveto{90.589}{53.592}
\emlineto{90.724}{53.520}
\emmoveto{90.724}{53.510}
\emlineto{90.859}{53.439}
\emmoveto{90.859}{53.429}
\emlineto{90.993}{53.357}
\emmoveto{90.993}{53.347}
\emlineto{91.127}{53.275}
\emmoveto{91.127}{53.265}
\emlineto{91.261}{53.193}
\emmoveto{91.261}{53.183}
\emlineto{91.393}{53.111}
\emmoveto{91.393}{53.101}
\emlineto{91.526}{53.030}
\emmoveto{91.526}{53.020}
\emlineto{91.658}{52.948}
\emmoveto{91.658}{52.938}
\emlineto{91.790}{52.866}
\emmoveto{91.790}{52.856}
\emlineto{91.921}{52.784}
\emmoveto{91.921}{52.774}
\emlineto{92.052}{52.702}
\emmoveto{92.052}{52.692}
\emlineto{92.183}{52.620}
\emmoveto{92.183}{52.610}
\emlineto{92.313}{52.538}
\emmoveto{92.313}{52.528}
\emlineto{92.442}{52.456}
\emmoveto{92.442}{52.446}
\emlineto{92.571}{52.375}
\emmoveto{92.571}{52.365}
\emlineto{92.700}{52.292}
\emmoveto{92.700}{52.282}
\emlineto{92.828}{52.211}
\emmoveto{92.828}{52.201}
\emlineto{92.956}{52.128}
\emmoveto{92.956}{52.118}
\emlineto{93.084}{52.046}
\emmoveto{93.084}{52.036}
\emlineto{93.211}{51.965}
\emmoveto{93.211}{51.955}
\emlineto{93.337}{51.882}
\emmoveto{93.337}{51.872}
\emlineto{93.464}{51.800}
\emmoveto{93.464}{51.790}
\emlineto{93.589}{51.718}
\emmoveto{93.589}{51.708}
\emlineto{93.715}{51.636}
\emmoveto{93.715}{51.626}
\emlineto{93.840}{51.554}
\emmoveto{93.840}{51.544}
\emlineto{93.964}{51.472}
\emmoveto{93.964}{51.462}
\emlineto{94.088}{51.390}
\emmoveto{94.088}{51.380}
\emlineto{94.212}{51.308}
\emmoveto{94.212}{51.298}
\emlineto{94.335}{51.225}
\emmoveto{94.335}{51.215}
\emlineto{94.458}{51.143}
\emmoveto{94.458}{51.133}
\emlineto{94.580}{51.061}
\emmoveto{94.580}{51.051}
\emlineto{94.702}{50.979}
\emmoveto{94.702}{50.969}
\emlineto{94.824}{50.896}
\emmoveto{94.824}{50.886}
\emlineto{94.945}{50.814}
\emmoveto{94.945}{50.804}
\emlineto{95.066}{50.732}
\emmoveto{95.066}{50.722}
\emlineto{95.186}{50.650}
\emmoveto{95.186}{50.640}
\emlineto{95.306}{50.567}
\emmoveto{95.306}{50.557}
\emlineto{95.425}{50.485}
\emmoveto{95.425}{50.475}
\emlineto{95.544}{50.403}
\emmoveto{95.544}{50.393}
\emlineto{95.662}{50.320}
\emmoveto{95.662}{50.310}
\emlineto{95.781}{50.238}
\emmoveto{95.781}{50.228}
\emlineto{95.898}{50.156}
\emmoveto{95.898}{50.146}
\emlineto{96.016}{50.073}
\emmoveto{96.016}{50.063}
\emlineto{96.132}{49.991}
\emmoveto{96.132}{49.981}
\emlineto{96.249}{49.908}
\emmoveto{96.249}{49.898}
\emlineto{96.365}{49.826}
\emmoveto{96.365}{49.816}
\emlineto{96.480}{49.743}
\emmoveto{96.480}{49.733}
\emlineto{96.595}{49.661}
\emmoveto{96.595}{49.651}
\emlineto{96.710}{49.578}
\emmoveto{96.710}{49.568}
\emlineto{96.824}{49.496}
\emmoveto{96.824}{49.486}
\emlineto{96.938}{49.413}
\emmoveto{96.938}{49.403}
\emlineto{97.051}{49.331}
\emmoveto{97.051}{49.321}
\emlineto{97.164}{49.248}
\emmoveto{97.164}{49.238}
\emlineto{97.277}{49.166}
\emmoveto{97.277}{49.156}
\emlineto{97.389}{49.083}
\emmoveto{97.389}{49.073}
\emlineto{97.501}{49.001}
\emmoveto{97.501}{48.991}
\emlineto{97.612}{48.918}
\emmoveto{97.612}{48.908}
\emlineto{97.723}{48.835}
\emmoveto{97.723}{48.825}
\emlineto{97.833}{48.753}
\emmoveto{97.833}{48.743}
\emlineto{97.943}{48.670}
\emmoveto{97.943}{48.660}
\emlineto{98.053}{48.587}
\emmoveto{98.162}{48.495}
\emlineto{98.379}{48.339}
\emmoveto{98.487}{48.247}
\emlineto{98.701}{48.091}
\emmoveto{98.808}{47.998}
\emlineto{99.019}{47.843}
\emmoveto{99.125}{47.750}
\emlineto{99.334}{47.594}
\emmoveto{99.438}{47.501}
\emlineto{99.644}{47.346}
\emmoveto{99.747}{47.253}
\emlineto{99.951}{47.097}
\emmoveto{100.052}{47.004}
\emlineto{100.254}{46.848}
\emmoveto{100.354}{46.755}
\emlineto{100.553}{46.599}
\emmoveto{100.652}{46.506}
\emlineto{100.848}{46.350}
\emmoveto{100.945}{46.257}
\emlineto{101.139}{46.101}
\emmoveto{101.235}{46.008}
\emlineto{101.427}{45.852}
\emmoveto{101.521}{45.758}
\emlineto{101.710}{45.602}
\emmoveto{101.804}{45.509}
\emlineto{101.990}{45.353}
\emmoveto{102.082}{45.260}
\emlineto{102.265}{45.103}
\emmoveto{102.356}{45.010}
\emlineto{102.537}{44.853}
\emmoveto{102.627}{44.760}
\emlineto{102.805}{44.603}
\emmoveto{102.893}{44.510}
\emlineto{103.069}{44.354}
\emmoveto{103.156}{44.260}
\emlineto{103.329}{44.104}
\emmoveto{103.415}{44.010}
\emlineto{103.585}{43.853}
\emmoveto{103.670}{43.760}
\emlineto{103.838}{43.603}
\emmoveto{103.921}{43.510}
\emlineto{104.086}{43.353}
\emmoveto{104.168}{43.259}
\emlineto{104.331}{43.102}
\emmoveto{104.412}{43.009}
\emlineto{104.572}{42.852}
\emmoveto{104.651}{42.758}
\emlineto{104.808}{42.601}
\emmoveto{104.886}{42.507}
\emlineto{105.041}{42.350}
\emmoveto{105.118}{42.257}
\emlineto{105.270}{42.099}
\emmoveto{105.346}{42.006}
\emlineto{105.495}{41.848}
\emmoveto{105.569}{41.755}
\emlineto{105.716}{41.598}
\emmoveto{105.789}{41.504}
\emlineto{105.934}{41.346}
\emmoveto{106.005}{41.253}
\emlineto{106.147}{41.095}
\emmoveto{106.217}{41.001}
\emlineto{106.356}{40.844}
\emmoveto{106.425}{40.750}
\emlineto{106.562}{40.592}
\emmoveto{106.630}{40.499}
\emlineto{106.764}{40.341}
\emmoveto{106.830}{40.247}
\emlineto{106.961}{40.090}
\emmoveto{107.026}{39.996}
\emlineto{107.155}{39.838}
\emmoveto{107.219}{39.744}
\emlineto{107.345}{39.586}
\emmoveto{107.407}{39.492}
\emlineto{107.531}{39.335}
\emmoveto{107.592}{39.241}
\emlineto{107.713}{39.083}
\emmoveto{107.773}{38.989}
\emlineto{107.891}{38.831}
\emmoveto{107.950}{38.737}
\emlineto{108.065}{38.579}
\emmoveto{108.122}{38.485}
\emlineto{108.235}{38.327}
\emmoveto{108.291}{38.233}
\emlineto{108.402}{38.075}
\emmoveto{108.456}{37.981}
\emlineto{108.564}{37.823}
\emmoveto{108.617}{37.728}
\emlineto{108.723}{37.570}
\emmoveto{108.775}{37.476}
\emlineto{108.877}{37.318}
\emmoveto{108.928}{37.224}
\emlineto{109.028}{37.066}
\emmoveto{109.077}{36.972}
\emlineto{109.175}{36.813}
\emmoveto{109.223}{36.719}
\emlineto{109.318}{36.561}
\emmoveto{109.364}{36.467}
\emlineto{109.456}{36.308}
\emmoveto{109.502}{36.214}
\emlineto{109.591}{36.056}
\emmoveto{109.635}{35.961}
\emlineto{109.722}{35.803}
\emmoveto{109.765}{35.709}
\emlineto{109.849}{35.550}
\emmoveto{109.891}{35.456}
\emlineto{109.973}{35.298}
\emmoveto{110.013}{35.203}
\emlineto{110.092}{35.045}
\emmoveto{110.131}{34.950}
\emlineto{110.207}{34.792}
\emmoveto{110.245}{34.698}
\emlineto{110.318}{34.539}
\emmoveto{110.355}{34.445}
\emlineto{110.426}{34.286}
\emmoveto{110.461}{34.192}
\emlineto{110.529}{34.033}
\emmoveto{110.563}{33.939}
\emlineto{110.629}{33.780}
\emmoveto{110.661}{33.686}
\emlineto{110.724}{33.527}
\emmoveto{110.755}{33.433}
\emlineto{110.816}{33.274}
\emmoveto{110.846}{33.180}
\emlineto{110.904}{33.021}
\emmoveto{110.932}{32.926}
\emlineto{110.988}{32.768}
\emmoveto{111.015}{32.673}
\emlineto{111.067}{32.514}
\emmoveto{111.093}{32.420}
\emlineto{111.143}{32.261}
\emmoveto{111.168}{32.167}
\emlineto{111.215}{32.008}
\emmoveto{111.238}{31.914}
\emlineto{111.283}{31.755}
\emmoveto{111.305}{31.660}
\emlineto{111.347}{31.501}
\emmoveto{111.368}{31.407}
\emlineto{111.407}{31.248}
\emmoveto{111.427}{31.154}
\emlineto{111.464}{30.995}
\emmoveto{111.482}{30.900}
\emlineto{111.516}{30.741}
\emmoveto{111.532}{30.647}
\emlineto{111.564}{30.488}
\emmoveto{111.579}{30.393}
\emlineto{111.609}{30.234}
\emmoveto{111.622}{30.140}
\emlineto{111.649}{29.981}
\emmoveto{111.662}{29.886}
\emlineto{111.685}{29.728}
\emmoveto{111.697}{29.633}
\emlineto{111.718}{29.474}
\emmoveto{111.728}{29.380}
\emlineto{111.747}{29.221}
\emmoveto{111.755}{29.126}
\emlineto{111.771}{28.967}
\emmoveto{111.779}{28.873}
\emlineto{111.792}{28.714}
\emmoveto{111.798}{28.619}
\emlineto{111.809}{28.460}
\emmoveto{111.813}{28.366}
\emlineto{111.822}{28.207}
\emmoveto{111.825}{28.112}
\emlineto{111.830}{27.953}
\emmoveto{111.833}{27.859}
\emlineto{111.835}{27.700}
\emmoveto{111.836}{27.605}
\emlineto{111.836}{27.446}
\emmoveto{111.836}{27.352}
\emlineto{111.833}{27.193}
\emmoveto{111.831}{27.098}
\emlineto{111.826}{26.939}
\emmoveto{111.823}{26.844}
\emlineto{111.816}{26.686}
\emmoveto{111.811}{26.591}
\emlineto{111.801}{26.432}
\emmoveto{111.795}{26.337}
\emlineto{111.782}{26.178}
\emmoveto{111.775}{26.084}
\emlineto{111.759}{25.925}
\emmoveto{111.751}{25.831}
\emlineto{111.733}{25.671}
\emmoveto{111.723}{25.577}
\emlineto{111.702}{25.418}
\emmoveto{111.691}{25.324}
\emlineto{111.668}{25.165}
\emmoveto{111.655}{25.070}
\emlineto{111.629}{24.911}
\emmoveto{111.615}{24.817}
\emlineto{111.587}{24.658}
\emmoveto{111.572}{24.563}
\emlineto{111.540}{24.404}
\emmoveto{111.524}{24.310}
\emlineto{111.490}{24.151}
\emmoveto{111.472}{24.056}
\emlineto{111.436}{23.897}
\emmoveto{111.417}{23.803}
\emlineto{111.378}{23.644}
\emmoveto{111.357}{23.550}
\emlineto{111.316}{23.391}
\emmoveto{111.294}{23.296}
\emlineto{111.249}{23.138}
\emmoveto{111.227}{23.043}
\emlineto{111.179}{22.884}
\emmoveto{111.155}{22.790}
\emlineto{111.105}{22.631}
\emmoveto{111.080}{22.537}
\emlineto{111.028}{22.378}
\emmoveto{111.001}{22.283}
\emlineto{110.946}{22.125}
\emmoveto{110.918}{22.030}
\emlineto{110.860}{21.872}
\emmoveto{110.831}{21.777}
\emlineto{110.770}{21.618}
\emmoveto{110.739}{21.524}
\emlineto{110.677}{21.365}
\emmoveto{110.645}{21.271}
\emlineto{110.579}{21.112}
\emmoveto{110.546}{21.018}
\emlineto{110.478}{20.859}
\emmoveto{110.443}{20.765}
\emlineto{110.372}{20.606}
\emmoveto{110.336}{20.512}
\emlineto{110.263}{20.353}
\emmoveto{110.225}{20.259}
\emlineto{110.149}{20.101}
\emmoveto{110.111}{20.006}
\emlineto{110.032}{19.848}
\emmoveto{109.992}{19.753}
\emlineto{109.911}{19.595}
\emmoveto{109.870}{19.501}
\emlineto{109.786}{19.342}
\emmoveto{109.743}{19.248}
\emlineto{109.657}{19.090}
\emmoveto{109.613}{18.995}
\emlineto{109.524}{18.837}
\emmoveto{109.479}{18.743}
\emlineto{109.387}{18.584}
\emmoveto{109.340}{18.490}
\emlineto{109.246}{18.332}
\emmoveto{109.198}{18.238}
\emlineto{109.101}{18.079}
\emmoveto{109.052}{17.985}
\emlineto{108.953}{17.827}
\emmoveto{108.902}{17.733}
\emlineto{108.800}{17.575}
\emmoveto{108.748}{17.481}
\emlineto{108.643}{17.323}
\emmoveto{108.590}{17.228}
\emlineto{108.483}{17.070}
\emmoveto{108.428}{16.976}
\emlineto{108.318}{16.818}
\emmoveto{108.263}{16.724}
\emlineto{108.150}{16.566}
\emmoveto{108.093}{16.472}
\emlineto{107.978}{16.314}
\emmoveto{107.920}{16.220}
\emlineto{107.802}{16.062}
\emmoveto{107.742}{15.968}
\emlineto{107.622}{15.810}
\emmoveto{107.561}{15.716}
\emlineto{107.438}{15.559}
\emmoveto{107.375}{15.465}
\emlineto{107.250}{15.307}
\emmoveto{107.186}{15.213}
\emlineto{107.058}{15.055}
\emmoveto{106.993}{14.961}
\emlineto{106.862}{14.804}
\emmoveto{106.796}{14.710}
\emlineto{106.662}{14.552}
\emmoveto{106.595}{14.458}
\emlineto{106.459}{14.301}
\emmoveto{106.390}{14.207}
\emlineto{106.251}{14.049}
\emmoveto{106.181}{13.956}
\emlineto{106.040}{13.798}
\emmoveto{105.968}{13.704}
\emlineto{105.825}{13.547}
\emmoveto{105.752}{13.453}
\emlineto{105.605}{13.296}
\emmoveto{105.531}{13.202}
\emlineto{105.382}{13.045}
\emmoveto{105.307}{12.951}
\emlineto{105.155}{12.794}
\emmoveto{105.079}{12.700}
\emlineto{104.924}{12.543}
\emmoveto{104.846}{12.450}
\emlineto{104.689}{12.293}
\emmoveto{104.610}{12.199}
\emlineto{104.451}{12.042}
\emmoveto{104.370}{11.948}
\emlineto{104.208}{11.792}
\emmoveto{104.126}{11.698}
\emlineto{103.961}{11.541}
\emmoveto{103.878}{11.448}
\emlineto{103.711}{11.291}
\emmoveto{103.627}{11.197}
\emlineto{103.457}{11.041}
\emmoveto{103.371}{10.947}
\emlineto{103.198}{10.791}
\emmoveto{103.112}{10.697}
\emlineto{102.936}{10.540}
\emshow{83.980}{31.700}{Classical particle}
\emmoveto{12.000}{80.000}
\emlineto{12.163}{80.010}
\emmoveto{12.245}{79.999}
\emlineto{12.408}{80.009}
\emmoveto{12.490}{79.998}
\emlineto{12.654}{80.007}
\emmoveto{12.735}{79.996}
\emlineto{12.899}{80.004}
\emmoveto{12.980}{79.993}
\emlineto{13.144}{80.000}
\emmoveto{13.225}{79.989}
\emlineto{13.389}{79.996}
\emmoveto{13.470}{79.984}
\emlineto{13.634}{79.991}
\emmoveto{13.715}{79.979}
\emlineto{13.879}{79.984}
\emmoveto{13.960}{79.972}
\emlineto{14.124}{79.978}
\emmoveto{14.205}{79.965}
\emlineto{14.368}{79.970}
\emmoveto{14.450}{79.957}
\emlineto{14.613}{79.961}
\emmoveto{14.695}{79.948}
\emlineto{14.858}{79.952}
\emmoveto{14.940}{79.939}
\emlineto{15.103}{79.942}
\emmoveto{15.185}{79.928}
\emlineto{15.348}{79.931}
\emmoveto{15.429}{79.917}
\emlineto{15.592}{79.919}
\emmoveto{15.674}{79.905}
\emlineto{15.837}{79.907}
\emmoveto{15.919}{79.892}
\emlineto{16.082}{79.893}
\emmoveto{16.163}{79.879}
\emlineto{16.326}{79.879}
\emmoveto{16.408}{79.865}
\emlineto{16.571}{79.865}
\emmoveto{16.652}{79.850}
\emlineto{16.815}{79.850}
\emmoveto{16.896}{79.834}
\emlineto{17.059}{79.833}
\emmoveto{17.141}{79.818}
\emlineto{17.303}{79.816}
\emmoveto{17.385}{79.801}
\emlineto{17.547}{79.799}
\emmoveto{17.629}{79.783}
\emlineto{17.792}{79.781}
\emmoveto{17.873}{79.764}
\emlineto{18.036}{79.762}
\emmoveto{18.117}{79.745}
\emlineto{18.279}{79.742}
\emmoveto{18.361}{79.725}
\emlineto{18.523}{79.722}
\emmoveto{18.604}{79.705}
\emlineto{18.767}{79.701}
\emmoveto{18.848}{79.684}
\emlineto{19.010}{79.679}
\emmoveto{19.092}{79.662}
\emlineto{19.254}{79.657}
\emmoveto{19.335}{79.639}
\emlineto{19.497}{79.634}
\emmoveto{19.578}{79.616}
\emlineto{19.741}{79.610}
\emmoveto{19.822}{79.592}
\emlineto{19.984}{79.585}
\emmoveto{20.065}{79.567}
\emlineto{20.227}{79.560}
\emmoveto{20.308}{79.542}
\emlineto{20.470}{79.534}
\emmoveto{20.551}{79.515}
\emlineto{20.712}{79.508}
\emmoveto{20.793}{79.488}
\emlineto{20.955}{79.480}
\emmoveto{21.036}{79.461}
\emlineto{21.198}{79.452}
\emmoveto{21.278}{79.433}
\emlineto{21.440}{79.423}
\emmoveto{21.521}{79.404}
\emlineto{21.682}{79.394}
\emmoveto{21.763}{79.374}
\emlineto{21.924}{79.364}
\emmoveto{22.005}{79.344}
\emlineto{22.166}{79.333}
\emmoveto{22.247}{79.313}
\emlineto{22.408}{79.301}
\emmoveto{22.489}{79.281}
\emlineto{22.650}{79.269}
\emmoveto{22.730}{79.248}
\emlineto{22.891}{79.236}
\emmoveto{22.972}{79.215}
\emlineto{23.133}{79.203}
\emmoveto{23.213}{79.181}
\emlineto{23.374}{79.168}
\emmoveto{23.455}{79.147}
\emlineto{23.615}{79.133}
\emmoveto{23.696}{79.111}
\emlineto{23.856}{79.097}
\emmoveto{23.936}{79.075}
\emlineto{24.097}{79.061}
\emmoveto{24.177}{79.039}
\emlineto{24.337}{79.024}
\emmoveto{24.418}{79.001}
\emlineto{24.578}{78.986}
\emmoveto{24.658}{78.963}
\emlineto{24.818}{78.947}
\emmoveto{24.898}{78.924}
\emlineto{25.058}{78.908}
\emmoveto{25.138}{78.885}
\emlineto{25.298}{78.868}
\emmoveto{25.378}{78.844}
\emlineto{25.538}{78.827}
\emmoveto{25.617}{78.803}
\emlineto{25.777}{78.786}
\emmoveto{25.857}{78.762}
\emlineto{26.016}{78.744}
\emmoveto{26.096}{78.719}
\emlineto{26.255}{78.701}
\emmoveto{26.335}{78.676}
\emlineto{26.494}{78.657}
\emmoveto{26.574}{78.633}
\emlineto{26.733}{78.613}
\emmoveto{26.812}{78.588}
\emlineto{26.971}{78.568}
\emmoveto{27.051}{78.543}
\emlineto{27.209}{78.522}
\emmoveto{27.289}{78.497}
\emlineto{27.447}{78.476}
\emmoveto{27.527}{78.450}
\emlineto{27.685}{78.429}
\emmoveto{27.765}{78.403}
\emlineto{27.923}{78.381}
\emmoveto{28.002}{78.355}
\emlineto{28.160}{78.333}
\emmoveto{28.239}{78.306}
\emlineto{28.397}{78.284}
\emmoveto{28.476}{78.257}
\emlineto{28.634}{78.234}
\emmoveto{28.713}{78.207}
\emlineto{28.871}{78.183}
\emmoveto{28.950}{78.156}
\emlineto{29.107}{78.132}
\emmoveto{29.186}{78.105}
\emlineto{29.344}{78.080}
\emmoveto{29.422}{78.052}
\emlineto{29.580}{78.027}
\emmoveto{29.658}{78.000}
\emlineto{29.815}{77.974}
\emmoveto{29.894}{77.946}
\emlineto{30.051}{77.920}
\emmoveto{30.129}{77.892}
\emlineto{30.286}{77.865}
\emmoveto{30.364}{77.837}
\emlineto{30.521}{77.811}
\emmoveto{30.599}{77.783}
\emlineto{30.755}{77.756}
\emmoveto{30.834}{77.728}
\emlineto{30.990}{77.702}
\emmoveto{31.068}{77.674}
\emlineto{31.224}{77.648}
\emmoveto{31.302}{77.620}
\emlineto{31.458}{77.594}
\emmoveto{31.536}{77.566}
\emlineto{31.692}{77.540}
\emmoveto{31.770}{77.512}
\emlineto{31.925}{77.486}
\emmoveto{32.003}{77.458}
\emlineto{32.158}{77.432}
\emmoveto{32.236}{77.405}
\emlineto{32.391}{77.379}
\emmoveto{32.469}{77.351}
\emlineto{32.624}{77.325}
\emmoveto{32.701}{77.297}
\emlineto{32.856}{77.272}
\emmoveto{32.934}{77.244}
\emlineto{33.088}{77.218}
\emmoveto{33.166}{77.191}
\emlineto{33.320}{77.165}
\emmoveto{33.398}{77.137}
\emlineto{33.552}{77.112}
\emmoveto{33.629}{77.084}
\emlineto{33.783}{77.058}
\emmoveto{33.861}{77.031}
\emlineto{34.015}{77.005}
\emmoveto{34.092}{76.978}
\emlineto{34.246}{76.952}
\emmoveto{34.323}{76.924}
\emlineto{34.476}{76.899}
\emmoveto{34.553}{76.871}
\emlineto{34.707}{76.846}
\emmoveto{34.783}{76.818}
\emlineto{34.937}{76.792}
\emmoveto{35.014}{76.765}
\emlineto{35.167}{76.739}
\emmoveto{35.243}{76.711}
\emlineto{35.396}{76.686}
\emmoveto{35.473}{76.658}
\emlineto{35.626}{76.633}
\emmoveto{35.702}{76.605}
\emlineto{35.855}{76.579}
\emmoveto{35.931}{76.552}
\emlineto{36.084}{76.526}
\emmoveto{36.160}{76.498}
\emlineto{36.313}{76.473}
\emmoveto{36.389}{76.445}
\emlineto{36.541}{76.419}
\emmoveto{36.617}{76.392}
\emlineto{36.769}{76.366}
\emmoveto{36.845}{76.338}
\emlineto{36.997}{76.313}
\emmoveto{37.073}{76.285}
\emlineto{37.225}{76.259}
\emmoveto{37.301}{76.231}
\emlineto{37.452}{76.206}
\emmoveto{37.528}{76.178}
\emlineto{37.680}{76.152}
\emmoveto{37.755}{76.124}
\emlineto{37.907}{76.099}
\emmoveto{37.982}{76.071}
\emlineto{38.133}{76.045}
\emmoveto{38.209}{76.017}
\emlineto{38.360}{75.992}
\emmoveto{38.435}{75.964}
\emlineto{38.586}{75.938}
\emmoveto{38.661}{75.910}
\emlineto{38.812}{75.885}
\emmoveto{38.887}{75.857}
\emlineto{39.038}{75.831}
\emmoveto{39.113}{75.803}
\emlineto{39.263}{75.777}
\emmoveto{39.338}{75.750}
\emlineto{39.488}{75.724}
\emmoveto{39.563}{75.696}
\emlineto{39.713}{75.670}
\emmoveto{39.788}{75.642}
\emlineto{39.938}{75.617}
\emmoveto{40.013}{75.589}
\emlineto{40.162}{75.563}
\emmoveto{40.237}{75.535}
\emlineto{40.386}{75.509}
\emmoveto{40.461}{75.481}
\emlineto{40.610}{75.456}
\emmoveto{40.685}{75.428}
\emlineto{40.834}{75.402}
\emmoveto{40.909}{75.374}
\emlineto{41.058}{75.348}
\emmoveto{41.132}{75.320}
\emlineto{41.281}{75.294}
\emmoveto{41.355}{75.267}
\emlineto{41.504}{75.241}
\emmoveto{41.578}{75.213}
\emlineto{41.726}{75.187}
\emmoveto{41.801}{75.159}
\emlineto{41.949}{75.133}
\emmoveto{42.023}{75.105}
\emlineto{42.171}{75.080}
\emmoveto{42.245}{75.052}
\emlineto{42.393}{75.026}
\emmoveto{42.467}{74.998}
\emlineto{42.615}{74.972}
\emmoveto{42.688}{74.944}
\emlineto{42.836}{74.918}
\emmoveto{42.910}{74.890}
\emlineto{43.057}{74.864}
\emmoveto{43.131}{74.836}
\emlineto{43.278}{74.810}
\emmoveto{43.352}{74.783}
\emlineto{43.499}{74.757}
\emmoveto{43.572}{74.729}
\emlineto{43.719}{74.703}
\emmoveto{43.793}{74.675}
\emlineto{43.939}{74.649}
\emmoveto{44.013}{74.621}
\emlineto{44.159}{74.595}
\emmoveto{44.233}{74.567}
\emlineto{44.379}{74.541}
\emmoveto{44.452}{74.513}
\emlineto{44.598}{74.487}
\emmoveto{44.671}{74.459}
\emlineto{44.818}{74.433}
\emmoveto{44.891}{74.405}
\emlineto{45.037}{74.379}
\emmoveto{45.109}{74.351}
\emlineto{45.255}{74.325}
\emmoveto{45.328}{74.297}
\emlineto{45.474}{74.271}
\emmoveto{45.546}{74.243}
\emlineto{45.692}{74.217}
\emmoveto{45.764}{74.189}
\emlineto{45.910}{74.163}
\emmoveto{45.982}{74.135}
\emlineto{46.127}{74.109}
\emmoveto{46.200}{74.081}
\emlineto{46.345}{74.055}
\emmoveto{46.417}{74.027}
\emlineto{46.562}{74.001}
\emmoveto{46.634}{73.973}
\emlineto{46.779}{73.947}
\emmoveto{46.851}{73.919}
\emlineto{46.995}{73.893}
\emmoveto{47.068}{73.865}
\emlineto{47.212}{73.839}
\emmoveto{47.284}{73.811}
\emlineto{47.428}{73.785}
\emmoveto{47.500}{73.757}
\emlineto{47.644}{73.731}
\emmoveto{47.716}{73.703}
\emlineto{47.859}{73.677}
\emmoveto{47.931}{73.649}
\emlineto{48.075}{73.623}
\emmoveto{48.147}{73.595}
\emlineto{48.290}{73.568}
\emmoveto{48.362}{73.540}
\emlineto{48.505}{73.514}
\emmoveto{48.576}{73.486}
\emlineto{48.720}{73.460}
\emmoveto{48.791}{73.432}
\emlineto{48.934}{73.406}
\emmoveto{49.005}{73.378}
\emlineto{49.148}{73.352}
\emmoveto{49.219}{73.324}
\emlineto{49.362}{73.298}
\emmoveto{49.433}{73.269}
\emlineto{49.576}{73.243}
\emmoveto{49.647}{73.215}
\emlineto{49.789}{73.189}
\emmoveto{49.860}{73.161}
\emlineto{50.002}{73.135}
\emmoveto{50.073}{73.107}
\emlineto{50.215}{73.081}
\emmoveto{50.286}{73.052}
\emlineto{50.427}{73.026}
\emmoveto{50.498}{72.998}
\emlineto{50.640}{72.972}
\emmoveto{50.711}{72.944}
\emlineto{50.852}{72.918}
\emmoveto{50.923}{72.889}
\emlineto{51.064}{72.863}
\emmoveto{51.134}{72.835}
\emlineto{51.275}{72.809}
\emmoveto{51.346}{72.781}
\emlineto{51.487}{72.755}
\emmoveto{51.557}{72.726}
\emlineto{51.698}{72.700}
\emmoveto{51.768}{72.672}
\emlineto{51.909}{72.646}
\emmoveto{51.979}{72.618}
\emlineto{52.119}{72.591}
\emmoveto{52.189}{72.563}
\emlineto{52.329}{72.537}
\emmoveto{52.399}{72.509}
\emlineto{52.540}{72.483}
\emmoveto{52.610}{72.454}
\emlineto{52.749}{72.428}
\emmoveto{52.819}{72.400}
\emlineto{52.959}{72.374}
\emmoveto{53.029}{72.346}
\emlineto{53.168}{72.319}
\emmoveto{53.238}{72.291}
\emlineto{53.377}{72.265}
\emmoveto{53.447}{72.237}
\emlineto{53.586}{72.210}
\emmoveto{53.656}{72.182}
\emlineto{53.795}{72.156}
\emmoveto{53.864}{72.128}
\emlineto{54.003}{72.101}
\emmoveto{54.072}{72.073}
\emlineto{54.211}{72.047}
\emmoveto{54.280}{72.019}
\emlineto{54.419}{71.992}
\emmoveto{54.488}{71.964}
\emlineto{54.626}{71.938}
\emmoveto{54.695}{71.909}
\emlineto{54.833}{71.883}
\emmoveto{54.902}{71.855}
\emlineto{55.041}{71.828}
\emmoveto{55.109}{71.800}
\emlineto{55.247}{71.774}
\emmoveto{55.316}{71.746}
\emlineto{55.454}{71.719}
\emmoveto{55.522}{71.691}
\emlineto{55.660}{71.665}
\emmoveto{55.729}{71.636}
\emlineto{55.866}{71.610}
\emmoveto{55.935}{71.582}
\emlineto{56.072}{71.555}
\emmoveto{56.140}{71.527}
\emlineto{56.277}{71.501}
\emmoveto{56.346}{71.472}
\emlineto{56.482}{71.446}
\emmoveto{56.551}{71.418}
\emlineto{56.687}{71.391}
\emmoveto{56.756}{71.363}
\emlineto{56.892}{71.337}
\emmoveto{56.960}{71.308}
\emlineto{57.097}{71.282}
\emmoveto{57.165}{71.254}
\emlineto{57.301}{71.227}
\emmoveto{57.369}{71.199}
\emlineto{57.505}{71.173}
\emmoveto{57.573}{71.144}
\emlineto{57.708}{71.118}
\emmoveto{57.776}{71.090}
\emlineto{57.912}{71.063}
\emmoveto{57.980}{71.035}
\emlineto{58.115}{71.008}
\emmoveto{58.183}{70.980}
\emlineto{58.318}{70.953}
\emmoveto{58.386}{70.925}
\emlineto{58.521}{70.899}
\emmoveto{58.588}{70.870}
\emlineto{58.723}{70.844}
\emmoveto{58.790}{70.815}
\emlineto{58.925}{70.789}
\emmoveto{58.993}{70.761}
\emlineto{59.127}{70.734}
\emmoveto{59.194}{70.706}
\emlineto{59.329}{70.679}
\emmoveto{59.396}{70.651}
\emlineto{59.530}{70.624}
\emmoveto{59.597}{70.596}
\emlineto{59.731}{70.569}
\emmoveto{59.798}{70.541}
\emlineto{59.932}{70.515}
\emmoveto{59.999}{70.486}
\emlineto{60.133}{70.460}
\emmoveto{60.200}{70.431}
\emlineto{60.333}{70.405}
\emmoveto{60.400}{70.376}
\emlineto{60.533}{70.350}
\emmoveto{60.600}{70.321}
\emlineto{60.733}{70.295}
\emmoveto{60.800}{70.266}
\emlineto{60.933}{70.240}
\emmoveto{60.999}{70.212}
\emlineto{61.132}{70.185}
\emmoveto{61.198}{70.157}
\emlineto{61.331}{70.130}
\emmoveto{61.397}{70.102}
\emlineto{61.530}{70.075}
\emmoveto{61.596}{70.047}
\emlineto{61.729}{70.020}
\emmoveto{61.795}{69.992}
\emlineto{61.993}{69.947}
\emmoveto{62.125}{69.900}
\emlineto{62.323}{69.855}
\emmoveto{62.454}{69.808}
\emlineto{62.652}{69.763}
\emmoveto{62.783}{69.716}
\emlineto{62.980}{69.671}
\emmoveto{63.111}{69.624}
\emlineto{63.308}{69.579}
\emmoveto{63.439}{69.533}
\emlineto{63.635}{69.487}
\emmoveto{63.765}{69.441}
\emlineto{63.961}{69.396}
\emmoveto{64.091}{69.349}
\emlineto{64.287}{69.304}
\emmoveto{64.417}{69.257}
\emlineto{64.611}{69.212}
\emmoveto{64.741}{69.165}
\emlineto{64.935}{69.120}
\emmoveto{65.065}{69.073}
\emlineto{65.259}{69.028}
\emmoveto{65.388}{68.981}
\emlineto{65.581}{68.935}
\emmoveto{65.710}{68.889}
\emlineto{65.903}{68.843}
\emmoveto{66.032}{68.796}
\emlineto{66.225}{68.751}
\emmoveto{66.353}{68.704}
\emlineto{66.545}{68.659}
\emmoveto{66.673}{68.612}
\emlineto{66.865}{68.567}
\emmoveto{66.993}{68.520}
\emlineto{67.184}{68.474}
\emmoveto{67.311}{68.427}
\emlineto{67.502}{68.382}
\emmoveto{67.629}{68.335}
\emlineto{67.820}{68.290}
\emmoveto{67.947}{68.243}
\emlineto{68.137}{68.197}
\emmoveto{68.263}{68.150}
\emlineto{68.453}{68.105}
\emmoveto{68.579}{68.058}
\emlineto{68.769}{68.013}
\emmoveto{68.895}{67.966}
\emlineto{69.083}{67.920}
\emmoveto{69.209}{67.873}
\emlineto{69.397}{67.828}
\emmoveto{69.523}{67.780}
\emlineto{69.711}{67.735}
\emmoveto{69.836}{67.688}
\emlineto{70.023}{67.642}
\emmoveto{70.148}{67.595}
\emlineto{70.335}{67.550}
\emmoveto{70.460}{67.503}
\emlineto{70.646}{67.457}
\emmoveto{70.771}{67.410}
\emlineto{70.957}{67.365}
\emmoveto{71.081}{67.317}
\emlineto{71.266}{67.272}
\emmoveto{71.390}{67.225}
\emlineto{71.575}{67.179}
\emmoveto{71.699}{67.132}
\emlineto{71.884}{67.086}
\emmoveto{72.007}{67.039}
\emlineto{72.191}{66.993}
\emmoveto{72.314}{66.946}
\emlineto{72.498}{66.901}
\emmoveto{72.621}{66.853}
\emlineto{72.804}{66.808}
\emmoveto{72.926}{66.761}
\emlineto{73.110}{66.715}
\emmoveto{73.232}{66.668}
\emlineto{73.414}{66.622}
\emmoveto{73.536}{66.575}
\emlineto{73.718}{66.529}
\emmoveto{73.840}{66.482}
\emlineto{74.021}{66.436}
\emmoveto{74.143}{66.389}
\emlineto{74.324}{66.343}
\emmoveto{74.445}{66.296}
\emlineto{74.626}{66.250}
\emmoveto{74.746}{66.202}
\emlineto{74.927}{66.156}
\emmoveto{75.047}{66.109}
\emlineto{75.227}{66.063}
\emmoveto{75.347}{66.016}
\emlineto{75.527}{65.970}
\emmoveto{75.646}{65.923}
\emlineto{75.825}{65.877}
\emmoveto{75.945}{65.830}
\emlineto{76.124}{65.784}
\emmoveto{76.243}{65.736}
\emlineto{76.421}{65.690}
\emmoveto{76.540}{65.643}
\emlineto{76.718}{65.597}
\emmoveto{76.836}{65.550}
\emlineto{77.014}{65.504}
\emmoveto{77.132}{65.456}
\emlineto{77.309}{65.410}
\emmoveto{77.427}{65.363}
\emlineto{77.603}{65.317}
\emmoveto{77.721}{65.270}
\emlineto{77.897}{65.224}
\emmoveto{78.015}{65.176}
\emlineto{78.190}{65.130}
\emmoveto{78.307}{65.083}
\emlineto{78.483}{65.036}
\emmoveto{78.599}{64.989}
\emlineto{78.774}{64.943}
\emmoveto{78.891}{64.896}
\emlineto{79.065}{64.849}
\emmoveto{79.181}{64.802}
\emlineto{79.355}{64.756}
\emmoveto{79.471}{64.708}
\emlineto{79.645}{64.662}
\emmoveto{79.760}{64.615}
\emlineto{79.933}{64.569}
\emmoveto{80.049}{64.521}
\emlineto{80.221}{64.475}
\emmoveto{80.336}{64.427}
\emlineto{80.509}{64.381}
\emmoveto{80.623}{64.333}
\emlineto{80.795}{64.287}
\emmoveto{80.909}{64.240}
\emlineto{81.081}{64.194}
\emmoveto{81.195}{64.146}
\emlineto{81.366}{64.100}
\emmoveto{81.480}{64.052}
\emlineto{81.650}{64.006}
\emmoveto{81.764}{63.958}
\emlineto{81.934}{63.912}
\emmoveto{82.047}{63.864}
\emlineto{82.217}{63.818}
\emmoveto{82.330}{63.770}
\emlineto{82.499}{63.724}
\emmoveto{82.611}{63.676}
\emlineto{82.780}{63.630}
\emmoveto{82.892}{63.582}
\emlineto{83.061}{63.536}
\emmoveto{83.173}{63.488}
\emlineto{83.341}{63.442}
\emmoveto{83.452}{63.394}
\emlineto{83.620}{63.348}
\emmoveto{83.731}{63.300}
\emlineto{83.898}{63.254}
\emmoveto{84.010}{63.206}
\emlineto{84.176}{63.160}
\emmoveto{84.287}{63.112}
\emlineto{84.453}{63.065}
\emmoveto{84.564}{63.018}
\emlineto{84.729}{62.971}
\emmoveto{84.840}{62.924}
\emlineto{85.005}{62.877}
\emmoveto{85.115}{62.829}
\emlineto{85.280}{62.783}
\emmoveto{85.389}{62.735}
\emlineto{85.554}{62.689}
\emmoveto{85.663}{62.641}
\emlineto{85.827}{62.594}
\emmoveto{85.936}{62.546}
\emlineto{86.100}{62.500}
\emmoveto{86.208}{62.452}
\emlineto{86.371}{62.405}
\emmoveto{86.480}{62.358}
\emlineto{86.643}{62.311}
\emmoveto{86.751}{62.263}
\emlineto{86.913}{62.217}
\emmoveto{87.021}{62.169}
\emlineto{87.183}{62.122}
\emmoveto{87.290}{62.074}
\emlineto{87.452}{62.028}
\emmoveto{87.559}{61.980}
\emlineto{87.720}{61.933}
\emmoveto{87.827}{61.885}
\emlineto{87.987}{61.838}
\emmoveto{88.094}{61.791}
\emlineto{88.254}{61.744}
\emmoveto{88.360}{61.696}
\emlineto{88.520}{61.649}
\emmoveto{88.626}{61.601}
\emlineto{88.785}{61.555}
\emmoveto{88.891}{61.507}
\emlineto{89.050}{61.460}
\emmoveto{89.155}{61.412}
\emlineto{89.313}{61.365}
\emmoveto{89.419}{61.317}
\emlineto{89.576}{61.270}
\emmoveto{89.681}{61.222}
\emlineto{89.839}{61.175}
\emmoveto{89.943}{61.128}
\emlineto{90.100}{61.081}
\emmoveto{90.205}{61.033}
\emlineto{90.361}{60.986}
\emmoveto{90.465}{60.938}
\emlineto{90.621}{60.891}
\emmoveto{90.725}{60.843}
\emlineto{90.881}{60.796}
\emmoveto{90.984}{60.748}
\emlineto{91.139}{60.701}
\emmoveto{91.242}{60.653}
\emlineto{91.397}{60.606}
\emmoveto{91.500}{60.558}
\emlineto{91.654}{60.511}
\emmoveto{91.757}{60.463}
\emlineto{91.911}{60.416}
\emmoveto{92.013}{60.368}
\emlineto{92.166}{60.321}
\emmoveto{92.268}{60.273}
\emlineto{92.421}{60.226}
\emmoveto{92.523}{60.178}
\emlineto{92.675}{60.131}
\emmoveto{92.777}{60.083}
\emlineto{92.929}{60.036}
\emmoveto{93.030}{59.988}
\emlineto{93.181}{59.941}
\emmoveto{93.282}{59.893}
\emlineto{93.433}{59.846}
\emmoveto{93.534}{59.797}
\emlineto{93.685}{59.750}
\emmoveto{93.785}{59.702}
\emlineto{93.935}{59.655}
\emmoveto{94.035}{59.607}
\emlineto{94.185}{59.560}
\emmoveto{94.284}{59.512}
\emlineto{94.434}{59.464}
\emmoveto{94.533}{59.416}
\emlineto{94.682}{59.369}
\emmoveto{94.781}{59.321}
\emlineto{94.929}{59.274}
\emmoveto{95.028}{59.226}
\emlineto{95.176}{59.179}
\emmoveto{95.275}{59.130}
\emlineto{95.422}{59.083}
\emmoveto{95.520}{59.035}
\emlineto{95.668}{58.988}
\emmoveto{95.765}{58.939}
\emlineto{95.912}{58.892}
\emmoveto{96.010}{58.844}
\emlineto{96.156}{58.797}
\emmoveto{96.253}{58.749}
\emlineto{96.399}{58.701}
\emmoveto{96.496}{58.653}
\emlineto{96.641}{58.606}
\emmoveto{96.738}{58.557}
\emlineto{96.883}{58.510}
\emmoveto{96.979}{58.462}
\emlineto{97.124}{58.414}
\emmoveto{97.220}{58.366}
\emlineto{97.364}{58.319}
\emmoveto{97.459}{58.270}
\emlineto{97.603}{58.223}
\emmoveto{97.698}{58.175}
\emlineto{97.842}{58.127}
\emmoveto{97.937}{58.079}
\emlineto{98.079}{58.032}
\emmoveto{98.174}{57.983}
\emlineto{98.316}{57.936}
\emmoveto{98.411}{57.888}
\emlineto{98.553}{57.840}
\emmoveto{98.647}{57.792}
\emlineto{98.788}{57.744}
\emmoveto{98.882}{57.696}
\emlineto{99.023}{57.649}
\emmoveto{99.117}{57.600}
\emlineto{99.257}{57.553}
\emmoveto{99.351}{57.504}
\emlineto{99.491}{57.457}
\emmoveto{99.584}{57.409}
\emlineto{99.723}{57.361}
\emmoveto{99.816}{57.313}
\emlineto{99.955}{57.265}
\emmoveto{100.048}{57.217}
\emlineto{100.186}{57.169}
\emmoveto{100.279}{57.121}
\emlineto{100.417}{57.073}
\emmoveto{100.509}{57.025}
\emlineto{100.646}{56.977}
\emmoveto{100.738}{56.929}
\emlineto{100.875}{56.881}
\emmoveto{100.967}{56.833}
\emlineto{101.103}{56.785}
\emmoveto{101.194}{56.737}
\emlineto{101.331}{56.689}
\emmoveto{101.421}{56.640}
\emlineto{101.557}{56.593}
\emmoveto{101.648}{56.544}
\emlineto{101.783}{56.497}
\emmoveto{101.874}{56.448}
\emlineto{102.008}{56.400}
\emmoveto{102.098}{56.352}
\emlineto{102.233}{56.304}
\emmoveto{102.322}{56.256}
\emlineto{102.457}{56.208}
\emmoveto{102.546}{56.160}
\emlineto{102.679}{56.112}
\emmoveto{102.768}{56.063}
\emlineto{102.902}{56.016}
\emmoveto{102.990}{55.968}
\emlineto{103.123}{55.921}
\emmoveto{103.211}{55.874}
\emlineto{103.344}{55.828}
\emmoveto{103.432}{55.781}
\emlineto{103.564}{55.736}
\emmoveto{103.651}{55.690}
\emlineto{103.783}{55.645}
\emmoveto{103.870}{55.600}
\emlineto{104.001}{55.556}
\emmoveto{104.089}{55.511}
\emlineto{104.219}{55.468}
\emmoveto{104.306}{55.424}
\emlineto{104.436}{55.382}
\emmoveto{104.523}{55.337}
\emlineto{104.653}{55.296}
\emmoveto{104.739}{55.252}
\emlineto{104.869}{55.212}
\emmoveto{104.955}{55.169}
\emlineto{105.084}{55.129}
\emmoveto{105.170}{55.086}
\emlineto{105.299}{55.047}
\emmoveto{105.384}{55.005}
\emlineto{105.555}{54.950}
\emmoveto{105.683}{54.893}
\emlineto{105.854}{54.839}
\emmoveto{105.981}{54.783}
\emlineto{106.151}{54.731}
\emmoveto{106.278}{54.675}
\emlineto{106.447}{54.624}
\emmoveto{106.573}{54.569}
\emlineto{106.741}{54.519}
\emmoveto{106.867}{54.465}
\emlineto{107.035}{54.417}
\emmoveto{107.160}{54.364}
\emlineto{107.327}{54.317}
\emmoveto{107.452}{54.265}
\emlineto{107.619}{54.219}
\emmoveto{107.743}{54.167}
\emlineto{107.909}{54.123}
\emmoveto{108.033}{54.072}
\emlineto{108.199}{54.029}
\emmoveto{108.322}{53.979}
\emlineto{108.487}{53.937}
\emmoveto{108.610}{53.888}
\emlineto{108.774}{53.848}
\emmoveto{108.897}{53.800}
\emlineto{109.061}{53.760}
\emmoveto{109.183}{53.713}
\emlineto{109.346}{53.675}
\emmoveto{109.468}{53.629}
\emlineto{109.631}{53.591}
\emmoveto{109.752}{53.546}
\emlineto{109.914}{53.510}
\emmoveto{110.036}{53.466}
\emlineto{110.197}{53.431}
\emmoveto{110.318}{53.388}
\emlineto{110.479}{53.354}
\emmoveto{110.600}{53.312}
\emlineto{110.760}{53.279}
\emmoveto{110.880}{53.238}
\emlineto{111.040}{53.206}
\emmoveto{111.160}{53.166}
\emlineto{111.320}{53.136}
\emmoveto{111.440}{53.096}
\emlineto{111.599}{53.067}
\emmoveto{111.718}{53.028}
\emlineto{111.877}{53.001}
\emmoveto{111.996}{52.963}
\emlineto{112.154}{52.936}
\emmoveto{112.273}{52.899}
\emlineto{112.431}{52.873}
\emmoveto{112.549}{52.837}
\emlineto{112.707}{52.813}
\emmoveto{112.825}{52.778}
\emlineto{112.982}{52.755}
\emmoveto{113.100}{52.720}
\emlineto{113.257}{52.698}
\emmoveto{113.374}{52.665}
\emlineto{113.531}{52.644}
\emmoveto{113.648}{52.611}
\emlineto{113.804}{52.592}
\emmoveto{113.921}{52.560}
\emlineto{114.077}{52.541}
\emmoveto{114.194}{52.511}
\emlineto{114.349}{52.493}
\emmoveto{114.466}{52.463}
\emlineto{114.621}{52.447}
\emmoveto{114.738}{52.418}
\emlineto{114.893}{52.403}
\emmoveto{115.009}{52.375}
\emlineto{115.164}{52.361}
\emmoveto{115.280}{52.334}
\emlineto{115.434}{52.321}
\emmoveto{115.550}{52.294}
\emlineto{115.704}{52.283}
\emmoveto{115.820}{52.257}
\emlineto{115.974}{52.247}
\emmoveto{116.089}{52.222}
\emlineto{116.243}{52.213}
\emmoveto{116.358}{52.189}
\emlineto{116.512}{52.181}
\emmoveto{116.627}{52.158}
\emlineto{116.780}{52.151}
\emmoveto{116.895}{52.128}
\emlineto{117.049}{52.123}
\emmoveto{117.164}{52.101}
\emlineto{117.317}{52.097}
\emmoveto{117.431}{52.076}
\emlineto{117.584}{52.072}
\emmoveto{117.699}{52.053}
\emlineto{117.852}{52.051}
\emmoveto{117.966}{52.032}
\emlineto{118.119}{52.030}
\emmoveto{118.233}{52.013}
\emlineto{118.386}{52.013}
\emmoveto{118.500}{51.995}
\emlineto{118.653}{51.997}
\emmoveto{118.767}{51.980}
\emlineto{118.919}{51.983}
\emmoveto{119.034}{51.967}
\emlineto{119.186}{51.971}
\emmoveto{119.300}{51.956}
\emlineto{119.452}{51.960}
\emmoveto{119.566}{51.947}
\emlineto{119.718}{51.953}
\emmoveto{119.833}{51.940}
\emlineto{119.985}{51.946}
\emmoveto{120.099}{51.935}
\emlineto{120.251}{51.942}
\emmoveto{120.365}{51.931}
\emlineto{120.517}{51.941}
\emmoveto{120.631}{51.930}
\emlineto{120.783}{51.941}
\emmoveto{120.897}{51.931}
\emlineto{121.049}{51.942}
\emmoveto{121.163}{51.933}
\emlineto{121.315}{51.944}
\emmoveto{121.429}{51.934}
\emlineto{121.581}{51.945}
\emmoveto{121.695}{51.935}
\emlineto{121.848}{51.946}
\emmoveto{121.962}{51.936}
\emlineto{122.114}{51.946}
\emmoveto{122.228}{51.936}
\emlineto{122.380}{51.946}
\emmoveto{122.494}{51.936}
\emlineto{122.646}{51.946}
\emmoveto{122.760}{51.936}
\emlineto{122.912}{51.946}
\emmoveto{123.026}{51.936}
\emlineto{123.178}{51.946}
\emmoveto{123.292}{51.936}
\emlineto{123.445}{51.946}
\emmoveto{123.559}{51.936}
\emlineto{123.711}{51.946}
\emmoveto{123.825}{51.936}
\emlineto{123.977}{51.946}
\emmoveto{124.091}{51.936}
\emlineto{124.243}{51.946}
\emmoveto{124.357}{51.936}
\emlineto{124.509}{51.946}
\emmoveto{124.623}{51.936}
\emlineto{124.775}{51.946}
\emmoveto{124.889}{51.936}
\emlineto{125.042}{51.946}
\emmoveto{125.156}{51.936}
\emlineto{125.308}{51.946}
\emmoveto{125.422}{51.936}
\emlineto{125.574}{51.946}
\emmoveto{125.688}{51.936}
\emlineto{125.840}{51.946}
\emmoveto{125.954}{51.936}
\emlineto{126.106}{51.946}
\emmoveto{126.220}{51.936}
\emlineto{126.372}{51.946}
\emshow{75.780}{69.700}{Sommerfeld particle:  a=0.2;}
\emshow{1.000}{10.000}{-1.00e-1}
\emshow{1.000}{17.000}{-6.00e-2}
\emshow{1.000}{24.000}{-2.00e-2}
\emshow{1.000}{31.000}{2.00e-2}
\emshow{1.000}{38.000}{6.00e-2}
\emshow{1.000}{45.000}{1.00e-1}
\emshow{1.000}{52.000}{1.40e-1}
\emshow{1.000}{59.000}{1.80e-1}
\emshow{1.000}{66.000}{2.20e-1}
\emshow{1.000}{73.000}{2.60e-1}
\emshow{1.000}{80.000}{3.00e-1}
\emshow{12.000}{5.000}{-1.00e-1}
\emshow{23.800}{5.000}{3.00e-2}
\emshow{35.600}{5.000}{1.60e-1}
\emshow{47.400}{5.000}{2.90e-1}
\emshow{59.200}{5.000}{4.20e-1}
\emshow{71.000}{5.000}{5.50e-1}
\emshow{82.800}{5.000}{6.80e-1}
\emshow{94.600}{5.000}{8.10e-1}
\emshow{106.400}{5.000}{9.40e-1}
\emshow{118.200}{5.000}{1.07e0}
\emshow{130.000}{5.000}{1.20e0}
\centerline{ \bf {Fig.A.1}}
\eject
\newcount\numpoint
\newcount\numpointo
\numpoint=1 \numpointo=1
\def\emmoveto#1#2{\offinterlineskip
\hbox to 0 true cm{\vbox to 0
true cm{\vskip - #2 true mm
\hskip #1 true mm \special{em:point
\the\numpoint}\vss}\hss}
\numpointo=\numpoint
\global\advance \numpoint by 1}
\def\emlineto#1#2{\offinterlineskip
\hbox to 0 true cm{\vbox to 0
true cm{\vskip - #2 true mm
\hskip #1 true mm \special{em:point
\the\numpoint}\vss}\hss}
\special{em:line
\the\numpointo,\the\numpoint}
\numpointo=\numpoint
\global\advance \numpoint by 1}
\def\emshow#1#2#3{\offinterlineskip
\hbox to 0 true cm{\vbox to 0
true cm{\vskip - #2 true mm
\hskip #1 true mm \vbox to 0
true cm{\vss\hbox{#3\hss
}}\vss}\hss}}
\special{em:linewidth 0.8pt}

\vrule width 0 mm height                0 mm depth 90.000 true mm

\special{em:linewidth 0.8pt}
\emmoveto{130.000}{10.000}
\emlineto{12.000}{10.000}
\emlineto{12.000}{80.000}
\emmoveto{71.000}{10.000}
\emlineto{71.000}{80.000}
\emmoveto{12.000}{45.000}
\emlineto{130.000}{45.000}
\emmoveto{130.000}{10.000}
\emlineto{130.000}{80.000}
\emlineto{12.000}{80.000}
\emlineto{12.000}{10.000}
\emlineto{130.000}{10.000}
\special{em:linewidth 0.4pt}
\emmoveto{12.000}{17.000}
\emlineto{130.000}{17.000}
\emmoveto{12.000}{24.000}
\emlineto{130.000}{24.000}
\emmoveto{12.000}{31.000}
\emlineto{130.000}{31.000}
\emmoveto{12.000}{38.000}
\emlineto{130.000}{38.000}
\emmoveto{12.000}{45.000}
\emlineto{130.000}{45.000}
\emmoveto{12.000}{52.000}
\emlineto{130.000}{52.000}
\emmoveto{12.000}{59.000}
\emlineto{130.000}{59.000}
\emmoveto{12.000}{66.000}
\emlineto{130.000}{66.000}
\emmoveto{12.000}{73.000}
\emlineto{130.000}{73.000}
\emmoveto{23.800}{10.000}
\emlineto{23.800}{80.000}
\emmoveto{35.600}{10.000}
\emlineto{35.600}{80.000}
\emmoveto{47.400}{10.000}
\emlineto{47.400}{80.000}
\emmoveto{59.200}{10.000}
\emlineto{59.200}{80.000}
\emmoveto{71.000}{10.000}
\emlineto{71.000}{80.000}
\emmoveto{82.800}{10.000}
\emlineto{82.800}{80.000}
\emmoveto{94.600}{10.000}
\emlineto{94.600}{80.000}
\emmoveto{106.400}{10.000}
\emlineto{106.400}{80.000}
\emmoveto{118.200}{10.000}
\emlineto{118.200}{80.000}
\special{em:linewidth 0.8pt}
\emmoveto{12.000}{80.000}
\emlineto{12.295}{79.949}
\emmoveto{12.295}{79.939}
\emlineto{12.589}{79.876}
\emmoveto{12.589}{79.866}
\emlineto{12.883}{79.803}
\emmoveto{12.883}{79.793}
\emlineto{13.176}{79.729}
\emmoveto{13.176}{79.719}
\emlineto{13.470}{79.656}
\emmoveto{13.470}{79.646}
\emlineto{13.762}{79.582}
\emmoveto{13.762}{79.572}
\emlineto{14.055}{79.508}
\emmoveto{14.055}{79.498}
\emlineto{14.346}{79.434}
\emmoveto{14.346}{79.424}
\emlineto{14.638}{79.361}
\emmoveto{14.638}{79.351}
\emlineto{14.929}{79.287}
\emmoveto{14.929}{79.277}
\emlineto{15.219}{79.213}
\emmoveto{15.219}{79.203}
\emlineto{15.510}{79.139}
\emmoveto{15.510}{79.129}
\emlineto{15.799}{79.066}
\emmoveto{15.799}{79.056}
\emlineto{16.089}{78.992}
\emmoveto{16.089}{78.982}
\emlineto{16.378}{78.918}
\emmoveto{16.378}{78.908}
\emlineto{16.666}{78.844}
\emmoveto{16.666}{78.834}
\emlineto{16.954}{78.770}
\emmoveto{16.954}{78.760}
\emlineto{17.242}{78.695}
\emmoveto{17.242}{78.685}
\emlineto{17.529}{78.621}
\emmoveto{17.529}{78.611}
\emlineto{17.816}{78.547}
\emmoveto{17.816}{78.537}
\emlineto{18.103}{78.473}
\emmoveto{18.103}{78.463}
\emlineto{18.389}{78.399}
\emmoveto{18.389}{78.389}
\emlineto{18.674}{78.325}
\emmoveto{18.674}{78.315}
\emlineto{18.960}{78.253}
\emmoveto{18.960}{78.243}
\emlineto{19.244}{78.181}
\emmoveto{19.244}{78.171}
\emlineto{19.529}{78.108}
\emmoveto{19.529}{78.098}
\emlineto{19.813}{78.036}
\emmoveto{19.813}{78.026}
\emlineto{20.096}{77.963}
\emmoveto{20.096}{77.953}
\emlineto{20.380}{77.890}
\emmoveto{20.380}{77.880}
\emlineto{20.662}{77.817}
\emmoveto{20.662}{77.807}
\emlineto{20.945}{77.743}
\emmoveto{20.945}{77.733}
\emlineto{21.227}{77.670}
\emmoveto{21.227}{77.660}
\emlineto{21.508}{77.596}
\emmoveto{21.508}{77.586}
\emlineto{21.789}{77.523}
\emmoveto{21.789}{77.513}
\emlineto{22.070}{77.449}
\emmoveto{22.070}{77.439}
\emlineto{22.350}{77.375}
\emmoveto{22.350}{77.365}
\emlineto{22.630}{77.302}
\emmoveto{22.630}{77.292}
\emlineto{22.910}{77.228}
\emmoveto{22.910}{77.218}
\emlineto{23.189}{77.153}
\emmoveto{23.189}{77.143}
\emlineto{23.467}{77.079}
\emmoveto{23.467}{77.069}
\emlineto{23.745}{77.005}
\emmoveto{23.745}{76.995}
\emlineto{24.023}{76.930}
\emmoveto{24.023}{76.920}
\emlineto{24.301}{76.855}
\emmoveto{24.301}{76.845}
\emlineto{24.577}{76.781}
\emmoveto{24.577}{76.771}
\emlineto{24.854}{76.706}
\emmoveto{24.854}{76.696}
\emlineto{25.130}{76.631}
\emmoveto{25.130}{76.621}
\emlineto{25.406}{76.556}
\emmoveto{25.406}{76.546}
\emlineto{25.681}{76.481}
\emmoveto{25.681}{76.471}
\emlineto{25.956}{76.407}
\emmoveto{25.956}{76.397}
\emlineto{26.230}{76.332}
\emmoveto{26.230}{76.322}
\emlineto{26.504}{76.257}
\emmoveto{26.504}{76.247}
\emlineto{26.778}{76.182}
\emmoveto{26.778}{76.172}
\emlineto{27.051}{76.106}
\emmoveto{27.051}{76.096}
\emlineto{27.324}{76.031}
\emmoveto{27.324}{76.021}
\emlineto{27.596}{75.956}
\emmoveto{27.596}{75.946}
\emlineto{27.868}{75.881}
\emmoveto{27.868}{75.871}
\emlineto{28.140}{75.805}
\emmoveto{28.140}{75.795}
\emlineto{28.411}{75.730}
\emmoveto{28.411}{75.720}
\emlineto{28.681}{75.655}
\emmoveto{28.681}{75.645}
\emlineto{28.951}{75.580}
\emmoveto{28.951}{75.570}
\emlineto{29.221}{75.505}
\emmoveto{29.221}{75.495}
\emlineto{29.491}{75.430}
\emmoveto{29.491}{75.420}
\emlineto{29.760}{75.357}
\emmoveto{29.760}{75.347}
\emlineto{30.028}{75.284}
\emmoveto{30.028}{75.274}
\emlineto{30.296}{75.210}
\emmoveto{30.296}{75.200}
\emlineto{30.564}{75.136}
\emmoveto{30.564}{75.126}
\emlineto{30.831}{75.062}
\emmoveto{30.831}{75.052}
\emlineto{31.098}{74.988}
\emmoveto{31.098}{74.978}
\emlineto{31.365}{74.914}
\emmoveto{31.365}{74.904}
\emlineto{31.631}{74.839}
\emmoveto{31.631}{74.829}
\emlineto{31.896}{74.765}
\emmoveto{31.896}{74.755}
\emlineto{32.161}{74.691}
\emmoveto{32.161}{74.681}
\emlineto{32.426}{74.616}
\emmoveto{32.426}{74.606}
\emlineto{32.691}{74.541}
\emmoveto{32.691}{74.531}
\emlineto{32.955}{74.466}
\emmoveto{32.955}{74.456}
\emlineto{33.218}{74.391}
\emmoveto{33.218}{74.381}
\emlineto{33.481}{74.316}
\emmoveto{33.481}{74.306}
\emlineto{33.744}{74.241}
\emmoveto{33.744}{74.231}
\emlineto{34.006}{74.165}
\emmoveto{34.006}{74.155}
\emlineto{34.268}{74.089}
\emmoveto{34.268}{74.079}
\emlineto{34.529}{74.013}
\emmoveto{34.529}{74.003}
\emlineto{34.790}{73.938}
\emmoveto{34.790}{73.928}
\emlineto{35.051}{73.862}
\emmoveto{35.051}{73.852}
\emlineto{35.311}{73.786}
\emmoveto{35.311}{73.776}
\emlineto{35.571}{73.710}
\emmoveto{35.571}{73.700}
\emlineto{35.830}{73.634}
\emmoveto{35.830}{73.624}
\emlineto{36.089}{73.558}
\emmoveto{36.089}{73.548}
\emlineto{36.347}{73.482}
\emmoveto{36.347}{73.472}
\emlineto{36.605}{73.406}
\emmoveto{36.605}{73.396}
\emlineto{36.863}{73.330}
\emmoveto{36.863}{73.320}
\emlineto{37.120}{73.253}
\emmoveto{37.120}{73.243}
\emlineto{37.377}{73.177}
\emmoveto{37.377}{73.167}
\emlineto{37.633}{73.100}
\emmoveto{37.633}{73.090}
\emlineto{37.889}{73.024}
\emmoveto{37.889}{73.014}
\emlineto{38.144}{72.948}
\emmoveto{38.144}{72.938}
\emlineto{38.399}{72.872}
\emmoveto{38.399}{72.862}
\emlineto{38.654}{72.795}
\emmoveto{38.654}{72.785}
\emlineto{38.908}{72.719}
\emmoveto{38.908}{72.709}
\emlineto{39.162}{72.643}
\emmoveto{39.162}{72.633}
\emlineto{39.415}{72.569}
\emmoveto{39.415}{72.559}
\emlineto{39.668}{72.495}
\emmoveto{39.668}{72.485}
\emlineto{39.920}{72.421}
\emmoveto{39.920}{72.411}
\emlineto{40.172}{72.346}
\emmoveto{40.172}{72.336}
\emlineto{40.424}{72.271}
\emmoveto{40.424}{72.261}
\emlineto{40.675}{72.196}
\emmoveto{40.675}{72.186}
\emlineto{40.926}{72.121}
\emmoveto{40.926}{72.111}
\emlineto{41.176}{72.046}
\emmoveto{41.176}{72.036}
\emlineto{41.426}{71.970}
\emmoveto{41.426}{71.960}
\emlineto{41.676}{71.895}
\emmoveto{41.676}{71.885}
\emlineto{41.925}{71.819}
\emmoveto{41.925}{71.809}
\emlineto{42.173}{71.743}
\emmoveto{42.173}{71.733}
\emlineto{42.422}{71.667}
\emmoveto{42.422}{71.657}
\emlineto{42.670}{71.591}
\emmoveto{42.670}{71.581}
\emlineto{42.917}{71.515}
\emmoveto{42.917}{71.505}
\emlineto{43.164}{71.439}
\emmoveto{43.164}{71.429}
\emlineto{43.410}{71.362}
\emmoveto{43.410}{71.352}
\emlineto{43.656}{71.285}
\emmoveto{43.656}{71.275}
\emlineto{43.902}{71.208}
\emmoveto{43.902}{71.198}
\emlineto{44.147}{71.131}
\emmoveto{44.147}{71.121}
\emlineto{44.392}{71.055}
\emmoveto{44.392}{71.045}
\emlineto{44.637}{70.978}
\emmoveto{44.637}{70.968}
\emlineto{44.880}{70.901}
\emmoveto{44.880}{70.891}
\emlineto{45.124}{70.824}
\emmoveto{45.124}{70.814}
\emlineto{45.367}{70.747}
\emmoveto{45.367}{70.737}
\emlineto{45.610}{70.670}
\emmoveto{45.610}{70.660}
\emlineto{45.852}{70.593}
\emmoveto{45.852}{70.583}
\emlineto{46.094}{70.515}
\emmoveto{46.094}{70.505}
\emlineto{46.335}{70.438}
\emmoveto{46.335}{70.428}
\emlineto{46.576}{70.361}
\emmoveto{46.576}{70.351}
\emlineto{46.816}{70.283}
\emmoveto{46.816}{70.273}
\emlineto{47.056}{70.205}
\emmoveto{47.056}{70.195}
\emlineto{47.296}{70.128}
\emmoveto{47.296}{70.118}
\emlineto{47.535}{70.051}
\emmoveto{47.535}{70.041}
\emlineto{47.774}{69.974}
\emmoveto{47.774}{69.964}
\emlineto{48.012}{69.897}
\emmoveto{48.012}{69.887}
\emlineto{48.250}{69.822}
\emmoveto{48.250}{69.812}
\emlineto{48.487}{69.747}
\emmoveto{48.487}{69.737}
\emlineto{48.724}{69.672}
\emmoveto{48.724}{69.662}
\emlineto{48.961}{69.597}
\emmoveto{48.961}{69.587}
\emlineto{49.197}{69.522}
\emmoveto{49.197}{69.512}
\emlineto{49.433}{69.446}
\emmoveto{49.433}{69.436}
\emlineto{49.668}{69.370}
\emmoveto{49.668}{69.360}
\emlineto{49.903}{69.293}
\emmoveto{49.903}{69.283}
\emlineto{50.138}{69.217}
\emmoveto{50.138}{69.207}
\emlineto{50.372}{69.140}
\emmoveto{50.372}{69.130}
\emlineto{50.605}{69.063}
\emmoveto{50.605}{69.053}
\emlineto{50.838}{68.986}
\emmoveto{50.838}{68.976}
\emlineto{51.071}{68.909}
\emmoveto{51.071}{68.899}
\emlineto{51.303}{68.832}
\emmoveto{51.303}{68.822}
\emlineto{51.535}{68.755}
\emmoveto{51.535}{68.745}
\emlineto{51.767}{68.678}
\emmoveto{51.767}{68.668}
\emlineto{51.998}{68.601}
\emmoveto{51.998}{68.591}
\emlineto{52.228}{68.523}
\emmoveto{52.228}{68.513}
\emlineto{52.459}{68.446}
\emmoveto{52.459}{68.436}
\emlineto{52.688}{68.368}
\emmoveto{52.688}{68.358}
\emlineto{52.918}{68.289}
\emmoveto{52.918}{68.279}
\emlineto{53.146}{68.211}
\emmoveto{53.146}{68.201}
\emlineto{53.375}{68.133}
\emmoveto{53.375}{68.123}
\emlineto{53.603}{68.055}
\emmoveto{53.603}{68.045}
\emlineto{53.830}{67.976}
\emmoveto{53.830}{67.966}
\emlineto{54.057}{67.898}
\emmoveto{54.057}{67.888}
\emlineto{54.284}{67.820}
\emmoveto{54.284}{67.810}
\emlineto{54.510}{67.742}
\emmoveto{54.510}{67.732}
\emlineto{54.736}{67.664}
\emmoveto{54.736}{67.654}
\emlineto{54.961}{67.586}
\emmoveto{54.961}{67.576}
\emlineto{55.186}{67.508}
\emmoveto{55.186}{67.498}
\emlineto{55.410}{67.430}
\emmoveto{55.410}{67.420}
\emlineto{55.634}{67.352}
\emmoveto{55.634}{67.342}
\emlineto{55.858}{67.273}
\emmoveto{55.858}{67.263}
\emlineto{56.081}{67.196}
\emmoveto{56.081}{67.186}
\emlineto{56.304}{67.120}
\emmoveto{56.304}{67.110}
\emlineto{56.526}{67.045}
\emmoveto{56.526}{67.035}
\emlineto{56.748}{66.969}
\emmoveto{56.748}{66.959}
\emlineto{56.969}{66.892}
\emmoveto{56.969}{66.882}
\emlineto{57.190}{66.816}
\emmoveto{57.190}{66.806}
\emlineto{57.411}{66.740}
\emmoveto{57.411}{66.730}
\emlineto{57.631}{66.663}
\emmoveto{57.631}{66.653}
\emlineto{57.851}{66.586}
\emmoveto{57.851}{66.576}
\emlineto{58.070}{66.509}
\emmoveto{58.070}{66.499}
\emlineto{58.289}{66.432}
\emmoveto{58.289}{66.422}
\emlineto{58.507}{66.354}
\emmoveto{58.507}{66.344}
\emlineto{58.725}{66.276}
\emmoveto{58.725}{66.266}
\emlineto{58.943}{66.198}
\emmoveto{58.943}{66.188}
\emlineto{59.160}{66.120}
\emmoveto{59.160}{66.110}
\emlineto{59.376}{66.041}
\emmoveto{59.376}{66.031}
\emlineto{59.593}{65.963}
\emmoveto{59.593}{65.953}
\emlineto{59.808}{65.884}
\emmoveto{59.808}{65.874}
\emlineto{60.024}{65.805}
\emmoveto{60.024}{65.795}
\emlineto{60.239}{65.727}
\emmoveto{60.239}{65.717}
\emlineto{60.453}{65.648}
\emmoveto{60.453}{65.638}
\emlineto{60.667}{65.570}
\emmoveto{60.667}{65.560}
\emlineto{60.881}{65.491}
\emmoveto{60.881}{65.481}
\emlineto{61.094}{65.412}
\emmoveto{61.094}{65.402}
\emlineto{61.306}{65.333}
\emmoveto{61.306}{65.323}
\emlineto{61.519}{65.254}
\emmoveto{61.519}{65.244}
\emlineto{61.730}{65.175}
\emmoveto{61.730}{65.165}
\emlineto{61.942}{65.095}
\emmoveto{61.942}{65.085}
\emlineto{62.153}{65.016}
\emmoveto{62.153}{65.006}
\emlineto{62.363}{64.936}
\emmoveto{62.363}{64.926}
\emlineto{62.573}{64.857}
\emmoveto{62.573}{64.847}
\emlineto{62.782}{64.778}
\emmoveto{62.782}{64.768}
\emlineto{62.992}{64.699}
\emmoveto{62.992}{64.689}
\emlineto{63.200}{64.621}
\emmoveto{63.200}{64.611}
\emlineto{63.408}{64.545}
\emmoveto{63.408}{64.535}
\emlineto{63.616}{64.469}
\emmoveto{63.616}{64.459}
\emlineto{63.824}{64.393}
\emmoveto{63.824}{64.383}
\emlineto{64.031}{64.317}
\emmoveto{64.031}{64.307}
\emlineto{64.237}{64.240}
\emmoveto{64.237}{64.230}
\emlineto{64.443}{64.163}
\emmoveto{64.443}{64.153}
\emlineto{64.649}{64.085}
\emmoveto{64.649}{64.075}
\emlineto{64.854}{64.006}
\emmoveto{64.854}{63.996}
\emlineto{65.059}{63.928}
\emmoveto{65.059}{63.918}
\emlineto{65.263}{63.850}
\emmoveto{65.263}{63.840}
\emlineto{65.467}{63.771}
\emmoveto{65.467}{63.761}
\emlineto{65.670}{63.693}
\emmoveto{65.670}{63.683}
\emlineto{65.873}{63.614}
\emmoveto{65.873}{63.604}
\emlineto{66.076}{63.536}
\emmoveto{66.076}{63.526}
\emlineto{66.278}{63.457}
\emmoveto{66.278}{63.447}
\emlineto{66.480}{63.378}
\emmoveto{66.480}{63.368}
\emlineto{66.681}{63.299}
\emmoveto{66.681}{63.289}
\emlineto{66.882}{63.220}
\emmoveto{66.882}{63.210}
\emlineto{67.082}{63.140}
\emmoveto{67.082}{63.130}
\emlineto{67.282}{63.061}
\emmoveto{67.282}{63.051}
\emlineto{67.482}{62.980}
\emmoveto{67.482}{62.970}
\emlineto{67.681}{62.900}
\emmoveto{67.681}{62.890}
\emlineto{67.879}{62.820}
\emmoveto{67.879}{62.810}
\emlineto{68.077}{62.740}
\emmoveto{68.077}{62.730}
\emlineto{68.275}{62.660}
\emmoveto{68.275}{62.650}
\emlineto{68.472}{62.580}
\emmoveto{68.472}{62.570}
\emlineto{68.669}{62.500}
\emmoveto{68.669}{62.490}
\emlineto{68.865}{62.420}
\emmoveto{68.865}{62.410}
\emlineto{69.061}{62.341}
\emmoveto{69.061}{62.331}
\emlineto{69.256}{62.261}
\emmoveto{69.256}{62.251}
\emlineto{69.451}{62.182}
\emmoveto{69.451}{62.172}
\emlineto{69.646}{62.102}
\emmoveto{69.646}{62.092}
\emlineto{69.840}{62.025}
\emmoveto{69.840}{62.015}
\emlineto{70.034}{61.949}
\emmoveto{70.034}{61.939}
\emlineto{70.227}{61.872}
\emmoveto{70.227}{61.862}
\emlineto{70.420}{61.794}
\emmoveto{70.420}{61.784}
\emlineto{70.612}{61.716}
\emmoveto{70.612}{61.706}
\emlineto{70.804}{61.638}
\emmoveto{70.804}{61.628}
\emlineto{70.995}{61.560}
\emmoveto{70.995}{61.550}
\emlineto{71.186}{61.482}
\emmoveto{71.186}{61.472}
\emlineto{71.377}{61.403}
\emmoveto{71.377}{61.393}
\emlineto{71.567}{61.324}
\emmoveto{71.567}{61.314}
\emlineto{71.757}{61.246}
\emmoveto{71.757}{61.236}
\emlineto{71.946}{61.167}
\emmoveto{71.946}{61.157}
\emlineto{72.135}{61.088}
\emmoveto{72.135}{61.078}
\emlineto{72.323}{61.008}
\emmoveto{72.323}{60.998}
\emlineto{72.511}{60.929}
\emmoveto{72.511}{60.919}
\emlineto{72.699}{60.849}
\emmoveto{72.699}{60.839}
\emlineto{72.886}{60.769}
\emmoveto{72.886}{60.759}
\emlineto{73.072}{60.688}
\emmoveto{73.072}{60.678}
\emlineto{73.258}{60.607}
\emmoveto{73.258}{60.597}
\emlineto{73.444}{60.526}
\emmoveto{73.444}{60.516}
\emlineto{73.629}{60.445}
\emmoveto{73.629}{60.435}
\emlineto{73.814}{60.365}
\emmoveto{73.814}{60.355}
\emlineto{73.998}{60.284}
\emmoveto{73.998}{60.274}
\emlineto{74.182}{60.204}
\emmoveto{74.182}{60.194}
\emlineto{74.366}{60.123}
\emmoveto{74.366}{60.113}
\emlineto{74.549}{60.043}
\emmoveto{74.549}{60.033}
\emlineto{74.731}{59.962}
\emmoveto{74.731}{59.952}
\emlineto{74.913}{59.882}
\emmoveto{74.913}{59.872}
\emlineto{75.095}{59.802}
\emmoveto{75.095}{59.792}
\emlineto{75.276}{59.722}
\emmoveto{75.276}{59.712}
\emlineto{75.457}{59.641}
\emmoveto{75.457}{59.631}
\emlineto{75.637}{59.562}
\emmoveto{75.637}{59.552}
\emlineto{75.817}{59.485}
\emmoveto{75.817}{59.475}
\emlineto{75.996}{59.407}
\emmoveto{75.996}{59.397}
\emlineto{76.175}{59.329}
\emmoveto{76.175}{59.319}
\emlineto{76.353}{59.251}
\emmoveto{76.353}{59.241}
\emlineto{76.532}{59.173}
\emmoveto{76.532}{59.163}
\emlineto{76.709}{59.095}
\emmoveto{76.709}{59.085}
\emlineto{76.886}{59.016}
\emmoveto{76.886}{59.006}
\emlineto{77.063}{58.937}
\emmoveto{77.063}{58.927}
\emlineto{77.239}{58.858}
\emmoveto{77.239}{58.848}
\emlineto{77.415}{58.779}
\emmoveto{77.415}{58.769}
\emlineto{77.591}{58.700}
\emmoveto{77.591}{58.690}
\emlineto{77.766}{58.620}
\emmoveto{77.766}{58.610}
\emlineto{77.940}{58.540}
\emmoveto{77.940}{58.530}
\emlineto{78.114}{58.460}
\emmoveto{78.114}{58.450}
\emlineto{78.288}{58.378}
\emmoveto{78.288}{58.368}
\emlineto{78.461}{58.296}
\emmoveto{78.461}{58.286}
\emlineto{78.634}{58.215}
\emmoveto{78.634}{58.205}
\emlineto{78.806}{58.133}
\emmoveto{78.806}{58.123}
\emlineto{78.978}{58.052}
\emmoveto{78.978}{58.042}
\emlineto{79.149}{57.971}
\emmoveto{79.149}{57.961}
\emlineto{79.320}{57.890}
\emmoveto{79.320}{57.880}
\emlineto{79.490}{57.808}
\emmoveto{79.490}{57.798}
\emlineto{79.660}{57.727}
\emmoveto{79.660}{57.717}
\emlineto{79.830}{57.646}
\emmoveto{79.830}{57.636}
\emlineto{79.999}{57.565}
\emmoveto{79.999}{57.555}
\emlineto{80.167}{57.485}
\emmoveto{80.167}{57.475}
\emlineto{80.335}{57.404}
\emmoveto{80.335}{57.394}
\emlineto{80.503}{57.323}
\emmoveto{80.503}{57.313}
\emlineto{80.670}{57.241}
\emmoveto{80.670}{57.231}
\emlineto{80.837}{57.160}
\emmoveto{80.837}{57.150}
\emlineto{81.003}{57.083}
\emmoveto{81.003}{57.073}
\emlineto{81.169}{57.005}
\emmoveto{81.169}{56.995}
\emlineto{81.335}{56.927}
\emmoveto{81.335}{56.917}
\emlineto{81.500}{56.849}
\emmoveto{81.500}{56.839}
\emlineto{81.664}{56.770}
\emmoveto{81.664}{56.760}
\emlineto{81.828}{56.692}
\emmoveto{81.828}{56.682}
\emlineto{81.992}{56.613}
\emmoveto{81.992}{56.603}
\emlineto{82.155}{56.534}
\emmoveto{82.155}{56.524}
\emlineto{82.318}{56.454}
\emmoveto{82.318}{56.444}
\emlineto{82.480}{56.375}
\emmoveto{82.480}{56.365}
\emlineto{82.642}{56.295}
\emmoveto{82.642}{56.285}
\emlineto{82.804}{56.215}
\emmoveto{82.804}{56.205}
\emlineto{82.965}{56.134}
\emmoveto{82.965}{56.124}
\emlineto{83.125}{56.053}
\emmoveto{83.125}{56.043}
\emlineto{83.285}{55.971}
\emmoveto{83.285}{55.961}
\emlineto{83.445}{55.888}
\emmoveto{83.445}{55.878}
\emlineto{83.604}{55.806}
\emmoveto{83.604}{55.796}
\emlineto{83.763}{55.724}
\emmoveto{83.763}{55.714}
\emlineto{83.921}{55.641}
\emmoveto{83.921}{55.631}
\emlineto{84.079}{55.559}
\emmoveto{84.079}{55.549}
\emlineto{84.236}{55.477}
\emmoveto{84.236}{55.467}
\emlineto{84.393}{55.396}
\emmoveto{84.393}{55.386}
\emlineto{84.550}{55.314}
\emmoveto{84.550}{55.304}
\emlineto{84.705}{55.232}
\emmoveto{84.705}{55.222}
\emlineto{84.861}{55.151}
\emmoveto{84.861}{55.141}
\emlineto{85.016}{55.069}
\emmoveto{85.016}{55.059}
\emlineto{85.171}{54.988}
\emmoveto{85.171}{54.978}
\emlineto{85.325}{54.906}
\emmoveto{85.325}{54.896}
\emlineto{85.478}{54.825}
\emmoveto{85.478}{54.815}
\emlineto{85.631}{54.746}
\emmoveto{85.631}{54.736}
\emlineto{85.784}{54.668}
\emmoveto{85.784}{54.658}
\emlineto{85.937}{54.589}
\emmoveto{85.937}{54.580}
\emlineto{86.088}{54.511}
\emmoveto{86.088}{54.501}
\emlineto{86.240}{54.432}
\emmoveto{86.240}{54.422}
\emlineto{86.391}{54.353}
\emmoveto{86.391}{54.343}
\emlineto{86.541}{54.274}
\emmoveto{86.541}{54.264}
\emlineto{86.691}{54.194}
\emmoveto{86.691}{54.184}
\emlineto{86.841}{54.115}
\emmoveto{86.841}{54.105}
\emlineto{86.990}{54.035}
\emmoveto{86.990}{54.025}
\emlineto{87.139}{53.954}
\emmoveto{87.139}{53.944}
\emlineto{87.287}{53.874}
\emmoveto{87.287}{53.864}
\emlineto{87.435}{53.793}
\emmoveto{87.435}{53.783}
\emlineto{87.583}{53.712}
\emmoveto{87.583}{53.702}
\emlineto{87.730}{53.631}
\emmoveto{87.730}{53.621}
\emlineto{87.876}{53.548}
\emmoveto{87.876}{53.538}
\emlineto{88.022}{53.464}
\emmoveto{88.022}{53.454}
\emlineto{88.168}{53.381}
\emmoveto{88.168}{53.371}
\emlineto{88.313}{53.297}
\emmoveto{88.313}{53.287}
\emlineto{88.457}{53.214}
\emmoveto{88.457}{53.204}
\emlineto{88.601}{53.131}
\emmoveto{88.601}{53.121}
\emlineto{88.745}{53.048}
\emmoveto{88.745}{53.038}
\emlineto{88.888}{52.965}
\emmoveto{88.888}{52.955}
\emlineto{89.031}{52.883}
\emmoveto{89.031}{52.873}
\emlineto{89.173}{52.801}
\emmoveto{89.173}{52.791}
\emlineto{89.315}{52.719}
\emmoveto{89.315}{52.709}
\emlineto{89.457}{52.637}
\emmoveto{89.457}{52.627}
\emlineto{89.597}{52.555}
\emmoveto{89.597}{52.545}
\emlineto{89.738}{52.476}
\emmoveto{89.738}{52.466}
\emlineto{89.878}{52.400}
\emmoveto{89.878}{52.390}
\emlineto{90.017}{52.323}
\emmoveto{90.017}{52.313}
\emlineto{90.157}{52.245}
\emmoveto{90.157}{52.235}
\emlineto{90.295}{52.165}
\emmoveto{90.295}{52.155}
\emlineto{90.434}{52.084}
\emmoveto{90.434}{52.074}
\emlineto{90.571}{52.004}
\emmoveto{90.571}{51.994}
\emlineto{90.709}{51.924}
\emmoveto{90.709}{51.914}
\emlineto{90.846}{51.843}
\emmoveto{90.846}{51.833}
\emlineto{90.982}{51.762}
\emmoveto{90.982}{51.752}
\emlineto{91.118}{51.681}
\emmoveto{91.118}{51.671}
\emlineto{91.254}{51.600}
\emmoveto{91.254}{51.590}
\emlineto{91.389}{51.519}
\emmoveto{91.389}{51.509}
\emlineto{91.523}{51.438}
\emmoveto{91.523}{51.428}
\emlineto{91.657}{51.356}
\emmoveto{91.657}{51.346}
\emlineto{91.791}{51.274}
\emmoveto{91.791}{51.264}
\emlineto{91.924}{51.192}
\emmoveto{91.924}{51.182}
\emlineto{92.057}{51.110}
\emmoveto{92.057}{51.100}
\emlineto{92.189}{51.027}
\emmoveto{92.189}{51.017}
\emlineto{92.321}{50.943}
\emmoveto{92.321}{50.933}
\emlineto{92.453}{50.857}
\emmoveto{92.453}{50.847}
\emlineto{92.584}{50.772}
\emmoveto{92.584}{50.762}
\emlineto{92.714}{50.688}
\emmoveto{92.714}{50.678}
\emlineto{92.844}{50.604}
\emmoveto{92.844}{50.594}
\emlineto{92.973}{50.520}
\emmoveto{92.973}{50.510}
\emlineto{93.103}{50.437}
\emmoveto{93.103}{50.427}
\emlineto{93.231}{50.354}
\emmoveto{93.231}{50.344}
\emlineto{93.359}{50.271}
\emmoveto{93.359}{50.261}
\emlineto{93.487}{50.194}
\emmoveto{93.487}{50.184}
\emlineto{93.614}{50.119}
\emmoveto{93.614}{50.109}
\emlineto{93.741}{50.042}
\emmoveto{93.741}{50.032}
\emlineto{93.867}{49.966}
\emmoveto{93.867}{49.956}
\emlineto{93.993}{49.888}
\emmoveto{93.993}{49.878}
\emlineto{94.118}{49.810}
\emmoveto{94.118}{49.800}
\emlineto{94.243}{49.732}
\emmoveto{94.243}{49.722}
\emlineto{94.368}{49.651}
\emmoveto{94.368}{49.641}
\emlineto{94.492}{49.568}
\emmoveto{94.492}{49.558}
\emlineto{94.616}{49.485}
\emmoveto{94.616}{49.475}
\emlineto{94.739}{49.402}
\emmoveto{94.739}{49.392}
\emlineto{94.862}{49.319}
\emmoveto{94.862}{49.309}
\emlineto{94.984}{49.236}
\emmoveto{94.984}{49.226}
\emlineto{95.106}{49.153}
\emmoveto{95.106}{49.143}
\emlineto{95.227}{49.070}
\emmoveto{95.227}{49.060}
\emlineto{95.348}{48.987}
\emmoveto{95.348}{48.977}
\emlineto{95.468}{48.904}
\emmoveto{95.468}{48.894}
\emlineto{95.588}{48.821}
\emmoveto{95.588}{48.811}
\emlineto{95.708}{48.738}
\emmoveto{95.708}{48.728}
\emlineto{95.827}{48.655}
\emmoveto{95.827}{48.645}
\emlineto{95.945}{48.572}
\emmoveto{95.945}{48.562}
\emlineto{96.063}{48.489}
\emmoveto{96.063}{48.479}
\emlineto{96.181}{48.406}
\emmoveto{96.181}{48.396}
\emlineto{96.298}{48.322}
\emmoveto{96.298}{48.312}
\emlineto{96.415}{48.235}
\emmoveto{96.415}{48.225}
\emlineto{96.531}{48.149}
\emmoveto{96.531}{48.139}
\emlineto{96.647}{48.063}
\emmoveto{96.647}{48.053}
\emlineto{96.762}{47.985}
\emmoveto{96.762}{47.975}
\emlineto{96.877}{47.908}
\emmoveto{96.877}{47.898}
\emlineto{96.991}{47.831}
\emmoveto{96.991}{47.821}
\emlineto{97.105}{47.754}
\emmoveto{97.105}{47.744}
\emlineto{97.218}{47.676}
\emmoveto{97.218}{47.666}
\emlineto{97.331}{47.598}
\emmoveto{97.331}{47.588}
\emlineto{97.444}{47.520}
\emmoveto{97.444}{47.510}
\emlineto{97.556}{47.441}
\emmoveto{97.668}{47.351}
\emlineto{97.890}{47.200}
\emmoveto{98.000}{47.109}
\emlineto{98.220}{46.955}
\emmoveto{98.220}{46.945}
\emlineto{98.329}{46.867}
\emmoveto{98.329}{46.857}
\emlineto{98.437}{46.780}
\emmoveto{98.437}{46.770}
\emlineto{98.545}{46.694}
\emmoveto{98.545}{46.684}
\emlineto{98.653}{46.608}
\emmoveto{98.760}{46.512}
\emlineto{98.972}{46.352}
\emmoveto{99.078}{46.258}
\emlineto{99.288}{46.100}
\emmoveto{99.392}{46.007}
\emlineto{99.599}{45.855}
\emmoveto{99.702}{45.772}
\emlineto{99.906}{45.634}
\emmoveto{100.008}{45.548}
\emlineto{100.209}{45.393}
\emmoveto{100.310}{45.300}
\emlineto{100.509}{45.145}
\emmoveto{100.608}{45.052}
\emlineto{100.804}{44.896}
\emmoveto{100.901}{44.803}
\emlineto{101.095}{44.647}
\emmoveto{101.191}{44.554}
\emlineto{101.381}{44.396}
\emmoveto{101.476}{44.302}
\emlineto{101.664}{44.144}
\emmoveto{101.757}{44.043}
\emlineto{101.942}{43.875}
\emmoveto{102.033}{43.778}
\emlineto{102.215}{43.637}
\emmoveto{102.306}{43.553}
\emlineto{102.485}{43.412}
\emmoveto{102.574}{43.326}
\emlineto{102.751}{43.181}
\emmoveto{102.839}{43.093}
\emlineto{103.013}{42.943}
\emmoveto{103.100}{42.852}
\emlineto{103.271}{42.697}
\emmoveto{103.356}{42.596}
\emlineto{103.525}{42.425}
\emmoveto{103.608}{42.325}
\emlineto{103.774}{42.159}
\emmoveto{103.856}{42.062}
\emlineto{104.019}{41.901}
\emmoveto{104.099}{41.807}
\emlineto{104.259}{41.660}
\emmoveto{104.338}{41.583}
\emlineto{104.496}{41.454}
\emmoveto{104.574}{41.372}
\emlineto{104.729}{41.234}
\emmoveto{104.806}{41.139}
\emlineto{104.958}{40.975}
\emmoveto{105.033}{40.878}
\emlineto{105.183}{40.716}
\emmoveto{105.257}{40.620}
\emlineto{105.403}{40.459}
\emmoveto{105.475}{40.364}
\emlineto{105.619}{40.204}
\emmoveto{105.690}{40.109}
\emlineto{105.831}{39.950}
\emmoveto{105.900}{39.856}
\emlineto{106.038}{39.694}
\emmoveto{106.106}{39.604}
\emlineto{106.242}{39.462}
\emmoveto{106.308}{39.375}
\emlineto{106.441}{39.231}
\emmoveto{106.507}{39.143}
\emlineto{106.637}{38.996}
\emmoveto{106.701}{38.906}
\emlineto{106.828}{38.755}
\emmoveto{106.891}{38.664}
\emlineto{107.016}{38.509}
\emmoveto{107.077}{38.415}
\emlineto{107.199}{38.254}
\emmoveto{107.259}{38.153}
\emlineto{107.377}{37.957}
\emmoveto{107.436}{37.850}
\emlineto{107.551}{37.680}
\emmoveto{107.608}{37.611}
\emlineto{107.721}{37.498}
\emmoveto{107.777}{37.425}
\emlineto{107.887}{37.304}
\emmoveto{107.942}{37.227}
\emlineto{108.051}{37.096}
\emmoveto{108.104}{37.013}
\emlineto{108.210}{36.872}
\emmoveto{108.263}{36.783}
\emlineto{108.366}{36.628}
\emmoveto{108.417}{36.506}
\emlineto{108.517}{36.303}
\emmoveto{108.566}{36.192}
\emlineto{108.663}{36.009}
\emmoveto{108.710}{35.908}
\emlineto{108.804}{35.750}
\emmoveto{108.896}{35.661}
\emlineto{109.032}{35.540}
\emmoveto{109.122}{35.431}
\emlineto{109.210}{35.331}
\emmoveto{109.254}{35.261}
\emlineto{109.341}{35.141}
\emmoveto{109.383}{35.060}
\emlineto{109.467}{34.855}
\emmoveto{109.508}{34.735}
\emlineto{109.588}{34.533}
\emmoveto{109.627}{34.422}
\emlineto{109.704}{34.241}
\emmoveto{109.741}{34.140}
\emlineto{109.815}{33.977}
\emmoveto{109.851}{33.886}
\emlineto{109.994}{33.844}
\emmoveto{110.100}{33.824}
\emlineto{110.242}{33.815}
\emmoveto{110.348}{33.824}
\emlineto{110.491}{33.875}
\emmoveto{110.598}{33.880}
\emlineto{110.742}{33.909}
\emmoveto{110.850}{33.910}
\emlineto{110.994}{33.934}
\emmoveto{111.102}{33.933}
\emlineto{111.247}{33.954}
\emmoveto{111.356}{33.951}
\emlineto{111.501}{33.969}
\emmoveto{111.610}{33.964}
\emlineto{111.755}{33.980}
\emmoveto{111.864}{33.974}
\emlineto{112.009}{33.989}
\emmoveto{112.119}{33.982}
\emlineto{112.264}{33.996}
\emmoveto{112.374}{33.988}
\emlineto{112.519}{34.001}
\emmoveto{112.629}{33.993}
\emlineto{112.775}{34.005}
\emmoveto{112.884}{33.996}
\emlineto{113.030}{34.008}
\emmoveto{113.140}{33.999}
\emlineto{113.286}{34.010}
\emmoveto{113.395}{34.001}
\emlineto{113.541}{34.012}
\emmoveto{113.651}{34.003}
\emlineto{113.797}{34.014}
\emmoveto{113.907}{34.004}
\emlineto{114.053}{34.015}
\emmoveto{114.162}{34.005}
\emlineto{114.309}{34.015}
\emmoveto{114.418}{34.006}
\emlineto{114.564}{34.016}
\emmoveto{114.674}{34.006}
\emlineto{114.820}{34.017}
\emmoveto{114.930}{34.007}
\emlineto{115.076}{34.017}
\emmoveto{115.185}{34.007}
\emlineto{115.332}{34.017}
\emmoveto{115.441}{34.007}
\emlineto{115.588}{34.017}
\emmoveto{115.697}{34.007}
\emlineto{115.843}{34.018}
\emmoveto{115.953}{34.008}
\emlineto{116.099}{34.018}
\emmoveto{116.209}{34.008}
\emlineto{116.355}{34.018}
\emmoveto{116.465}{34.008}
\emlineto{116.611}{34.018}
\emmoveto{116.721}{34.008}
\emlineto{116.867}{34.018}
\emmoveto{116.977}{34.008}
\emlineto{117.123}{34.018}
\emmoveto{117.232}{34.008}
\emlineto{117.379}{34.018}
\emmoveto{117.488}{34.008}
\emlineto{117.634}{34.018}
\emmoveto{117.744}{34.008}
\emlineto{117.890}{34.018}
\emmoveto{118.000}{34.008}
\emlineto{118.146}{34.018}
\emmoveto{118.256}{34.008}
\emlineto{118.402}{34.018}
\emmoveto{118.512}{34.008}
\emlineto{118.658}{34.018}
\emmoveto{118.768}{34.008}
\emlineto{118.914}{34.018}
\emmoveto{119.023}{34.008}
\emlineto{119.170}{34.018}
\emmoveto{119.279}{34.008}
\emlineto{119.426}{34.018}
\emmoveto{119.535}{34.008}
\emlineto{119.681}{34.018}
\emmoveto{119.791}{34.008}
\emlineto{119.937}{34.018}
\emmoveto{120.047}{34.008}
\emlineto{120.193}{34.018}
\emmoveto{120.303}{34.008}
\emlineto{120.449}{34.018}
\emmoveto{120.559}{34.008}
\emlineto{120.705}{34.018}
\emmoveto{120.815}{34.008}
\emlineto{120.961}{34.018}
\emmoveto{121.071}{34.008}
\emlineto{121.217}{34.018}
\emmoveto{121.326}{34.008}
\emlineto{121.473}{34.018}
\emmoveto{121.582}{34.008}
\emlineto{121.692}{34.018}
\emshow{72.180}{38.700}{Sommerfeld particle: \ \ \     a=0.001}
\emmoveto{12.049}{80.000}
\emlineto{12.344}{79.937}
\emmoveto{12.344}{79.927}
\emlineto{12.638}{79.863}
\emmoveto{12.638}{79.853}
\emlineto{12.932}{79.790}
\emmoveto{12.932}{79.780}
\emlineto{13.225}{79.716}
\emmoveto{13.225}{79.706}
\emlineto{13.518}{79.643}
\emmoveto{13.518}{79.633}
\emlineto{13.811}{79.569}
\emmoveto{13.811}{79.559}
\emlineto{14.103}{79.496}
\emmoveto{14.103}{79.486}
\emlineto{14.395}{79.422}
\emmoveto{14.395}{79.412}
\emlineto{14.686}{79.349}
\emmoveto{14.686}{79.339}
\emlineto{14.977}{79.275}
\emmoveto{14.977}{79.265}
\emlineto{15.268}{79.201}
\emmoveto{15.268}{79.191}
\emlineto{15.558}{79.127}
\emmoveto{15.558}{79.117}
\emlineto{15.848}{79.054}
\emmoveto{15.848}{79.044}
\emlineto{16.137}{78.980}
\emmoveto{16.137}{78.970}
\emlineto{16.426}{78.906}
\emmoveto{16.426}{78.896}
\emlineto{16.714}{78.832}
\emmoveto{16.714}{78.822}
\emlineto{17.002}{78.759}
\emmoveto{17.002}{78.749}
\emlineto{17.290}{78.685}
\emmoveto{17.290}{78.675}
\emlineto{17.577}{78.611}
\emmoveto{17.577}{78.601}
\emlineto{17.864}{78.537}
\emmoveto{17.864}{78.527}
\emlineto{18.150}{78.463}
\emmoveto{18.150}{78.453}
\emlineto{18.436}{78.389}
\emmoveto{18.436}{78.379}
\emlineto{18.722}{78.315}
\emmoveto{18.722}{78.305}
\emlineto{19.007}{78.241}
\emmoveto{19.007}{78.231}
\emlineto{19.292}{78.167}
\emmoveto{19.292}{78.157}
\emlineto{19.576}{78.093}
\emmoveto{19.576}{78.083}
\emlineto{19.860}{78.018}
\emmoveto{19.860}{78.008}
\emlineto{20.144}{77.944}
\emmoveto{20.144}{77.934}
\emlineto{20.427}{77.870}
\emmoveto{20.427}{77.860}
\emlineto{20.709}{77.796}
\emmoveto{20.709}{77.786}
\emlineto{20.991}{77.721}
\emmoveto{20.991}{77.711}
\emlineto{21.273}{77.647}
\emmoveto{21.273}{77.637}
\emlineto{21.555}{77.573}
\emmoveto{21.555}{77.563}
\emlineto{21.836}{77.498}
\emmoveto{21.836}{77.488}
\emlineto{22.116}{77.424}
\emmoveto{22.116}{77.414}
\emlineto{22.396}{77.349}
\emmoveto{22.396}{77.339}
\emlineto{22.676}{77.275}
\emmoveto{22.676}{77.265}
\emlineto{22.955}{77.201}
\emmoveto{22.955}{77.191}
\emlineto{23.234}{77.126}
\emmoveto{23.234}{77.116}
\emlineto{23.513}{77.052}
\emmoveto{23.513}{77.042}
\emlineto{23.791}{76.977}
\emmoveto{23.791}{76.967}
\emlineto{24.068}{76.902}
\emmoveto{24.068}{76.892}
\emlineto{24.346}{76.828}
\emmoveto{24.346}{76.818}
\emlineto{24.623}{76.753}
\emmoveto{24.623}{76.743}
\emlineto{24.899}{76.678}
\emmoveto{24.899}{76.668}
\emlineto{25.175}{76.604}
\emmoveto{25.175}{76.594}
\emlineto{25.450}{76.529}
\emmoveto{25.450}{76.519}
\emlineto{25.725}{76.454}
\emmoveto{25.725}{76.444}
\emlineto{26.000}{76.379}
\emmoveto{26.000}{76.369}
\emlineto{26.274}{76.305}
\emmoveto{26.274}{76.295}
\emlineto{26.548}{76.230}
\emmoveto{26.548}{76.220}
\emlineto{26.822}{76.155}
\emmoveto{26.822}{76.145}
\emlineto{27.095}{76.080}
\emmoveto{27.095}{76.070}
\emlineto{27.367}{76.005}
\emmoveto{27.367}{75.995}
\emlineto{27.639}{75.930}
\emmoveto{27.639}{75.920}
\emlineto{27.911}{75.855}
\emmoveto{27.911}{75.845}
\emlineto{28.183}{75.780}
\emmoveto{28.183}{75.770}
\emlineto{28.453}{75.705}
\emmoveto{28.453}{75.695}
\emlineto{28.724}{75.630}
\emmoveto{28.724}{75.620}
\emlineto{28.994}{75.555}
\emmoveto{28.994}{75.545}
\emlineto{29.264}{75.479}
\emmoveto{29.264}{75.469}
\emlineto{29.533}{75.404}
\emmoveto{29.533}{75.394}
\emlineto{29.802}{75.329}
\emmoveto{29.802}{75.319}
\emlineto{30.070}{75.254}
\emmoveto{30.070}{75.244}
\emlineto{30.338}{75.179}
\emmoveto{30.338}{75.169}
\emlineto{30.606}{75.103}
\emmoveto{30.606}{75.093}
\emlineto{30.873}{75.028}
\emmoveto{30.873}{75.018}
\emlineto{31.139}{74.953}
\emmoveto{31.139}{74.943}
\emlineto{31.406}{74.877}
\emmoveto{31.406}{74.867}
\emlineto{31.671}{74.802}
\emmoveto{31.671}{74.792}
\emlineto{31.937}{74.726}
\emmoveto{31.937}{74.716}
\emlineto{32.202}{74.651}
\emmoveto{32.202}{74.641}
\emlineto{32.466}{74.575}
\emmoveto{32.466}{74.565}
\emlineto{32.731}{74.500}
\emmoveto{32.731}{74.490}
\emlineto{32.994}{74.424}
\emmoveto{32.994}{74.414}
\emlineto{33.258}{74.349}
\emmoveto{33.258}{74.339}
\emlineto{33.520}{74.273}
\emmoveto{33.520}{74.263}
\emlineto{33.783}{74.198}
\emmoveto{33.783}{74.188}
\emlineto{34.045}{74.122}
\emmoveto{34.045}{74.112}
\emlineto{34.306}{74.046}
\emmoveto{34.306}{74.036}
\emlineto{34.568}{73.971}
\emmoveto{34.568}{73.961}
\emlineto{34.828}{73.895}
\emmoveto{34.828}{73.885}
\emlineto{35.089}{73.819}
\emmoveto{35.089}{73.809}
\emlineto{35.349}{73.743}
\emmoveto{35.349}{73.733}
\emlineto{35.608}{73.667}
\emmoveto{35.608}{73.657}
\emlineto{35.867}{73.592}
\emmoveto{35.867}{73.582}
\emlineto{36.126}{73.516}
\emmoveto{36.126}{73.506}
\emlineto{36.384}{73.440}
\emmoveto{36.384}{73.430}
\emlineto{36.642}{73.364}
\emmoveto{36.642}{73.354}
\emlineto{36.899}{73.288}
\emmoveto{36.899}{73.278}
\emlineto{37.156}{73.212}
\emmoveto{37.156}{73.202}
\emlineto{37.412}{73.136}
\emmoveto{37.412}{73.126}
\emlineto{37.668}{73.060}
\emmoveto{37.668}{73.050}
\emlineto{37.924}{72.984}
\emmoveto{37.924}{72.974}
\emlineto{38.179}{72.907}
\emmoveto{38.179}{72.897}
\emlineto{38.434}{72.831}
\emmoveto{38.434}{72.821}
\emlineto{38.688}{72.755}
\emmoveto{38.688}{72.745}
\emlineto{38.942}{72.679}
\emmoveto{38.942}{72.669}
\emlineto{39.196}{72.603}
\emmoveto{39.196}{72.593}
\emlineto{39.449}{72.526}
\emmoveto{39.449}{72.516}
\emlineto{39.701}{72.450}
\emmoveto{39.701}{72.440}
\emlineto{39.954}{72.374}
\emmoveto{39.954}{72.364}
\emlineto{40.205}{72.298}
\emmoveto{40.205}{72.288}
\emlineto{40.457}{72.221}
\emmoveto{40.457}{72.211}
\emlineto{40.708}{72.145}
\emmoveto{40.708}{72.135}
\emlineto{40.958}{72.069}
\emmoveto{40.958}{72.059}
\emlineto{41.208}{71.992}
\emmoveto{41.208}{71.982}
\emlineto{41.458}{71.916}
\emmoveto{41.458}{71.906}
\emlineto{41.707}{71.839}
\emmoveto{41.707}{71.829}
\emlineto{41.956}{71.763}
\emmoveto{41.956}{71.753}
\emlineto{42.204}{71.686}
\emmoveto{42.204}{71.676}
\emlineto{42.452}{71.609}
\emmoveto{42.452}{71.599}
\emlineto{42.700}{71.533}
\emmoveto{42.700}{71.523}
\emlineto{42.947}{71.456}
\emmoveto{42.947}{71.446}
\emlineto{43.193}{71.380}
\emmoveto{43.193}{71.370}
\emlineto{43.439}{71.303}
\emmoveto{43.439}{71.293}
\emlineto{43.685}{71.226}
\emmoveto{43.685}{71.216}
\emlineto{43.930}{71.149}
\emmoveto{43.930}{71.139}
\emlineto{44.175}{71.073}
\emmoveto{44.175}{71.063}
\emlineto{44.420}{70.996}
\emmoveto{44.420}{70.986}
\emlineto{44.664}{70.919}
\emmoveto{44.664}{70.909}
\emlineto{44.908}{70.842}
\emmoveto{44.908}{70.832}
\emlineto{45.151}{70.765}
\emmoveto{45.151}{70.755}
\emlineto{45.393}{70.688}
\emmoveto{45.393}{70.678}
\emlineto{45.636}{70.612}
\emmoveto{45.636}{70.602}
\emlineto{45.878}{70.535}
\emmoveto{45.878}{70.525}
\emlineto{46.119}{70.458}
\emmoveto{46.119}{70.448}
\emlineto{46.360}{70.381}
\emmoveto{46.360}{70.371}
\emlineto{46.601}{70.304}
\emmoveto{46.601}{70.294}
\emlineto{46.841}{70.227}
\emmoveto{46.841}{70.217}
\emlineto{47.080}{70.149}
\emmoveto{47.080}{70.139}
\emlineto{47.320}{70.072}
\emmoveto{47.320}{70.062}
\emlineto{47.559}{69.995}
\emmoveto{47.559}{69.985}
\emlineto{47.797}{69.918}
\emmoveto{47.797}{69.908}
\emlineto{48.035}{69.841}
\emmoveto{48.035}{69.831}
\emlineto{48.273}{69.764}
\emmoveto{48.273}{69.754}
\emlineto{48.510}{69.686}
\emmoveto{48.510}{69.676}
\emlineto{48.746}{69.609}
\emmoveto{48.746}{69.599}
\emlineto{48.983}{69.532}
\emmoveto{48.983}{69.522}
\emlineto{49.218}{69.455}
\emmoveto{49.218}{69.445}
\emlineto{49.454}{69.377}
\emmoveto{49.454}{69.367}
\emlineto{49.689}{69.300}
\emmoveto{49.689}{69.290}
\emlineto{49.923}{69.222}
\emmoveto{49.923}{69.212}
\emlineto{50.157}{69.145}
\emmoveto{50.157}{69.135}
\emlineto{50.391}{69.068}
\emmoveto{50.391}{69.058}
\emlineto{50.624}{68.990}
\emmoveto{50.624}{68.980}
\emlineto{50.857}{68.913}
\emmoveto{50.857}{68.903}
\emlineto{51.089}{68.835}
\emmoveto{51.089}{68.825}
\emlineto{51.321}{68.758}
\emmoveto{51.321}{68.748}
\emlineto{51.552}{68.680}
\emmoveto{51.552}{68.670}
\emlineto{51.784}{68.602}
\emmoveto{51.784}{68.592}
\emlineto{52.014}{68.525}
\emmoveto{52.014}{68.515}
\emlineto{52.244}{68.447}
\emmoveto{52.244}{68.437}
\emlineto{52.474}{68.369}
\emmoveto{52.474}{68.359}
\emlineto{52.703}{68.292}
\emmoveto{52.703}{68.282}
\emlineto{52.932}{68.214}
\emmoveto{52.932}{68.204}
\emlineto{53.161}{68.136}
\emmoveto{53.161}{68.126}
\emlineto{53.388}{68.058}
\emmoveto{53.388}{68.048}
\emlineto{53.616}{67.981}
\emmoveto{53.616}{67.971}
\emlineto{53.843}{67.903}
\emmoveto{53.843}{67.893}
\emlineto{54.070}{67.825}
\emmoveto{54.070}{67.815}
\emlineto{54.296}{67.747}
\emmoveto{54.296}{67.737}
\emlineto{54.522}{67.669}
\emmoveto{54.522}{67.659}
\emlineto{54.747}{67.591}
\emmoveto{54.747}{67.581}
\emlineto{54.972}{67.513}
\emmoveto{54.972}{67.503}
\emlineto{55.196}{67.435}
\emmoveto{55.196}{67.425}
\emlineto{55.420}{67.357}
\emmoveto{55.420}{67.347}
\emlineto{55.644}{67.279}
\emmoveto{55.644}{67.269}
\emlineto{55.867}{67.201}
\emmoveto{55.867}{67.191}
\emlineto{56.090}{67.123}
\emmoveto{56.090}{67.113}
\emlineto{56.312}{67.045}
\emmoveto{56.312}{67.035}
\emlineto{56.534}{66.967}
\emmoveto{56.534}{66.957}
\emlineto{56.755}{66.889}
\emmoveto{56.755}{66.879}
\emlineto{56.976}{66.810}
\emmoveto{56.976}{66.800}
\emlineto{57.197}{66.732}
\emmoveto{57.197}{66.722}
\emlineto{57.417}{66.654}
\emmoveto{57.417}{66.644}
\emlineto{57.637}{66.576}
\emmoveto{57.637}{66.566}
\emlineto{57.856}{66.498}
\emmoveto{57.856}{66.488}
\emlineto{58.075}{66.419}
\emmoveto{58.075}{66.409}
\emlineto{58.293}{66.341}
\emmoveto{58.293}{66.331}
\emlineto{58.511}{66.263}
\emmoveto{58.511}{66.253}
\emlineto{58.728}{66.184}
\emmoveto{58.728}{66.174}
\emlineto{58.945}{66.106}
\emmoveto{58.945}{66.096}
\emlineto{59.162}{66.027}
\emmoveto{59.162}{66.017}
\emlineto{59.378}{65.949}
\emmoveto{59.378}{65.939}
\emlineto{59.594}{65.870}
\emmoveto{59.594}{65.860}
\emlineto{59.809}{65.792}
\emmoveto{59.809}{65.782}
\emlineto{60.024}{65.713}
\emmoveto{60.024}{65.703}
\emlineto{60.238}{65.635}
\emmoveto{60.238}{65.625}
\emlineto{60.452}{65.556}
\emmoveto{60.452}{65.546}
\emlineto{60.665}{65.478}
\emmoveto{60.665}{65.468}
\emlineto{60.878}{65.399}
\emmoveto{60.878}{65.389}
\emlineto{61.091}{65.320}
\emmoveto{61.091}{65.310}
\emlineto{61.303}{65.242}
\emmoveto{61.303}{65.232}
\emlineto{61.515}{65.163}
\emmoveto{61.515}{65.153}
\emlineto{61.726}{65.084}
\emmoveto{61.726}{65.074}
\emlineto{61.937}{65.006}
\emmoveto{61.937}{64.996}
\emlineto{62.147}{64.927}
\emmoveto{62.147}{64.917}
\emlineto{62.357}{64.848}
\emmoveto{62.357}{64.838}
\emlineto{62.567}{64.769}
\emmoveto{62.567}{64.759}
\emlineto{62.776}{64.690}
\emmoveto{62.776}{64.680}
\emlineto{62.984}{64.611}
\emmoveto{62.984}{64.601}
\emlineto{63.193}{64.532}
\emmoveto{63.193}{64.522}
\emlineto{63.400}{64.454}
\emmoveto{63.400}{64.444}
\emlineto{63.608}{64.375}
\emmoveto{63.608}{64.365}
\emlineto{63.814}{64.296}
\emmoveto{63.814}{64.286}
\emlineto{64.021}{64.217}
\emmoveto{64.021}{64.207}
\emlineto{64.227}{64.138}
\emmoveto{64.227}{64.128}
\emlineto{64.432}{64.059}
\emmoveto{64.432}{64.049}
\emlineto{64.637}{63.980}
\emmoveto{64.637}{63.970}
\emlineto{64.842}{63.901}
\emmoveto{64.842}{63.891}
\emlineto{65.046}{63.821}
\emmoveto{65.046}{63.811}
\emlineto{65.250}{63.742}
\emmoveto{65.250}{63.732}
\emlineto{65.453}{63.663}
\emmoveto{65.453}{63.653}
\emlineto{65.656}{63.584}
\emmoveto{65.656}{63.574}
\emlineto{65.858}{63.505}
\emmoveto{65.858}{63.495}
\emlineto{66.060}{63.426}
\emmoveto{66.060}{63.416}
\emlineto{66.262}{63.347}
\emmoveto{66.262}{63.337}
\emlineto{66.463}{63.267}
\emmoveto{66.463}{63.257}
\emlineto{66.664}{63.188}
\emmoveto{66.664}{63.178}
\emlineto{66.864}{63.109}
\emmoveto{66.864}{63.099}
\emlineto{67.063}{63.029}
\emmoveto{67.063}{63.019}
\emlineto{67.263}{62.950}
\emmoveto{67.263}{62.940}
\emlineto{67.461}{62.871}
\emmoveto{67.461}{62.861}
\emlineto{67.660}{62.791}
\emmoveto{67.660}{62.781}
\emlineto{67.858}{62.712}
\emmoveto{67.858}{62.702}
\emlineto{68.055}{62.632}
\emmoveto{68.055}{62.622}
\emlineto{68.252}{62.553}
\emmoveto{68.252}{62.543}
\emlineto{68.449}{62.473}
\emmoveto{68.449}{62.463}
\emlineto{68.645}{62.394}
\emmoveto{68.645}{62.384}
\emlineto{68.841}{62.314}
\emmoveto{68.841}{62.304}
\emlineto{69.036}{62.235}
\emmoveto{69.036}{62.225}
\emlineto{69.231}{62.155}
\emmoveto{69.231}{62.145}
\emlineto{69.425}{62.075}
\emmoveto{69.425}{62.065}
\emlineto{69.619}{61.996}
\emmoveto{69.619}{61.986}
\emlineto{69.813}{61.916}
\emmoveto{69.813}{61.906}
\emlineto{70.006}{61.837}
\emmoveto{70.006}{61.827}
\emlineto{70.198}{61.757}
\emmoveto{70.198}{61.747}
\emlineto{70.390}{61.677}
\emmoveto{70.390}{61.667}
\emlineto{70.582}{61.598}
\emmoveto{70.582}{61.588}
\emlineto{70.773}{61.518}
\emmoveto{70.773}{61.508}
\emlineto{70.964}{61.438}
\emmoveto{70.964}{61.428}
\emlineto{71.155}{61.358}
\emmoveto{71.155}{61.348}
\emlineto{71.344}{61.278}
\emmoveto{71.344}{61.268}
\emlineto{71.534}{61.199}
\emmoveto{71.534}{61.189}
\emlineto{71.723}{61.119}
\emmoveto{71.723}{61.109}
\emlineto{71.911}{61.039}
\emmoveto{71.911}{61.029}
\emlineto{72.099}{60.959}
\emmoveto{72.099}{60.949}
\emlineto{72.287}{60.879}
\emmoveto{72.287}{60.869}
\emlineto{72.474}{60.799}
\emmoveto{72.474}{60.789}
\emlineto{72.661}{60.719}
\emmoveto{72.661}{60.709}
\emlineto{72.847}{60.639}
\emmoveto{72.847}{60.629}
\emlineto{73.033}{60.559}
\emmoveto{73.033}{60.549}
\emlineto{73.219}{60.479}
\emmoveto{73.219}{60.469}
\emlineto{73.404}{60.399}
\emmoveto{73.404}{60.389}
\emlineto{73.588}{60.319}
\emmoveto{73.588}{60.309}
\emlineto{73.772}{60.239}
\emmoveto{73.772}{60.229}
\emlineto{73.956}{60.159}
\emmoveto{73.956}{60.149}
\emlineto{74.139}{60.078}
\emmoveto{74.139}{60.068}
\emlineto{74.322}{59.998}
\emmoveto{74.322}{59.988}
\emlineto{74.504}{59.918}
\emmoveto{74.504}{59.908}
\emlineto{74.686}{59.838}
\emmoveto{74.686}{59.828}
\emlineto{74.867}{59.758}
\emmoveto{74.867}{59.748}
\emlineto{75.048}{59.677}
\emmoveto{75.048}{59.667}
\emlineto{75.228}{59.597}
\emmoveto{75.228}{59.587}
\emlineto{75.408}{59.517}
\emmoveto{75.408}{59.507}
\emlineto{75.588}{59.436}
\emmoveto{75.588}{59.426}
\emlineto{75.767}{59.356}
\emmoveto{75.767}{59.346}
\emlineto{75.946}{59.276}
\emmoveto{75.946}{59.266}
\emlineto{76.124}{59.196}
\emmoveto{76.124}{59.186}
\emlineto{76.301}{59.115}
\emmoveto{76.301}{59.105}
\emlineto{76.479}{59.035}
\emmoveto{76.479}{59.025}
\emlineto{76.656}{58.954}
\emmoveto{76.656}{58.944}
\emlineto{76.832}{58.874}
\emmoveto{76.832}{58.864}
\emlineto{77.008}{58.793}
\emmoveto{77.008}{58.783}
\emlineto{77.183}{58.713}
\emmoveto{77.183}{58.703}
\emlineto{77.359}{58.632}
\emmoveto{77.359}{58.622}
\emlineto{77.533}{58.552}
\emmoveto{77.533}{58.542}
\emlineto{77.707}{58.471}
\emmoveto{77.707}{58.461}
\emlineto{77.881}{58.391}
\emmoveto{77.881}{58.381}
\emlineto{78.054}{58.310}
\emmoveto{78.054}{58.300}
\emlineto{78.227}{58.229}
\emmoveto{78.227}{58.219}
\emlineto{78.399}{58.149}
\emmoveto{78.399}{58.139}
\emlineto{78.571}{58.068}
\emmoveto{78.571}{58.058}
\emlineto{78.742}{57.987}
\emmoveto{78.742}{57.977}
\emlineto{78.913}{57.907}
\emmoveto{78.913}{57.897}
\emlineto{79.084}{57.826}
\emmoveto{79.084}{57.816}
\emlineto{79.254}{57.745}
\emmoveto{79.254}{57.735}
\emlineto{79.424}{57.664}
\emmoveto{79.424}{57.654}
\emlineto{79.593}{57.584}
\emmoveto{79.593}{57.574}
\emlineto{79.761}{57.503}
\emmoveto{79.761}{57.493}
\emlineto{79.930}{57.422}
\emmoveto{79.930}{57.412}
\emlineto{80.097}{57.341}
\emmoveto{80.097}{57.331}
\emlineto{80.265}{57.260}
\emmoveto{80.265}{57.250}
\emlineto{80.432}{57.180}
\emmoveto{80.432}{57.170}
\emlineto{80.598}{57.098}
\emmoveto{80.598}{57.088}
\emlineto{80.764}{57.018}
\emmoveto{80.764}{57.008}
\emlineto{80.929}{56.937}
\emmoveto{80.929}{56.927}
\emlineto{81.095}{56.856}
\emmoveto{81.095}{56.846}
\emlineto{81.259}{56.775}
\emmoveto{81.259}{56.765}
\emlineto{81.423}{56.694}
\emmoveto{81.423}{56.684}
\emlineto{81.587}{56.613}
\emmoveto{81.587}{56.603}
\emlineto{81.750}{56.532}
\emmoveto{81.750}{56.522}
\emlineto{81.913}{56.451}
\emmoveto{81.913}{56.441}
\emlineto{82.075}{56.370}
\emmoveto{82.075}{56.360}
\emlineto{82.237}{56.288}
\emmoveto{82.237}{56.278}
\emlineto{82.399}{56.207}
\emmoveto{82.399}{56.197}
\emlineto{82.560}{56.126}
\emmoveto{82.560}{56.116}
\emlineto{82.720}{56.045}
\emmoveto{82.720}{56.035}
\emlineto{82.880}{55.964}
\emmoveto{82.880}{55.954}
\emlineto{83.040}{55.883}
\emmoveto{83.040}{55.873}
\emlineto{83.199}{55.802}
\emmoveto{83.199}{55.792}
\emlineto{83.358}{55.720}
\emmoveto{83.358}{55.710}
\emlineto{83.516}{55.639}
\emmoveto{83.516}{55.629}
\emlineto{83.674}{55.558}
\emmoveto{83.674}{55.548}
\emlineto{83.831}{55.476}
\emmoveto{83.831}{55.466}
\emlineto{83.988}{55.395}
\emmoveto{83.988}{55.385}
\emlineto{84.144}{55.314}
\emmoveto{84.144}{55.304}
\emlineto{84.300}{55.232}
\emmoveto{84.300}{55.222}
\emlineto{84.455}{55.151}
\emmoveto{84.455}{55.141}
\emlineto{84.610}{55.070}
\emmoveto{84.610}{55.060}
\emlineto{84.765}{54.988}
\emmoveto{84.765}{54.978}
\emlineto{84.919}{54.907}
\emmoveto{84.919}{54.897}
\emlineto{85.073}{54.826}
\emmoveto{85.073}{54.816}
\emlineto{85.226}{54.744}
\emmoveto{85.226}{54.734}
\emlineto{85.379}{54.663}
\emmoveto{85.379}{54.653}
\emlineto{85.531}{54.581}
\emmoveto{85.531}{54.571}
\emlineto{85.683}{54.500}
\emmoveto{85.683}{54.490}
\emlineto{85.834}{54.418}
\emmoveto{85.834}{54.408}
\emlineto{85.985}{54.337}
\emmoveto{85.985}{54.327}
\emlineto{86.136}{54.255}
\emmoveto{86.136}{54.245}
\emlineto{86.286}{54.174}
\emmoveto{86.286}{54.164}
\emlineto{86.435}{54.092}
\emmoveto{86.435}{54.082}
\emlineto{86.584}{54.010}
\emmoveto{86.584}{54.000}
\emlineto{86.733}{53.929}
\emmoveto{86.733}{53.919}
\emlineto{86.881}{53.847}
\emmoveto{86.881}{53.837}
\emlineto{87.029}{53.765}
\emmoveto{87.029}{53.755}
\emlineto{87.176}{53.684}
\emmoveto{87.176}{53.674}
\emlineto{87.323}{53.602}
\emmoveto{87.323}{53.592}
\emlineto{87.469}{53.520}
\emmoveto{87.469}{53.510}
\emlineto{87.615}{53.439}
\emmoveto{87.615}{53.429}
\emlineto{87.760}{53.357}
\emmoveto{87.760}{53.347}
\emlineto{87.905}{53.275}
\emmoveto{87.905}{53.265}
\emlineto{88.050}{53.193}
\emmoveto{88.050}{53.183}
\emlineto{88.194}{53.111}
\emmoveto{88.194}{53.101}
\emlineto{88.337}{53.030}
\emmoveto{88.337}{53.020}
\emlineto{88.480}{52.948}
\emmoveto{88.480}{52.938}
\emlineto{88.623}{52.866}
\emmoveto{88.623}{52.856}
\emlineto{88.765}{52.784}
\emmoveto{88.765}{52.774}
\emlineto{88.907}{52.702}
\emmoveto{88.907}{52.692}
\emlineto{89.048}{52.620}
\emmoveto{89.048}{52.610}
\emlineto{89.189}{52.538}
\emmoveto{89.189}{52.528}
\emlineto{89.329}{52.456}
\emmoveto{89.329}{52.446}
\emlineto{89.469}{52.375}
\emmoveto{89.469}{52.365}
\emlineto{89.609}{52.292}
\emmoveto{89.609}{52.282}
\emlineto{89.748}{52.211}
\emmoveto{89.748}{52.201}
\emlineto{89.886}{52.128}
\emmoveto{89.886}{52.118}
\emlineto{90.024}{52.046}
\emmoveto{90.024}{52.036}
\emlineto{90.162}{51.965}
\emmoveto{90.162}{51.955}
\emlineto{90.299}{51.882}
\emmoveto{90.299}{51.872}
\emlineto{90.436}{51.800}
\emmoveto{90.436}{51.790}
\emlineto{90.572}{51.718}
\emmoveto{90.572}{51.708}
\emlineto{90.707}{51.636}
\emmoveto{90.707}{51.626}
\emlineto{90.843}{51.554}
\emmoveto{90.843}{51.544}
\emlineto{90.978}{51.472}
\emmoveto{90.978}{51.462}
\emlineto{91.112}{51.390}
\emmoveto{91.112}{51.380}
\emlineto{91.246}{51.308}
\emmoveto{91.246}{51.298}
\emlineto{91.379}{51.225}
\emmoveto{91.379}{51.215}
\emlineto{91.512}{51.143}
\emmoveto{91.512}{51.133}
\emlineto{91.645}{51.061}
\emmoveto{91.645}{51.051}
\emlineto{91.777}{50.979}
\emmoveto{91.777}{50.969}
\emlineto{91.908}{50.896}
\emmoveto{91.908}{50.886}
\emlineto{92.039}{50.814}
\emmoveto{92.039}{50.804}
\emlineto{92.170}{50.732}
\emmoveto{92.170}{50.722}
\emlineto{92.300}{50.650}
\emmoveto{92.300}{50.640}
\emlineto{92.430}{50.567}
\emmoveto{92.430}{50.557}
\emlineto{92.559}{50.485}
\emmoveto{92.559}{50.475}
\emlineto{92.688}{50.403}
\emmoveto{92.688}{50.393}
\emlineto{92.816}{50.320}
\emmoveto{92.816}{50.310}
\emlineto{92.944}{50.238}
\emmoveto{92.944}{50.228}
\emlineto{93.072}{50.156}
\emmoveto{93.072}{50.146}
\emlineto{93.199}{50.073}
\emmoveto{93.199}{50.063}
\emlineto{93.325}{49.991}
\emmoveto{93.325}{49.981}
\emlineto{93.451}{49.908}
\emmoveto{93.451}{49.898}
\emlineto{93.577}{49.826}
\emmoveto{93.577}{49.816}
\emlineto{93.702}{49.743}
\emmoveto{93.702}{49.733}
\emlineto{93.827}{49.661}
\emmoveto{93.827}{49.651}
\emlineto{93.951}{49.578}
\emmoveto{93.951}{49.568}
\emlineto{94.075}{49.496}
\emmoveto{94.075}{49.486}
\emlineto{94.198}{49.413}
\emmoveto{94.198}{49.403}
\emlineto{94.321}{49.331}
\emmoveto{94.321}{49.321}
\emlineto{94.443}{49.248}
\emmoveto{94.443}{49.238}
\emlineto{94.565}{49.166}
\emmoveto{94.565}{49.156}
\emlineto{94.686}{49.083}
\emmoveto{94.686}{49.073}
\emlineto{94.807}{49.001}
\emmoveto{94.807}{48.991}
\emlineto{94.928}{48.918}
\emmoveto{94.928}{48.908}
\emlineto{95.048}{48.835}
\emmoveto{95.048}{48.825}
\emlineto{95.167}{48.753}
\emmoveto{95.167}{48.743}
\emlineto{95.286}{48.670}
\emmoveto{95.286}{48.660}
\emlineto{95.405}{48.587}
\emmoveto{95.405}{48.577}
\emlineto{95.523}{48.505}
\emmoveto{95.523}{48.495}
\emlineto{95.641}{48.422}
\emmoveto{95.641}{48.412}
\emlineto{95.758}{48.339}
\emmoveto{95.758}{48.329}
\emlineto{95.875}{48.257}
\emmoveto{95.875}{48.247}
\emlineto{95.991}{48.174}
\emmoveto{95.991}{48.164}
\emlineto{96.107}{48.091}
\emmoveto{96.107}{48.081}
\emlineto{96.222}{48.008}
\emmoveto{96.222}{47.998}
\emlineto{96.337}{47.926}
\emmoveto{96.337}{47.916}
\emlineto{96.452}{47.843}
\emmoveto{96.452}{47.833}
\emlineto{96.566}{47.760}
\emmoveto{96.566}{47.750}
\emlineto{96.679}{47.677}
\emmoveto{96.679}{47.667}
\emlineto{96.792}{47.594}
\emmoveto{96.792}{47.584}
\emlineto{96.905}{47.511}
\emmoveto{96.905}{47.501}
\emlineto{97.017}{47.429}
\emmoveto{97.017}{47.419}
\emlineto{97.128}{47.346}
\emmoveto{97.128}{47.336}
\emlineto{97.240}{47.263}
\emmoveto{97.240}{47.253}
\emlineto{97.350}{47.180}
\emmoveto{97.350}{47.170}
\emlineto{97.461}{47.097}
\emmoveto{97.461}{47.087}
\emlineto{97.570}{47.014}
\emmoveto{97.570}{47.004}
\emlineto{97.680}{46.931}
\emmoveto{97.789}{46.838}
\emlineto{98.005}{46.682}
\emmoveto{98.112}{46.589}
\emlineto{98.326}{46.433}
\emmoveto{98.432}{46.340}
\emlineto{98.642}{46.184}
\emmoveto{98.747}{46.091}
\emlineto{98.955}{45.935}
\emmoveto{99.058}{45.842}
\emlineto{99.263}{45.685}
\emmoveto{99.365}{45.592}
\emlineto{99.568}{45.436}
\emmoveto{99.668}{45.343}
\emlineto{99.867}{45.186}
\emmoveto{99.967}{45.093}
\emlineto{100.163}{44.937}
\emmoveto{100.261}{44.843}
\emlineto{100.455}{44.687}
\emmoveto{100.551}{44.593}
\emlineto{100.742}{44.437}
\emmoveto{100.837}{44.344}
\emlineto{101.025}{44.187}
\emmoveto{101.119}{44.094}
\emlineto{101.304}{43.937}
\emmoveto{101.396}{43.843}
\emlineto{101.579}{43.686}
\emmoveto{101.669}{43.593}
\emlineto{101.849}{43.436}
\emmoveto{101.939}{43.343}
\emlineto{102.116}{43.186}
\emmoveto{102.203}{43.092}
\emlineto{102.378}{42.935}
\emmoveto{102.464}{42.842}
\emlineto{102.635}{42.685}
\emmoveto{102.720}{42.591}
\emlineto{102.889}{42.434}
\emmoveto{102.973}{42.340}
\emlineto{103.138}{42.183}
\emmoveto{103.220}{42.089}
\emlineto{103.383}{41.932}
\emmoveto{103.464}{41.838}
\emlineto{103.624}{41.681}
\emmoveto{103.704}{41.588}
\emlineto{103.861}{41.430}
\emmoveto{103.939}{41.336}
\emlineto{104.093}{41.179}
\emmoveto{104.170}{41.085}
\emlineto{104.322}{40.928}
\emmoveto{104.397}{40.834}
\emlineto{104.546}{40.676}
\emmoveto{104.619}{40.582}
\emlineto{104.765}{40.425}
\emmoveto{104.838}{40.331}
\emlineto{104.981}{40.173}
\emmoveto{105.052}{40.080}
\emlineto{105.192}{39.922}
\emmoveto{105.262}{39.828}
\emlineto{105.399}{39.670}
\emmoveto{105.467}{39.576}
\emlineto{105.602}{39.418}
\emmoveto{105.669}{39.325}
\emlineto{105.800}{39.167}
\emmoveto{105.866}{39.073}
\emlineto{105.995}{38.915}
\emmoveto{106.059}{38.821}
\emlineto{106.185}{38.663}
\emmoveto{106.247}{38.569}
\emlineto{106.371}{38.411}
\emmoveto{106.432}{38.317}
\emlineto{106.552}{38.159}
\emmoveto{106.612}{38.065}
\emlineto{106.729}{37.907}
\emmoveto{106.788}{37.813}
\emlineto{106.902}{37.654}
\emmoveto{106.959}{37.560}
\emlineto{107.071}{37.402}
\emmoveto{107.127}{37.308}
\emlineto{107.236}{37.150}
\emmoveto{107.290}{37.056}
\emlineto{107.396}{36.897}
\emmoveto{107.449}{36.803}
\emlineto{107.552}{36.645}
\emmoveto{107.603}{36.551}
\emlineto{107.704}{36.392}
\emmoveto{107.754}{36.298}
\emlineto{107.852}{36.140}
\emmoveto{107.900}{36.046}
\emlineto{107.995}{35.887}
\emmoveto{108.042}{35.793}
\emlineto{108.134}{35.635}
\emmoveto{108.179}{35.540}
\emlineto{108.269}{35.382}
\emmoveto{108.313}{35.288}
\emlineto{108.399}{35.129}
\emmoveto{108.442}{35.035}
\emlineto{108.525}{34.876}
\emmoveto{108.567}{34.782}
\emlineto{108.647}{34.623}
\emmoveto{108.687}{34.529}
\emlineto{108.765}{34.370}
\emmoveto{108.803}{34.276}
\emlineto{108.879}{34.117}
\emmoveto{108.915}{34.023}
\emlineto{108.988}{33.864}
\emmoveto{109.023}{33.770}
\emlineto{109.093}{33.611}
\emmoveto{109.127}{33.517}
\emlineto{109.193}{33.358}
\emmoveto{109.226}{33.264}
\emlineto{109.290}{33.105}
\emmoveto{109.321}{33.011}
\emlineto{109.382}{32.852}
\emmoveto{109.412}{32.758}
\emlineto{109.470}{32.599}
\emmoveto{109.498}{32.504}
\emlineto{109.554}{32.346}
\emmoveto{109.580}{32.251}
\emlineto{109.633}{32.092}
\emmoveto{109.658}{31.998}
\emlineto{109.708}{31.839}
\emmoveto{109.732}{31.745}
\emlineto{109.779}{31.586}
\emmoveto{109.801}{31.491}
\emlineto{109.845}{31.333}
\emmoveto{109.866}{31.238}
\emlineto{109.908}{31.079}
\emmoveto{109.927}{30.985}
\emlineto{109.966}{30.826}
\emmoveto{109.984}{30.731}
\emlineto{110.019}{30.572}
\emmoveto{110.036}{30.478}
\emlineto{110.069}{30.319}
\emmoveto{110.084}{30.224}
\emlineto{110.114}{30.066}
\emmoveto{110.128}{29.971}
\emlineto{110.155}{29.812}
\emmoveto{110.168}{29.718}
\emlineto{110.192}{29.559}
\emmoveto{110.203}{29.464}
\emlineto{110.224}{29.305}
\emmoveto{110.234}{29.211}
\emlineto{110.252}{29.052}
\emmoveto{110.260}{28.957}
\emlineto{110.276}{28.798}
\emmoveto{110.283}{28.704}
\emlineto{110.295}{28.545}
\emmoveto{110.301}{28.450}
\emlineto{110.311}{28.291}
\emmoveto{110.315}{28.197}
\emlineto{110.322}{28.038}
\emmoveto{110.325}{27.943}
\emlineto{110.329}{27.784}
\emmoveto{110.330}{27.690}
\emlineto{110.331}{27.531}
\emmoveto{110.331}{27.436}
\emlineto{110.329}{27.277}
\emmoveto{110.328}{27.183}
\emlineto{110.323}{27.024}
\emmoveto{110.320}{26.929}
\emlineto{110.313}{26.770}
\emmoveto{110.308}{26.676}
\emlineto{110.298}{26.517}
\emmoveto{110.292}{26.422}
\emlineto{110.279}{26.263}
\emmoveto{110.272}{26.168}
\emlineto{110.256}{26.010}
\emmoveto{110.248}{25.915}
\emlineto{110.229}{25.756}
\emmoveto{110.219}{25.661}
\emlineto{110.197}{25.503}
\emmoveto{110.186}{25.408}
\emlineto{110.161}{25.249}
\emmoveto{110.148}{25.155}
\emlineto{110.121}{24.996}
\emmoveto{110.107}{24.901}
\emlineto{110.076}{24.742}
\emmoveto{110.061}{24.648}
\emlineto{110.028}{24.489}
\emmoveto{110.010}{24.394}
\emlineto{109.975}{24.235}
\emmoveto{109.956}{24.141}
\emlineto{109.917}{23.982}
\emmoveto{109.897}{23.887}
\emlineto{109.856}{23.729}
\emmoveto{109.834}{23.634}
\emlineto{109.790}{23.475}
\emmoveto{109.767}{23.381}
\emlineto{109.720}{23.222}
\emmoveto{109.695}{23.128}
\emlineto{109.645}{22.969}
\emmoveto{109.620}{22.874}
\emlineto{109.567}{22.716}
\emmoveto{109.539}{22.621}
\emlineto{109.484}{22.462}
\emmoveto{109.455}{22.368}
\emlineto{109.397}{22.209}
\emmoveto{109.367}{22.115}
\emlineto{109.305}{21.956}
\emmoveto{109.274}{21.862}
\emlineto{109.209}{21.703}
\emmoveto{109.177}{21.608}
\emlineto{109.109}{21.450}
\emmoveto{109.075}{21.355}
\emlineto{109.005}{21.197}
\emmoveto{108.969}{21.102}
\emlineto{108.897}{20.944}
\emmoveto{108.859}{20.849}
\emlineto{108.784}{20.691}
\emmoveto{108.745}{20.596}
\emlineto{108.667}{20.438}
\emmoveto{108.627}{20.343}
\emlineto{108.546}{20.185}
\emmoveto{108.504}{20.091}
\emlineto{108.420}{19.932}
\emmoveto{108.377}{19.838}
\emlineto{108.290}{19.679}
\emmoveto{108.246}{19.585}
\emlineto{108.156}{19.427}
\emmoveto{108.110}{19.332}
\emlineto{108.018}{19.174}
\emmoveto{107.971}{19.080}
\emlineto{107.875}{18.921}
\emmoveto{107.827}{18.827}
\emlineto{107.728}{18.669}
\emmoveto{107.678}{18.574}
\emlineto{107.577}{18.416}
\emmoveto{107.526}{18.322}
\emlineto{107.422}{18.164}
\emmoveto{107.369}{18.069}
\emlineto{107.262}{17.911}
\emmoveto{107.208}{17.817}
\emlineto{107.098}{17.659}
\emmoveto{107.043}{17.565}
\emlineto{106.930}{17.407}
\emmoveto{106.873}{17.313}
\emlineto{106.758}{17.154}
\emmoveto{106.699}{17.060}
\emlineto{106.581}{16.902}
\emmoveto{106.521}{16.808}
\emlineto{106.400}{16.650}
\emmoveto{106.339}{16.556}
\emlineto{106.215}{16.398}
\emmoveto{106.153}{16.304}
\emlineto{106.026}{16.146}
\emmoveto{105.962}{16.052}
\emlineto{105.832}{15.894}
\emmoveto{105.767}{15.800}
\emlineto{105.634}{15.642}
\emmoveto{105.568}{15.549}
\emlineto{105.432}{15.391}
\emmoveto{105.364}{15.297}
\emlineto{105.226}{15.139}
\emmoveto{105.156}{15.045}
\emlineto{105.015}{14.887}
\emmoveto{104.944}{14.794}
\emlineto{104.801}{14.636}
\emmoveto{104.728}{14.542}
\emlineto{104.582}{14.384}
\emmoveto{104.508}{14.291}
\emlineto{104.358}{14.133}
\emmoveto{104.283}{14.039}
\emlineto{104.131}{13.882}
\emmoveto{104.054}{13.788}
\emlineto{103.899}{13.631}
\emmoveto{103.821}{13.537}
\emlineto{103.663}{13.380}
\emmoveto{103.583}{13.286}
\emlineto{103.423}{13.129}
\emmoveto{103.342}{13.035}
\emlineto{103.178}{12.878}
\emmoveto{103.096}{12.784}
\emlineto{102.930}{12.627}
\emmoveto{102.846}{12.533}
\emlineto{102.677}{12.376}
\emmoveto{102.592}{12.283}
\emlineto{102.420}{12.126}
\emmoveto{102.333}{12.032}
\emlineto{102.158}{11.875}
\emmoveto{102.070}{11.782}
\emlineto{101.893}{11.625}
\emmoveto{101.803}{11.531}
\emlineto{101.623}{11.374}
\emmoveto{101.532}{11.281}
\emlineto{101.349}{11.124}
\emmoveto{101.257}{11.031}
\emlineto{101.071}{10.874}
\emmoveto{100.977}{10.781}
\emlineto{100.788}{10.624}
\emshow{83.980}{17.700}{Classical particle}
\emshow{1.000}{10.000}{-1.00e-1}
\emshow{1.000}{17.000}{-6.00e-2}
\emshow{1.000}{24.000}{-2.00e-2}
\emshow{1.000}{31.000}{2.00e-2}
\emshow{1.000}{38.000}{6.00e-2}
\emshow{1.000}{45.000}{1.00e-1}
\emshow{1.000}{52.000}{1.40e-1}
\emshow{1.000}{59.000}{1.80e-1}
\emshow{1.000}{66.000}{2.20e-1}
\emshow{1.000}{73.000}{2.60e-1}
\emshow{1.000}{80.000}{3.00e-1}
\emshow{12.000}{5.000}{-5.00e-4}
\emshow{23.800}{5.000}{1.20e-1}
\emshow{35.600}{5.000}{2.40e-1}
\emshow{47.400}{5.000}{3.60e-1}
\emshow{59.200}{5.000}{4.80e-1}
\emshow{71.000}{5.000}{6.00e-1}
\emshow{82.800}{5.000}{7.20e-1}
\emshow{94.600}{5.000}{8.40e-1}
\emshow{106.400}{5.000}{9.60e-1}
\emshow{118.200}{5.000}{1.08e0}
\emshow{130.000}{5.000}{1.20e0}
\centerline{\bf{Fig.A.2}}
\eject
\newcount\numpoint
\newcount\numpointo
\numpoint=1 \numpointo=1
\def\emmoveto#1#2{\offinterlineskip
\hbox to 0 true cm{\vbox to 0
true cm{\vskip - #2 true mm
\hskip #1 true mm \special{em:point
\the\numpoint}\vss}\hss}
\numpointo=\numpoint
\global\advance \numpoint by 1}
\def\emlineto#1#2{\offinterlineskip
\hbox to 0 true cm{\vbox to 0
true cm{\vskip - #2 true mm
\hskip #1 true mm \special{em:point
\the\numpoint}\vss}\hss}
\special{em:line
\the\numpointo,\the\numpoint}
\numpointo=\numpoint
\global\advance \numpoint by 1}
\def\emshow#1#2#3{\offinterlineskip
\hbox to 0 true cm{\vbox to 0
true cm{\vskip - #2 true mm
\hskip #1 true mm \vbox to 0
true cm{\vss\hbox{#3\hss
}}\vss}\hss}}
\special{em:linewidth 0.8pt}

\vrule width 0 mm height                0 mm depth 90.000 true mm

\special{em:linewidth 0.8pt}
\emmoveto{130.000}{10.000}
\emlineto{12.000}{10.000}
\emlineto{12.000}{80.000}
\emmoveto{71.000}{10.000}
\emlineto{71.000}{80.000}
\emmoveto{12.000}{45.000}
\emlineto{130.000}{45.000}
\emmoveto{130.000}{10.000}
\emlineto{130.000}{80.000}
\emlineto{12.000}{80.000}
\emlineto{12.000}{10.000}
\emlineto{130.000}{10.000}
\special{em:linewidth 0.4pt}
\emmoveto{12.000}{17.000}
\emlineto{130.000}{17.000}
\emmoveto{12.000}{24.000}
\emlineto{130.000}{24.000}
\emmoveto{12.000}{31.000}
\emlineto{130.000}{31.000}
\emmoveto{12.000}{38.000}
\emlineto{130.000}{38.000}
\emmoveto{12.000}{45.000}
\emlineto{130.000}{45.000}
\emmoveto{12.000}{52.000}
\emlineto{130.000}{52.000}
\emmoveto{12.000}{59.000}
\emlineto{130.000}{59.000}
\emmoveto{12.000}{66.000}
\emlineto{130.000}{66.000}
\emmoveto{12.000}{73.000}
\emlineto{130.000}{73.000}
\emmoveto{23.800}{10.000}
\emlineto{23.800}{80.000}
\emmoveto{35.600}{10.000}
\emlineto{35.600}{80.000}
\emmoveto{47.400}{10.000}
\emlineto{47.400}{80.000}
\emmoveto{59.200}{10.000}
\emlineto{59.200}{80.000}
\emmoveto{71.000}{10.000}
\emlineto{71.000}{80.000}
\emmoveto{82.800}{10.000}
\emlineto{82.800}{80.000}
\emmoveto{94.600}{10.000}
\emlineto{94.600}{80.000}
\emmoveto{106.400}{10.000}
\emlineto{106.400}{80.000}
\emmoveto{118.200}{10.000}
\emlineto{118.200}{80.000}
\special{em:linewidth 0.8pt}
\emmoveto{12.000}{80.000}
\emlineto{12.147}{79.968}
\emmoveto{12.221}{79.935}
\emlineto{12.368}{79.900}
\emmoveto{12.442}{79.867}
\emlineto{12.589}{79.832}
\emmoveto{12.663}{79.799}
\emlineto{12.810}{79.764}
\emmoveto{12.883}{79.731}
\emlineto{13.030}{79.695}
\emmoveto{13.103}{79.663}
\emlineto{13.250}{79.627}
\emmoveto{13.323}{79.594}
\emlineto{13.470}{79.559}
\emmoveto{13.543}{79.526}
\emlineto{13.690}{79.490}
\emmoveto{13.763}{79.457}
\emlineto{13.909}{79.422}
\emmoveto{13.982}{79.389}
\emlineto{14.128}{79.353}
\emmoveto{14.201}{79.320}
\emlineto{14.347}{79.284}
\emmoveto{14.420}{79.251}
\emlineto{14.566}{79.215}
\emmoveto{14.639}{79.182}
\emlineto{14.784}{79.146}
\emmoveto{14.857}{79.113}
\emlineto{15.003}{79.077}
\emmoveto{15.075}{79.044}
\emlineto{15.221}{79.008}
\emmoveto{15.293}{78.975}
\emlineto{15.438}{78.938}
\emmoveto{15.511}{78.905}
\emlineto{15.656}{78.869}
\emmoveto{15.728}{78.835}
\emlineto{15.873}{78.799}
\emmoveto{15.946}{78.765}
\emlineto{16.090}{78.729}
\emmoveto{16.163}{78.695}
\emlineto{16.307}{78.659}
\emmoveto{16.380}{78.625}
\emlineto{16.524}{78.589}
\emmoveto{16.596}{78.555}
\emlineto{16.740}{78.518}
\emmoveto{16.812}{78.485}
\emlineto{16.956}{78.448}
\emmoveto{17.029}{78.415}
\emlineto{17.172}{78.378}
\emmoveto{17.244}{78.345}
\emlineto{17.388}{78.308}
\emmoveto{17.460}{78.274}
\emlineto{17.604}{78.237}
\emmoveto{17.675}{78.204}
\emlineto{17.819}{78.167}
\emmoveto{17.891}{78.134}
\emlineto{18.034}{78.097}
\emmoveto{18.106}{78.063}
\emlineto{18.249}{78.026}
\emmoveto{18.320}{77.993}
\emlineto{18.463}{77.956}
\emmoveto{18.535}{77.922}
\emlineto{18.678}{77.885}
\emmoveto{18.749}{77.851}
\emlineto{18.892}{77.814}
\emmoveto{18.963}{77.781}
\emlineto{19.106}{77.743}
\emmoveto{19.177}{77.710}
\emlineto{19.319}{77.673}
\emmoveto{19.390}{77.639}
\emlineto{19.532}{77.602}
\emmoveto{19.604}{77.568}
\emlineto{19.746}{77.531}
\emmoveto{19.817}{77.497}
\emlineto{19.959}{77.459}
\emmoveto{20.030}{77.426}
\emlineto{20.171}{77.388}
\emmoveto{20.242}{77.354}
\emlineto{20.384}{77.317}
\emmoveto{20.454}{77.283}
\emlineto{20.596}{77.245}
\emmoveto{20.667}{77.211}
\emlineto{20.808}{77.174}
\emmoveto{20.878}{77.140}
\emlineto{21.020}{77.102}
\emmoveto{21.090}{77.068}
\emlineto{21.231}{77.031}
\emmoveto{21.302}{76.997}
\emlineto{21.442}{76.959}
\emmoveto{21.513}{76.926}
\emlineto{21.653}{76.888}
\emmoveto{21.724}{76.854}
\emlineto{21.864}{76.817}
\emmoveto{21.934}{76.783}
\emlineto{22.075}{76.745}
\emmoveto{22.145}{76.711}
\emlineto{22.285}{76.674}
\emmoveto{22.355}{76.640}
\emlineto{22.495}{76.602}
\emmoveto{22.565}{76.568}
\emlineto{22.705}{76.531}
\emmoveto{22.775}{76.497}
\emlineto{22.915}{76.459}
\emmoveto{22.984}{76.426}
\emlineto{23.124}{76.388}
\emmoveto{23.194}{76.354}
\emlineto{23.333}{76.316}
\emmoveto{23.403}{76.283}
\emlineto{23.542}{76.245}
\emmoveto{23.612}{76.211}
\emlineto{23.751}{76.174}
\emmoveto{23.820}{76.140}
\emlineto{23.959}{76.102}
\emmoveto{24.029}{76.068}
\emlineto{24.167}{76.030}
\emmoveto{24.237}{75.996}
\emlineto{24.375}{75.958}
\emmoveto{24.444}{75.924}
\emlineto{24.583}{75.886}
\emmoveto{24.652}{75.852}
\emlineto{24.790}{75.814}
\emmoveto{24.859}{75.781}
\emlineto{24.998}{75.743}
\emmoveto{25.067}{75.709}
\emlineto{25.205}{75.671}
\emmoveto{25.274}{75.637}
\emlineto{25.411}{75.600}
\emmoveto{25.480}{75.566}
\emlineto{25.618}{75.529}
\emmoveto{25.687}{75.496}
\emlineto{25.824}{75.460}
\emmoveto{25.893}{75.427}
\emlineto{26.030}{75.392}
\emmoveto{26.099}{75.359}
\emlineto{26.236}{75.323}
\emmoveto{26.305}{75.290}
\emlineto{26.442}{75.254}
\emmoveto{26.510}{75.221}
\emlineto{26.647}{75.185}
\emmoveto{26.715}{75.152}
\emlineto{26.852}{75.116}
\emmoveto{26.921}{75.083}
\emlineto{27.057}{75.047}
\emmoveto{27.125}{75.014}
\emlineto{27.262}{74.977}
\emmoveto{27.330}{74.944}
\emlineto{27.466}{74.908}
\emmoveto{27.534}{74.874}
\emlineto{27.671}{74.838}
\emmoveto{27.739}{74.805}
\emlineto{27.875}{74.768}
\emmoveto{27.943}{74.735}
\emlineto{28.078}{74.698}
\emmoveto{28.146}{74.665}
\emlineto{28.282}{74.628}
\emmoveto{28.350}{74.594}
\emlineto{28.485}{74.557}
\emmoveto{28.553}{74.523}
\emlineto{28.688}{74.486}
\emmoveto{28.756}{74.452}
\emlineto{28.891}{74.415}
\emmoveto{28.959}{74.381}
\emlineto{29.094}{74.344}
\emmoveto{29.161}{74.311}
\emlineto{29.296}{74.273}
\emmoveto{29.364}{74.240}
\emlineto{29.498}{74.202}
\emmoveto{29.566}{74.169}
\emlineto{29.700}{74.131}
\emmoveto{29.767}{74.097}
\emlineto{29.902}{74.060}
\emmoveto{29.969}{74.026}
\emlineto{30.103}{73.989}
\emmoveto{30.170}{73.955}
\emlineto{30.304}{73.918}
\emmoveto{30.371}{73.884}
\emlineto{30.505}{73.846}
\emmoveto{30.572}{73.812}
\emlineto{30.706}{73.775}
\emmoveto{30.773}{73.741}
\emlineto{30.907}{73.703}
\emmoveto{30.973}{73.669}
\emlineto{31.107}{73.632}
\emmoveto{31.174}{73.598}
\emlineto{31.307}{73.560}
\emmoveto{31.373}{73.526}
\emlineto{31.507}{73.488}
\emmoveto{31.573}{73.455}
\emlineto{31.706}{73.417}
\emmoveto{31.773}{73.383}
\emlineto{31.906}{73.345}
\emmoveto{31.972}{73.311}
\emlineto{32.105}{73.273}
\emmoveto{32.171}{73.239}
\emlineto{32.303}{73.201}
\emmoveto{32.370}{73.167}
\emlineto{32.502}{73.128}
\emmoveto{32.568}{73.094}
\emlineto{32.700}{73.055}
\emmoveto{32.766}{73.021}
\emlineto{32.898}{72.982}
\emmoveto{32.964}{72.948}
\emlineto{33.096}{72.910}
\emmoveto{33.162}{72.875}
\emlineto{33.294}{72.837}
\emmoveto{33.360}{72.803}
\emlineto{33.491}{72.764}
\emmoveto{33.557}{72.730}
\emlineto{33.688}{72.691}
\emmoveto{33.754}{72.657}
\emlineto{33.885}{72.619}
\emmoveto{33.951}{72.585}
\emlineto{34.082}{72.546}
\emmoveto{34.148}{72.512}
\emlineto{34.278}{72.473}
\emmoveto{34.344}{72.439}
\emlineto{34.475}{72.401}
\emmoveto{34.540}{72.366}
\emlineto{34.671}{72.328}
\emmoveto{34.736}{72.294}
\emlineto{34.866}{72.255}
\emmoveto{34.931}{72.221}
\emlineto{35.062}{72.183}
\emmoveto{35.127}{72.148}
\emlineto{35.257}{72.110}
\emmoveto{35.322}{72.076}
\emlineto{35.452}{72.037}
\emmoveto{35.517}{72.003}
\emlineto{35.647}{71.964}
\emmoveto{35.712}{71.930}
\emlineto{35.841}{71.892}
\emmoveto{35.906}{71.857}
\emlineto{36.035}{71.819}
\emmoveto{36.100}{71.785}
\emlineto{36.229}{71.746}
\emmoveto{36.294}{71.712}
\emlineto{36.423}{71.673}
\emmoveto{36.488}{71.639}
\emlineto{36.617}{71.600}
\emmoveto{36.681}{71.565}
\emlineto{36.810}{71.526}
\emmoveto{36.874}{71.492}
\emlineto{37.003}{71.453}
\emmoveto{37.067}{71.419}
\emlineto{37.196}{71.380}
\emmoveto{37.260}{71.345}
\emlineto{37.388}{71.307}
\emmoveto{37.517}{71.248}
\emlineto{37.709}{71.185}
\emmoveto{37.837}{71.126}
\emlineto{38.029}{71.064}
\emmoveto{38.156}{71.007}
\emlineto{38.348}{70.948}
\emmoveto{38.475}{70.891}
\emlineto{38.666}{70.832}
\emmoveto{38.793}{70.775}
\emlineto{38.984}{70.715}
\emmoveto{39.111}{70.658}
\emlineto{39.301}{70.598}
\emmoveto{39.428}{70.541}
\emlineto{39.617}{70.481}
\emmoveto{39.744}{70.424}
\emlineto{39.933}{70.363}
\emmoveto{40.060}{70.306}
\emlineto{40.249}{70.245}
\emmoveto{40.374}{70.188}
\emlineto{40.563}{70.126}
\emmoveto{40.689}{70.068}
\emlineto{40.877}{70.006}
\emmoveto{41.002}{69.948}
\emlineto{41.190}{69.886}
\emmoveto{41.315}{69.827}
\emlineto{41.503}{69.765}
\emmoveto{41.628}{69.707}
\emlineto{41.815}{69.645}
\emmoveto{41.939}{69.586}
\emlineto{42.126}{69.524}
\emmoveto{42.250}{69.466}
\emlineto{42.437}{69.403}
\emmoveto{42.561}{69.345}
\emlineto{42.746}{69.282}
\emmoveto{42.870}{69.224}
\emlineto{43.056}{69.161}
\emmoveto{43.179}{69.102}
\emlineto{43.364}{69.039}
\emmoveto{43.488}{68.981}
\emlineto{43.672}{68.918}
\emmoveto{43.795}{68.859}
\emlineto{43.979}{68.796}
\emmoveto{44.102}{68.737}
\emlineto{44.286}{68.673}
\emmoveto{44.408}{68.614}
\emlineto{44.592}{68.550}
\emmoveto{44.714}{68.491}
\emlineto{44.897}{68.427}
\emmoveto{45.019}{68.367}
\emlineto{45.202}{68.303}
\emmoveto{45.323}{68.243}
\emlineto{45.505}{68.179}
\emmoveto{45.627}{68.120}
\emlineto{45.809}{68.056}
\emmoveto{45.930}{67.996}
\emlineto{46.111}{67.932}
\emmoveto{46.232}{67.873}
\emlineto{46.413}{67.809}
\emmoveto{46.533}{67.750}
\emlineto{46.714}{67.686}
\emmoveto{46.834}{67.626}
\emlineto{47.014}{67.562}
\emmoveto{47.134}{67.503}
\emlineto{47.314}{67.439}
\emmoveto{47.434}{67.380}
\emlineto{47.613}{67.316}
\emmoveto{47.733}{67.256}
\emlineto{47.912}{67.192}
\emmoveto{48.031}{67.133}
\emlineto{48.209}{67.069}
\emmoveto{48.328}{67.010}
\emlineto{48.506}{66.945}
\emmoveto{48.625}{66.885}
\emlineto{48.803}{66.821}
\emmoveto{48.921}{66.761}
\emlineto{49.098}{66.697}
\emmoveto{49.216}{66.640}
\emlineto{49.393}{66.579}
\emmoveto{49.511}{66.522}
\emlineto{49.688}{66.462}
\emmoveto{49.805}{66.404}
\emlineto{49.981}{66.344}
\emmoveto{50.099}{66.286}
\emlineto{50.274}{66.225}
\emmoveto{50.391}{66.167}
\emlineto{50.567}{66.106}
\emmoveto{50.684}{66.048}
\emlineto{50.858}{65.987}
\emmoveto{50.975}{65.929}
\emlineto{51.150}{65.867}
\emmoveto{51.266}{65.809}
\emlineto{51.440}{65.747}
\emmoveto{51.556}{65.689}
\emlineto{51.730}{65.626}
\emmoveto{51.845}{65.568}
\emlineto{52.019}{65.505}
\emmoveto{52.134}{65.446}
\emlineto{52.307}{65.383}
\emmoveto{52.422}{65.324}
\emlineto{52.595}{65.260}
\emmoveto{52.710}{65.201}
\emlineto{52.882}{65.136}
\emmoveto{52.997}{65.077}
\emlineto{53.168}{65.013}
\emmoveto{53.283}{64.953}
\emlineto{53.454}{64.889}
\emmoveto{53.568}{64.830}
\emlineto{53.739}{64.766}
\emmoveto{53.853}{64.706}
\emlineto{54.023}{64.642}
\emmoveto{54.137}{64.582}
\emlineto{54.307}{64.518}
\emmoveto{54.420}{64.458}
\emlineto{54.590}{64.394}
\emmoveto{54.703}{64.334}
\emlineto{54.872}{64.270}
\emmoveto{54.985}{64.210}
\emlineto{55.154}{64.145}
\emmoveto{55.266}{64.085}
\emlineto{55.435}{64.021}
\emmoveto{55.547}{63.961}
\emlineto{55.715}{63.896}
\emmoveto{55.827}{63.836}
\emlineto{55.995}{63.771}
\emmoveto{56.106}{63.711}
\emlineto{56.273}{63.644}
\emmoveto{56.385}{63.584}
\emlineto{56.552}{63.518}
\emmoveto{56.663}{63.457}
\emlineto{56.829}{63.392}
\emmoveto{56.940}{63.331}
\emlineto{57.106}{63.266}
\emmoveto{57.216}{63.205}
\emlineto{57.382}{63.140}
\emmoveto{57.492}{63.080}
\emlineto{57.657}{63.014}
\emmoveto{57.767}{62.954}
\emlineto{57.932}{62.889}
\emmoveto{58.042}{62.829}
\emlineto{58.206}{62.764}
\emmoveto{58.315}{62.704}
\emlineto{58.479}{62.639}
\emmoveto{58.588}{62.579}
\emlineto{58.752}{62.515}
\emmoveto{58.861}{62.456}
\emlineto{59.024}{62.395}
\emmoveto{59.132}{62.338}
\emlineto{59.295}{62.277}
\emmoveto{59.403}{62.220}
\emlineto{59.566}{62.159}
\emmoveto{59.674}{62.101}
\emlineto{59.836}{62.039}
\emmoveto{59.943}{61.980}
\emlineto{60.105}{61.917}
\emmoveto{60.212}{61.859}
\emlineto{60.374}{61.796}
\emmoveto{60.481}{61.737}
\emlineto{60.642}{61.673}
\emmoveto{60.749}{61.615}
\emlineto{60.909}{61.551}
\emmoveto{61.016}{61.492}
\emlineto{61.175}{61.429}
\emmoveto{61.282}{61.369}
\emlineto{61.441}{61.306}
\emmoveto{61.548}{61.246}
\emlineto{61.707}{61.182}
\emmoveto{61.812}{61.123}
\emlineto{61.971}{61.059}
\emmoveto{62.077}{61.000}
\emlineto{62.235}{60.935}
\emmoveto{62.340}{60.876}
\emlineto{62.498}{60.811}
\emmoveto{62.603}{60.752}
\emlineto{62.761}{60.687}
\emmoveto{62.866}{60.627}
\emlineto{63.023}{60.562}
\emmoveto{63.127}{60.502}
\emlineto{63.284}{60.437}
\emmoveto{63.388}{60.377}
\emlineto{63.544}{60.311}
\emmoveto{63.648}{60.250}
\emlineto{63.804}{60.183}
\emmoveto{63.908}{60.122}
\emlineto{64.063}{60.055}
\emmoveto{64.167}{59.994}
\emlineto{64.321}{59.928}
\emmoveto{64.425}{59.867}
\emlineto{64.579}{59.801}
\emmoveto{64.682}{59.740}
\emlineto{64.836}{59.673}
\emmoveto{64.939}{59.613}
\emlineto{65.092}{59.546}
\emmoveto{65.195}{59.486}
\emlineto{65.348}{59.419}
\emmoveto{65.450}{59.359}
\emlineto{65.603}{59.293}
\emmoveto{65.704}{59.232}
\emlineto{65.857}{59.166}
\emmoveto{65.958}{59.105}
\emlineto{66.110}{59.039}
\emmoveto{66.211}{58.978}
\emlineto{66.363}{58.913}
\emmoveto{66.464}{58.852}
\emlineto{66.615}{58.786}
\emmoveto{66.716}{58.725}
\emlineto{66.867}{58.659}
\emmoveto{66.967}{58.599}
\emlineto{67.117}{58.531}
\emmoveto{67.217}{58.470}
\emlineto{67.367}{58.403}
\emmoveto{67.467}{58.344}
\emlineto{67.616}{58.282}
\emmoveto{67.716}{58.224}
\emlineto{67.865}{58.162}
\emmoveto{67.964}{58.104}
\emlineto{68.113}{58.042}
\emmoveto{68.212}{57.984}
\emlineto{68.360}{57.921}
\emmoveto{68.459}{57.863}
\emlineto{68.607}{57.800}
\emmoveto{68.705}{57.741}
\emlineto{68.853}{57.678}
\emmoveto{68.951}{57.620}
\emlineto{69.098}{57.556}
\emmoveto{69.196}{57.497}
\emlineto{69.343}{57.434}
\emmoveto{69.440}{57.375}
\emlineto{69.586}{57.311}
\emmoveto{69.684}{57.251}
\emlineto{69.830}{57.187}
\emmoveto{69.927}{57.127}
\emlineto{70.072}{57.063}
\emmoveto{70.169}{57.003}
\emlineto{70.314}{56.938}
\emmoveto{70.411}{56.878}
\emlineto{70.555}{56.813}
\emmoveto{70.652}{56.751}
\emlineto{70.796}{56.684}
\emmoveto{70.892}{56.623}
\emlineto{71.035}{56.556}
\emmoveto{71.131}{56.495}
\emlineto{71.275}{56.428}
\emmoveto{71.370}{56.367}
\emlineto{71.513}{56.300}
\emmoveto{71.608}{56.238}
\emlineto{71.751}{56.171}
\emmoveto{71.845}{56.110}
\emlineto{71.987}{56.043}
\emmoveto{72.082}{55.982}
\emlineto{72.224}{55.916}
\emmoveto{72.318}{55.854}
\emlineto{72.459}{55.788}
\emmoveto{72.553}{55.726}
\emlineto{72.694}{55.659}
\emmoveto{72.788}{55.598}
\emlineto{72.928}{55.531}
\emmoveto{73.022}{55.470}
\emlineto{73.162}{55.403}
\emmoveto{73.255}{55.342}
\emlineto{73.394}{55.275}
\emmoveto{73.487}{55.214}
\emlineto{73.626}{55.147}
\emmoveto{73.719}{55.085}
\emlineto{73.858}{55.018}
\emmoveto{73.950}{54.956}
\emlineto{74.088}{54.887}
\emmoveto{74.180}{54.825}
\emlineto{74.318}{54.757}
\emmoveto{74.410}{54.695}
\emlineto{74.547}{54.627}
\emmoveto{74.639}{54.565}
\emlineto{74.776}{54.497}
\emmoveto{74.867}{54.436}
\emlineto{75.003}{54.372}
\emmoveto{75.094}{54.315}
\emlineto{75.230}{54.253}
\emmoveto{75.321}{54.195}
\emlineto{75.457}{54.134}
\emmoveto{75.547}{54.075}
\emlineto{75.682}{54.013}
\emmoveto{75.773}{53.955}
\emlineto{75.908}{53.893}
\emmoveto{75.997}{53.834}
\emlineto{76.132}{53.771}
\emmoveto{76.221}{53.712}
\emlineto{76.356}{53.649}
\emmoveto{76.445}{53.590}
\emlineto{76.579}{53.526}
\emmoveto{76.668}{53.467}
\emlineto{76.801}{53.402}
\emmoveto{76.890}{53.343}
\emlineto{77.023}{53.278}
\emmoveto{77.111}{53.218}
\emlineto{77.244}{53.153}
\emmoveto{77.332}{53.092}
\emlineto{77.464}{53.024}
\emmoveto{77.552}{52.962}
\emlineto{77.684}{52.895}
\emmoveto{77.771}{52.833}
\emlineto{77.902}{52.766}
\emmoveto{77.990}{52.704}
\emlineto{78.121}{52.637}
\emmoveto{78.208}{52.575}
\emlineto{78.338}{52.508}
\emmoveto{78.425}{52.446}
\emlineto{78.555}{52.379}
\emmoveto{78.641}{52.317}
\emlineto{78.771}{52.250}
\emmoveto{78.857}{52.188}
\emlineto{78.986}{52.121}
\emmoveto{79.072}{52.059}
\emlineto{79.201}{51.992}
\emmoveto{79.286}{51.930}
\emlineto{79.415}{51.863}
\emmoveto{79.500}{51.801}
\emlineto{79.628}{51.733}
\emmoveto{79.713}{51.672}
\emlineto{79.840}{51.604}
\emmoveto{79.925}{51.542}
\emlineto{80.052}{51.474}
\emmoveto{80.137}{51.413}
\emlineto{80.263}{51.345}
\emmoveto{80.347}{51.283}
\emlineto{80.473}{51.214}
\emmoveto{80.557}{51.150}
\emlineto{80.683}{51.080}
\emmoveto{80.767}{51.017}
\emlineto{80.892}{50.948}
\emmoveto{80.975}{50.885}
\emlineto{81.100}{50.816}
\emmoveto{81.183}{50.753}
\emlineto{81.307}{50.685}
\emmoveto{81.390}{50.623}
\emlineto{81.514}{50.555}
\emmoveto{81.597}{50.498}
\emlineto{81.720}{50.437}
\emmoveto{81.802}{50.379}
\emlineto{81.925}{50.318}
\emmoveto{82.007}{50.260}
\emlineto{82.130}{50.198}
\emmoveto{82.212}{50.140}
\emlineto{82.334}{50.077}
\emmoveto{82.416}{50.019}
\emlineto{82.537}{49.956}
\emmoveto{82.619}{49.898}
\emlineto{82.740}{49.834}
\emmoveto{82.821}{49.775}
\emlineto{82.942}{49.711}
\emmoveto{83.023}{49.652}
\emlineto{83.144}{49.587}
\emmoveto{83.224}{49.527}
\emlineto{83.344}{49.463}
\emmoveto{83.424}{49.402}
\emlineto{83.544}{49.337}
\emmoveto{83.624}{49.275}
\emlineto{83.743}{49.207}
\emmoveto{83.823}{49.144}
\emlineto{83.942}{49.076}
\emmoveto{84.021}{49.013}
\emlineto{84.140}{48.945}
\emmoveto{84.219}{48.883}
\emlineto{84.337}{48.814}
\emmoveto{84.415}{48.752}
\emlineto{84.533}{48.684}
\emmoveto{84.612}{48.621}
\emlineto{84.729}{48.553}
\emmoveto{84.807}{48.491}
\emlineto{84.924}{48.423}
\emmoveto{85.001}{48.360}
\emlineto{85.118}{48.292}
\emmoveto{85.195}{48.230}
\emlineto{85.312}{48.162}
\emmoveto{85.389}{48.099}
\emlineto{85.504}{48.031}
\emmoveto{85.581}{47.969}
\emlineto{85.696}{47.901}
\emmoveto{85.773}{47.838}
\emlineto{85.888}{47.770}
\emmoveto{85.964}{47.707}
\emlineto{86.078}{47.639}
\emmoveto{86.154}{47.577}
\emlineto{86.268}{47.508}
\emmoveto{86.344}{47.446}
\emlineto{86.457}{47.377}
\emmoveto{86.533}{47.314}
\emlineto{86.646}{47.242}
\emmoveto{86.721}{47.178}
\emlineto{86.833}{47.107}
\emmoveto{86.908}{47.044}
\emlineto{87.020}{46.973}
\emmoveto{87.095}{46.910}
\emlineto{87.244}{46.814}
\emmoveto{87.355}{46.733}
\emlineto{87.503}{46.648}
\emmoveto{87.614}{46.566}
\emlineto{87.761}{46.480}
\emmoveto{87.871}{46.398}
\emlineto{88.018}{46.312}
\emmoveto{88.128}{46.229}
\emlineto{88.273}{46.141}
\emmoveto{88.382}{46.058}
\emlineto{88.527}{45.969}
\emmoveto{88.636}{45.885}
\emlineto{88.780}{45.796}
\emmoveto{88.888}{45.711}
\emlineto{89.032}{45.620}
\emmoveto{89.139}{45.535}
\emlineto{89.282}{45.443}
\emmoveto{89.389}{45.356}
\emlineto{89.531}{45.261}
\emmoveto{89.637}{45.170}
\emlineto{89.778}{45.072}
\emmoveto{89.883}{44.981}
\emlineto{90.024}{44.884}
\emmoveto{90.128}{44.794}
\emlineto{90.268}{44.697}
\emmoveto{90.372}{44.607}
\emlineto{90.511}{44.511}
\emmoveto{90.615}{44.421}
\emlineto{90.752}{44.325}
\emmoveto{90.855}{44.236}
\emlineto{90.992}{44.140}
\emmoveto{91.095}{44.051}
\emlineto{91.231}{43.955}
\emmoveto{91.333}{43.866}
\emlineto{91.468}{43.770}
\emmoveto{91.570}{43.681}
\emlineto{91.704}{43.586}
\emmoveto{91.805}{43.497}
\emlineto{91.938}{43.402}
\emmoveto{92.038}{43.314}
\emlineto{92.171}{43.216}
\emmoveto{92.271}{43.128}
\emlineto{92.403}{43.041}
\emmoveto{92.502}{42.958}
\emlineto{92.633}{42.871}
\emmoveto{92.731}{42.788}
\emlineto{92.862}{42.700}
\emmoveto{92.960}{42.616}
\emlineto{93.090}{42.528}
\emmoveto{93.187}{42.443}
\emlineto{93.316}{42.354}
\emmoveto{93.412}{42.270}
\emlineto{93.541}{42.179}
\emmoveto{93.637}{42.094}
\emlineto{93.764}{42.004}
\emmoveto{93.860}{41.918}
\emlineto{93.986}{41.826}
\emmoveto{94.081}{41.739}
\emlineto{94.207}{41.647}
\emmoveto{94.301}{41.559}
\emlineto{94.427}{41.466}
\emmoveto{94.520}{41.378}
\emlineto{94.645}{41.283}
\emmoveto{94.738}{41.194}
\emlineto{94.861}{41.095}
\emmoveto{94.953}{41.000}
\emlineto{95.076}{40.898}
\emmoveto{95.168}{40.804}
\emlineto{95.290}{40.704}
\emmoveto{95.381}{40.611}
\emlineto{95.502}{40.511}
\emmoveto{95.592}{40.419}
\emlineto{95.712}{40.321}
\emmoveto{95.802}{40.230}
\emlineto{95.922}{40.132}
\emmoveto{96.011}{40.042}
\emlineto{96.129}{39.946}
\emmoveto{96.218}{39.856}
\emlineto{96.336}{39.761}
\emmoveto{96.423}{39.672}
\emlineto{96.540}{39.586}
\emmoveto{96.628}{39.510}
\emlineto{96.744}{39.430}
\emmoveto{96.831}{39.352}
\emlineto{96.946}{39.271}
\emmoveto{97.033}{39.191}
\emlineto{97.147}{39.107}
\emmoveto{97.233}{39.025}
\emlineto{97.347}{38.934}
\emmoveto{97.433}{38.845}
\emlineto{97.546}{38.750}
\emmoveto{97.630}{38.661}
\emlineto{97.743}{38.566}
\emmoveto{97.827}{38.477}
\emlineto{97.938}{38.381}
\emmoveto{98.022}{38.292}
\emlineto{98.133}{38.197}
\emmoveto{98.215}{38.108}
\emlineto{98.325}{38.012}
\emmoveto{98.408}{37.923}
\emlineto{98.517}{37.828}
\emmoveto{98.598}{37.738}
\emlineto{98.707}{37.642}
\emmoveto{98.787}{37.553}
\emlineto{98.895}{37.457}
\emmoveto{98.975}{37.367}
\emlineto{99.082}{37.270}
\emmoveto{99.162}{37.180}
\emlineto{99.268}{37.083}
\emmoveto{99.347}{36.993}
\emlineto{99.452}{36.895}
\emmoveto{99.530}{36.804}
\emlineto{99.635}{36.698}
\emmoveto{99.712}{36.600}
\emlineto{99.816}{36.494}
\emmoveto{99.893}{36.399}
\emlineto{99.995}{36.296}
\emmoveto{100.072}{36.203}
\emlineto{100.173}{36.120}
\emmoveto{100.274}{36.021}
\emlineto{100.400}{35.918}
\emmoveto{100.501}{35.817}
\emlineto{100.626}{35.712}
\emmoveto{100.725}{35.609}
\emlineto{100.849}{35.501}
\emmoveto{100.947}{35.396}
\emlineto{101.045}{35.310}
\emmoveto{101.118}{35.227}
\emlineto{101.216}{35.140}
\emmoveto{101.288}{35.055}
\emlineto{101.385}{34.965}
\emmoveto{101.457}{34.880}
\emlineto{101.553}{34.788}
\emmoveto{101.624}{34.701}
\emlineto{101.719}{34.607}
\emmoveto{101.790}{34.516}
\emlineto{101.884}{34.406}
\emmoveto{101.954}{34.308}
\emlineto{102.048}{34.200}
\emmoveto{102.117}{34.103}
\emlineto{102.209}{33.998}
\emmoveto{102.278}{33.902}
\emlineto{102.370}{33.798}
\emmoveto{102.438}{33.704}
\emlineto{102.528}{33.602}
\emmoveto{102.596}{33.509}
\emlineto{102.686}{33.409}
\emmoveto{102.753}{33.318}
\emlineto{102.841}{33.219}
\emmoveto{102.908}{33.129}
\emlineto{102.996}{33.032}
\emmoveto{103.083}{32.917}
\emlineto{103.192}{32.809}
\emmoveto{103.279}{32.721}
\emlineto{103.387}{32.632}
\emmoveto{103.472}{32.540}
\emlineto{103.579}{32.444}
\emmoveto{103.664}{32.348}
\emlineto{103.770}{32.246}
\emmoveto{103.854}{32.128}
\emlineto{103.937}{32.029}
\emmoveto{104.021}{31.910}
\emlineto{104.124}{31.784}
\emmoveto{104.206}{31.666}
\emlineto{104.308}{31.541}
\emmoveto{104.389}{31.422}
\emlineto{104.489}{31.297}
\emmoveto{104.569}{31.179}
\emlineto{104.668}{31.055}
\emmoveto{104.747}{30.937}
\emlineto{104.845}{30.812}
\emmoveto{104.923}{30.694}
\emlineto{105.019}{30.569}
\emmoveto{105.096}{30.451}
\emlineto{105.191}{30.326}
\emmoveto{105.267}{30.207}
\emlineto{105.361}{30.081}
\emmoveto{105.435}{29.962}
\emlineto{105.527}{29.825}
\emmoveto{105.583}{29.717}
\emlineto{105.655}{29.600}
\emmoveto{105.728}{29.501}
\emlineto{105.818}{29.404}
\emmoveto{105.889}{29.308}
\emlineto{105.978}{29.209}
\emmoveto{106.049}{29.111}
\emlineto{106.137}{29.009}
\emmoveto{106.207}{28.909}
\emlineto{106.294}{28.805}
\emmoveto{106.362}{28.702}
\emlineto{106.448}{28.595}
\emmoveto{106.516}{28.490}
\emlineto{106.600}{28.379}
\emmoveto{106.667}{28.271}
\emlineto{106.751}{28.157}
\emmoveto{106.817}{28.046}
\emlineto{106.899}{27.927}
\emmoveto{106.964}{27.813}
\emlineto{107.044}{27.689}
\emmoveto{107.108}{27.570}
\emlineto{107.188}{27.432}
\emmoveto{107.235}{27.314}
\emlineto{107.297}{27.183}
\emmoveto{107.343}{27.070}
\emlineto{107.404}{26.948}
\emmoveto{107.450}{26.841}
\emlineto{107.510}{26.727}
\emmoveto{107.570}{26.596}
\emlineto{107.644}{26.462}
\emmoveto{107.717}{26.343}
\emlineto{107.804}{26.236}
\emmoveto{107.876}{26.125}
\emlineto{107.962}{26.010}
\emmoveto{108.032}{25.891}
\emlineto{108.116}{25.766}
\emmoveto{108.186}{25.638}
\emlineto{108.255}{25.525}
\emmoveto{108.309}{25.414}
\emlineto{108.376}{25.292}
\emmoveto{108.430}{25.172}
\emlineto{108.496}{25.039}
\emmoveto{108.535}{24.916}
\emlineto{108.587}{24.758}
\emmoveto{108.625}{24.624}
\emlineto{108.675}{24.473}
\emmoveto{108.713}{24.346}
\emlineto{108.762}{24.203}
\emmoveto{108.798}{24.082}
\emlineto{108.846}{23.947}
\emmoveto{108.882}{23.831}
\emlineto{108.929}{23.705}
\emmoveto{108.964}{23.595}
\emlineto{109.021}{23.444}
\emmoveto{109.055}{23.341}
\emlineto{109.088}{23.191}
\emmoveto{109.110}{23.061}
\emlineto{109.143}{22.892}
\emmoveto{109.164}{22.764}
\emlineto{109.196}{22.598}
\emmoveto{109.217}{22.472}
\emlineto{109.247}{22.309}
\emmoveto{109.268}{22.184}
\emlineto{109.297}{22.024}
\emmoveto{109.317}{21.902}
\emlineto{109.346}{21.744}
\emmoveto{109.364}{21.624}
\emlineto{109.392}{21.469}
\emmoveto{109.411}{21.350}
\emlineto{109.437}{21.198}
\emmoveto{109.455}{21.081}
\emlineto{109.481}{20.932}
\emmoveto{109.498}{20.817}
\emlineto{109.523}{20.670}
\emmoveto{109.540}{20.557}
\emlineto{109.564}{20.413}
\emmoveto{109.580}{20.301}
\emlineto{109.604}{20.159}
\emmoveto{109.619}{20.049}
\emlineto{109.642}{19.910}
\emmoveto{109.657}{19.801}
\emlineto{109.679}{19.665}
\emmoveto{109.693}{19.558}
\emlineto{109.714}{19.423}
\emmoveto{109.728}{19.318}
\emlineto{109.748}{19.186}
\emmoveto{109.761}{19.082}
\emlineto{109.781}{18.953}
\emmoveto{109.793}{18.851}
\emlineto{109.812}{18.723}
\emmoveto{109.830}{18.578}
\emlineto{109.854}{18.408}
\emmoveto{109.871}{18.266}
\emlineto{109.894}{18.101}
\emmoveto{109.910}{17.961}
\emlineto{109.931}{17.800}
\emmoveto{109.946}{17.663}
\emlineto{109.966}{17.506}
\emmoveto{109.980}{17.373}
\emlineto{109.998}{17.220}
\emmoveto{110.012}{17.089}
\emlineto{110.029}{16.939}
\emmoveto{110.041}{16.811}
\emlineto{110.057}{16.665}
\emmoveto{110.069}{16.540}
\emlineto{110.083}{16.398}
\emmoveto{110.094}{16.275}
\emlineto{110.108}{16.136}
\emmoveto{110.117}{16.016}
\emlineto{110.130}{15.881}
\emmoveto{110.139}{15.763}
\emlineto{110.150}{15.631}
\emmoveto{110.158}{15.516}
\emlineto{110.168}{15.387}
\emmoveto{110.176}{15.275}
\emlineto{110.187}{15.115}
\emmoveto{110.196}{14.972}
\emlineto{110.205}{14.818}
\emmoveto{110.213}{14.679}
\emlineto{110.221}{14.530}
\emmoveto{110.227}{14.394}
\emlineto{110.233}{14.250}
\emmoveto{110.238}{14.118}
\emlineto{110.243}{13.978}
\emmoveto{110.246}{13.850}
\emlineto{110.250}{13.714}
\emmoveto{110.252}{13.590}
\emlineto{110.255}{13.459}
\emmoveto{110.256}{13.337}
\emlineto{110.257}{13.183}
\emmoveto{110.257}{13.039}
\emlineto{110.255}{12.891}
\emmoveto{110.254}{12.751}
\emlineto{110.251}{12.609}
\emmoveto{110.248}{12.474}
\emlineto{110.243}{12.337}
\emmoveto{110.238}{12.207}
\emlineto{110.232}{12.075}
\emmoveto{110.225}{11.949}
\emlineto{110.216}{11.800}
\emmoveto{110.207}{11.656}
\emlineto{110.195}{11.514}
\emmoveto{110.183}{11.376}
\emlineto{110.169}{11.240}
\emmoveto{110.157}{11.108}
\emlineto{110.140}{10.978}
\emmoveto{110.126}{10.850}
\emlineto{110.105}{10.708}
\emmoveto{110.086}{10.568}
\emlineto{110.063}{10.432}
\emmoveto{110.041}{10.298}
\emlineto{110.016}{10.170}
\emshow{48.580}{66.700}{Sommerfeld particle: a=0.0002;}
\emmoveto{12.010}{80.000}
\emlineto{12.305}{79.917}
\emmoveto{12.305}{79.907}
\emlineto{12.599}{79.823}
\emmoveto{12.599}{79.813}
\emlineto{12.893}{79.730}
\emmoveto{12.893}{79.720}
\emlineto{13.186}{79.636}
\emmoveto{13.186}{79.626}
\emlineto{13.480}{79.542}
\emmoveto{13.480}{79.532}
\emlineto{13.772}{79.449}
\emmoveto{13.772}{79.439}
\emlineto{14.065}{79.355}
\emmoveto{14.065}{79.345}
\emlineto{14.356}{79.261}
\emmoveto{14.356}{79.251}
\emlineto{14.648}{79.167}
\emmoveto{14.648}{79.157}
\emlineto{14.939}{79.073}
\emmoveto{14.939}{79.063}
\emlineto{15.230}{78.980}
\emmoveto{15.230}{78.970}
\emlineto{15.520}{78.886}
\emmoveto{15.520}{78.876}
\emlineto{15.810}{78.792}
\emmoveto{15.810}{78.782}
\emlineto{16.099}{78.698}
\emmoveto{16.099}{78.688}
\emlineto{16.388}{78.604}
\emmoveto{16.388}{78.594}
\emlineto{16.677}{78.510}
\emmoveto{16.677}{78.500}
\emlineto{16.965}{78.416}
\emmoveto{16.965}{78.406}
\emlineto{17.252}{78.322}
\emmoveto{17.252}{78.312}
\emlineto{17.540}{78.227}
\emmoveto{17.540}{78.217}
\emlineto{17.827}{78.133}
\emmoveto{17.827}{78.123}
\emlineto{18.113}{78.039}
\emmoveto{18.113}{78.029}
\emlineto{18.399}{77.945}
\emmoveto{18.399}{77.935}
\emlineto{18.685}{77.850}
\emmoveto{18.685}{77.840}
\emlineto{18.970}{77.756}
\emmoveto{18.970}{77.746}
\emlineto{19.255}{77.662}
\emmoveto{19.255}{77.652}
\emlineto{19.539}{77.567}
\emmoveto{19.539}{77.557}
\emlineto{19.823}{77.473}
\emmoveto{19.823}{77.463}
\emlineto{20.107}{77.378}
\emmoveto{20.107}{77.368}
\emlineto{20.390}{77.284}
\emmoveto{20.390}{77.274}
\emlineto{20.673}{77.189}
\emmoveto{20.673}{77.179}
\emlineto{20.955}{77.095}
\emmoveto{20.955}{77.085}
\emlineto{21.237}{77.000}
\emmoveto{21.237}{76.990}
\emlineto{21.519}{76.905}
\emmoveto{21.519}{76.895}
\emlineto{21.800}{76.811}
\emmoveto{21.800}{76.801}
\emlineto{22.080}{76.716}
\emmoveto{22.080}{76.706}
\emlineto{22.361}{76.621}
\emmoveto{22.361}{76.611}
\emlineto{22.640}{76.526}
\emmoveto{22.640}{76.516}
\emlineto{22.920}{76.431}
\emmoveto{22.920}{76.421}
\emlineto{23.199}{76.336}
\emmoveto{23.199}{76.326}
\emlineto{23.477}{76.241}
\emmoveto{23.477}{76.231}
\emlineto{23.755}{76.146}
\emmoveto{23.755}{76.136}
\emlineto{24.033}{76.051}
\emmoveto{24.033}{76.041}
\emlineto{24.311}{75.956}
\emmoveto{24.311}{75.946}
\emlineto{24.587}{75.861}
\emmoveto{24.587}{75.851}
\emlineto{24.864}{75.766}
\emmoveto{24.864}{75.756}
\emlineto{25.140}{75.671}
\emmoveto{25.140}{75.661}
\emlineto{25.415}{75.575}
\emmoveto{25.415}{75.565}
\emlineto{25.691}{75.480}
\emmoveto{25.691}{75.470}
\emlineto{25.966}{75.385}
\emmoveto{25.966}{75.375}
\emlineto{26.240}{75.290}
\emmoveto{26.240}{75.280}
\emlineto{26.514}{75.194}
\emmoveto{26.514}{75.184}
\emlineto{26.787}{75.099}
\emmoveto{26.787}{75.089}
\emlineto{27.060}{75.003}
\emmoveto{27.060}{74.993}
\emlineto{27.333}{74.908}
\emmoveto{27.333}{74.898}
\emlineto{27.605}{74.812}
\emmoveto{27.605}{74.802}
\emlineto{27.877}{74.717}
\emmoveto{27.877}{74.707}
\emlineto{28.149}{74.621}
\emmoveto{28.149}{74.611}
\emlineto{28.420}{74.526}
\emmoveto{28.420}{74.516}
\emlineto{28.690}{74.430}
\emmoveto{28.690}{74.420}
\emlineto{28.960}{74.334}
\emmoveto{28.960}{74.324}
\emlineto{29.230}{74.239}
\emmoveto{29.230}{74.229}
\emlineto{29.499}{74.143}
\emmoveto{29.499}{74.133}
\emlineto{29.768}{74.047}
\emmoveto{29.768}{74.037}
\emlineto{30.037}{73.951}
\emmoveto{30.037}{73.941}
\emlineto{30.305}{73.855}
\emmoveto{30.305}{73.845}
\emlineto{30.572}{73.759}
\emmoveto{30.572}{73.749}
\emlineto{30.840}{73.663}
\emmoveto{30.840}{73.653}
\emlineto{31.106}{73.568}
\emmoveto{31.106}{73.558}
\emlineto{31.373}{73.471}
\emmoveto{31.373}{73.461}
\emlineto{31.639}{73.375}
\emmoveto{31.639}{73.365}
\emlineto{31.904}{73.279}
\emmoveto{31.904}{73.269}
\emlineto{32.169}{73.183}
\emmoveto{32.169}{73.173}
\emlineto{32.434}{73.087}
\emmoveto{32.434}{73.077}
\emlineto{32.698}{72.991}
\emmoveto{32.698}{72.981}
\emlineto{32.962}{72.895}
\emmoveto{32.962}{72.885}
\emlineto{33.225}{72.798}
\emmoveto{33.225}{72.788}
\emlineto{33.488}{72.702}
\emmoveto{33.488}{72.692}
\emlineto{33.751}{72.606}
\emmoveto{33.751}{72.596}
\emlineto{34.013}{72.509}
\emmoveto{34.013}{72.499}
\emlineto{34.275}{72.413}
\emmoveto{34.275}{72.403}
\emlineto{34.536}{72.316}
\emmoveto{34.536}{72.306}
\emlineto{34.797}{72.220}
\emmoveto{34.797}{72.210}
\emlineto{35.057}{72.123}
\emmoveto{35.057}{72.113}
\emlineto{35.317}{72.027}
\emmoveto{35.317}{72.017}
\emlineto{35.577}{71.930}
\emmoveto{35.577}{71.920}
\emlineto{35.836}{71.834}
\emmoveto{35.836}{71.824}
\emlineto{36.094}{71.737}
\emmoveto{36.094}{71.727}
\emlineto{36.353}{71.640}
\emmoveto{36.353}{71.630}
\emlineto{36.610}{71.543}
\emmoveto{36.610}{71.533}
\emlineto{36.868}{71.447}
\emmoveto{36.868}{71.437}
\emlineto{37.125}{71.350}
\emmoveto{37.125}{71.340}
\emlineto{37.382}{71.253}
\emmoveto{37.382}{71.243}
\emlineto{37.638}{71.156}
\emmoveto{37.638}{71.146}
\emlineto{37.893}{71.059}
\emmoveto{37.893}{71.049}
\emlineto{38.149}{70.962}
\emmoveto{38.149}{70.952}
\emlineto{38.404}{70.865}
\emmoveto{38.404}{70.855}
\emlineto{38.658}{70.768}
\emmoveto{38.658}{70.758}
\emlineto{38.912}{70.671}
\emmoveto{38.912}{70.661}
\emlineto{39.165}{70.574}
\emmoveto{39.165}{70.564}
\emlineto{39.419}{70.477}
\emmoveto{39.419}{70.467}
\emlineto{39.671}{70.380}
\emmoveto{39.671}{70.370}
\emlineto{39.924}{70.283}
\emmoveto{39.924}{70.273}
\emlineto{40.175}{70.185}
\emmoveto{40.175}{70.175}
\emlineto{40.427}{70.088}
\emmoveto{40.427}{70.078}
\emlineto{40.678}{69.991}
\emmoveto{40.678}{69.981}
\emlineto{40.929}{69.893}
\emmoveto{40.929}{69.883}
\emlineto{41.179}{69.796}
\emmoveto{41.179}{69.786}
\emlineto{41.428}{69.699}
\emmoveto{41.428}{69.689}
\emlineto{41.678}{69.601}
\emmoveto{41.678}{69.591}
\emlineto{41.927}{69.504}
\emmoveto{41.927}{69.494}
\emlineto{42.175}{69.406}
\emmoveto{42.175}{69.396}
\emlineto{42.423}{69.309}
\emmoveto{42.423}{69.299}
\emlineto{42.670}{69.211}
\emmoveto{42.670}{69.201}
\emlineto{42.918}{69.113}
\emmoveto{42.918}{69.103}
\emlineto{43.164}{69.016}
\emmoveto{43.164}{69.006}
\emlineto{43.411}{68.918}
\emmoveto{43.411}{68.908}
\emlineto{43.656}{68.820}
\emmoveto{43.656}{68.810}
\emlineto{43.902}{68.723}
\emmoveto{43.902}{68.713}
\emlineto{44.147}{68.625}
\emmoveto{44.147}{68.615}
\emlineto{44.391}{68.527}
\emmoveto{44.391}{68.517}
\emlineto{44.635}{68.429}
\emmoveto{44.635}{68.419}
\emlineto{44.879}{68.331}
\emmoveto{44.879}{68.321}
\emlineto{45.122}{68.234}
\emmoveto{45.122}{68.224}
\emlineto{45.365}{68.135}
\emmoveto{45.365}{68.125}
\emlineto{45.608}{68.038}
\emmoveto{45.608}{68.028}
\emlineto{45.850}{67.939}
\emmoveto{45.850}{67.929}
\emlineto{46.091}{67.841}
\emmoveto{46.091}{67.831}
\emlineto{46.332}{67.743}
\emmoveto{46.332}{67.733}
\emlineto{46.573}{67.645}
\emmoveto{46.573}{67.635}
\emlineto{46.813}{67.547}
\emmoveto{46.813}{67.537}
\emlineto{47.053}{67.449}
\emmoveto{47.053}{67.439}
\emlineto{47.292}{67.351}
\emmoveto{47.292}{67.341}
\emlineto{47.531}{67.252}
\emmoveto{47.531}{67.242}
\emlineto{47.770}{67.154}
\emmoveto{47.770}{67.144}
\emlineto{48.008}{67.056}
\emmoveto{48.008}{67.046}
\emlineto{48.245}{66.957}
\emmoveto{48.245}{66.947}
\emlineto{48.482}{66.859}
\emmoveto{48.482}{66.849}
\emlineto{48.719}{66.761}
\emmoveto{48.719}{66.751}
\emlineto{48.956}{66.662}
\emmoveto{48.956}{66.652}
\emlineto{49.191}{66.564}
\emmoveto{49.191}{66.554}
\emlineto{49.427}{66.465}
\emmoveto{49.427}{66.455}
\emlineto{49.662}{66.366}
\emmoveto{49.662}{66.356}
\emlineto{49.896}{66.268}
\emmoveto{49.896}{66.258}
\emlineto{50.131}{66.169}
\emmoveto{50.131}{66.159}
\emlineto{50.364}{66.071}
\emmoveto{50.364}{66.061}
\emlineto{50.598}{65.972}
\emmoveto{50.598}{65.962}
\emlineto{50.830}{65.873}
\emmoveto{50.830}{65.863}
\emlineto{51.063}{65.774}
\emmoveto{51.063}{65.764}
\emlineto{51.295}{65.676}
\emmoveto{51.295}{65.666}
\emlineto{51.526}{65.577}
\emmoveto{51.526}{65.567}
\emlineto{51.757}{65.478}
\emmoveto{51.757}{65.468}
\emlineto{51.988}{65.379}
\emmoveto{51.988}{65.369}
\emlineto{52.218}{65.280}
\emmoveto{52.218}{65.270}
\emlineto{52.448}{65.181}
\emmoveto{52.448}{65.171}
\emlineto{52.678}{65.082}
\emmoveto{52.678}{65.072}
\emlineto{52.906}{64.983}
\emmoveto{52.906}{64.973}
\emlineto{53.135}{64.884}
\emmoveto{53.135}{64.874}
\emlineto{53.363}{64.785}
\emmoveto{53.363}{64.775}
\emlineto{53.591}{64.686}
\emmoveto{53.591}{64.676}
\emlineto{53.818}{64.587}
\emmoveto{53.818}{64.577}
\emlineto{54.044}{64.488}
\emmoveto{54.044}{64.478}
\emlineto{54.271}{64.388}
\emmoveto{54.271}{64.378}
\emlineto{54.497}{64.289}
\emmoveto{54.497}{64.279}
\emlineto{54.722}{64.190}
\emmoveto{54.722}{64.180}
\emlineto{54.947}{64.091}
\emmoveto{54.947}{64.081}
\emlineto{55.172}{63.991}
\emmoveto{55.172}{63.981}
\emlineto{55.396}{63.892}
\emmoveto{55.396}{63.882}
\emlineto{55.619}{63.793}
\emmoveto{55.619}{63.783}
\emlineto{55.842}{63.693}
\emmoveto{55.842}{63.683}
\emlineto{56.065}{63.593}
\emmoveto{56.065}{63.583}
\emlineto{56.288}{63.494}
\emmoveto{56.288}{63.484}
\emlineto{56.510}{63.395}
\emmoveto{56.510}{63.385}
\emlineto{56.731}{63.295}
\emmoveto{56.731}{63.285}
\emlineto{56.952}{63.195}
\emmoveto{56.952}{63.185}
\emlineto{57.173}{63.096}
\emmoveto{57.173}{63.086}
\emlineto{57.393}{62.996}
\emmoveto{57.393}{62.986}
\emlineto{57.613}{62.896}
\emmoveto{57.613}{62.886}
\emlineto{57.832}{62.797}
\emmoveto{57.832}{62.787}
\emlineto{58.051}{62.697}
\emmoveto{58.051}{62.687}
\emlineto{58.269}{62.597}
\emmoveto{58.269}{62.587}
\emlineto{58.487}{62.497}
\emmoveto{58.487}{62.487}
\emlineto{58.705}{62.397}
\emmoveto{58.705}{62.387}
\emlineto{58.922}{62.298}
\emmoveto{58.922}{62.288}
\emlineto{59.138}{62.198}
\emmoveto{59.138}{62.188}
\emlineto{59.354}{62.098}
\emmoveto{59.354}{62.088}
\emlineto{59.570}{61.998}
\emmoveto{59.570}{61.988}
\emlineto{59.785}{61.898}
\emmoveto{59.785}{61.888}
\emlineto{60.000}{61.798}
\emmoveto{60.000}{61.788}
\emlineto{60.215}{61.698}
\emmoveto{60.215}{61.688}
\emlineto{60.429}{61.598}
\emmoveto{60.429}{61.588}
\emlineto{60.642}{61.497}
\emmoveto{60.642}{61.487}
\emlineto{60.855}{61.397}
\emmoveto{60.855}{61.387}
\emlineto{61.068}{61.297}
\emmoveto{61.068}{61.287}
\emlineto{61.280}{61.197}
\emmoveto{61.280}{61.187}
\emlineto{61.492}{61.097}
\emmoveto{61.492}{61.087}
\emlineto{61.703}{60.996}
\emmoveto{61.703}{60.986}
\emlineto{61.914}{60.896}
\emmoveto{61.914}{60.886}
\emlineto{62.125}{60.796}
\emmoveto{62.125}{60.786}
\emlineto{62.335}{60.695}
\emmoveto{62.335}{60.685}
\emlineto{62.544}{60.595}
\emmoveto{62.544}{60.585}
\emlineto{62.753}{60.494}
\emmoveto{62.753}{60.484}
\emlineto{62.962}{60.394}
\emmoveto{62.962}{60.384}
\emlineto{63.170}{60.293}
\emmoveto{63.170}{60.283}
\emlineto{63.378}{60.193}
\emmoveto{63.378}{60.183}
\emlineto{63.585}{60.092}
\emmoveto{63.585}{60.082}
\emlineto{63.792}{59.992}
\emmoveto{63.792}{59.982}
\emlineto{63.999}{59.891}
\emmoveto{63.999}{59.881}
\emlineto{64.205}{59.790}
\emmoveto{64.205}{59.780}
\emlineto{64.410}{59.690}
\emmoveto{64.410}{59.680}
\emlineto{64.616}{59.589}
\emmoveto{64.616}{59.579}
\emlineto{64.820}{59.488}
\emmoveto{64.820}{59.478}
\emlineto{65.024}{59.388}
\emmoveto{65.024}{59.378}
\emlineto{65.228}{59.287}
\emmoveto{65.228}{59.277}
\emlineto{65.432}{59.186}
\emmoveto{65.432}{59.176}
\emlineto{65.635}{59.085}
\emmoveto{65.635}{59.075}
\emlineto{65.837}{58.984}
\emmoveto{65.837}{58.974}
\emlineto{66.039}{58.883}
\emmoveto{66.039}{58.873}
\emlineto{66.241}{58.783}
\emmoveto{66.241}{58.773}
\emlineto{66.442}{58.682}
\emmoveto{66.442}{58.672}
\emlineto{66.642}{58.581}
\emmoveto{66.642}{58.571}
\emlineto{66.843}{58.479}
\emmoveto{66.843}{58.469}
\emlineto{67.042}{58.378}
\emmoveto{67.042}{58.368}
\emlineto{67.242}{58.277}
\emmoveto{67.242}{58.267}
\emlineto{67.441}{58.176}
\emmoveto{67.441}{58.166}
\emlineto{67.639}{58.075}
\emmoveto{67.639}{58.065}
\emlineto{67.837}{57.974}
\emmoveto{67.837}{57.964}
\emlineto{68.035}{57.873}
\emmoveto{68.035}{57.863}
\emlineto{68.232}{57.772}
\emmoveto{68.232}{57.762}
\emlineto{68.429}{57.670}
\emmoveto{68.429}{57.660}
\emlineto{68.625}{57.569}
\emmoveto{68.625}{57.559}
\emlineto{68.820}{57.468}
\emmoveto{68.820}{57.458}
\emlineto{69.016}{57.366}
\emmoveto{69.016}{57.356}
\emlineto{69.211}{57.265}
\emmoveto{69.211}{57.255}
\emlineto{69.405}{57.163}
\emmoveto{69.405}{57.153}
\emlineto{69.599}{57.062}
\emmoveto{69.599}{57.052}
\emlineto{69.793}{56.961}
\emmoveto{69.793}{56.951}
\emlineto{69.986}{56.859}
\emmoveto{69.986}{56.849}
\emlineto{70.178}{56.758}
\emmoveto{70.178}{56.748}
\emlineto{70.371}{56.656}
\emmoveto{70.371}{56.646}
\emlineto{70.562}{56.555}
\emmoveto{70.562}{56.545}
\emlineto{70.754}{56.453}
\emmoveto{70.754}{56.443}
\emlineto{70.944}{56.351}
\emmoveto{70.944}{56.341}
\emlineto{71.135}{56.250}
\emmoveto{71.135}{56.240}
\emlineto{71.325}{56.148}
\emmoveto{71.325}{56.138}
\emlineto{71.514}{56.046}
\emmoveto{71.514}{56.036}
\emlineto{71.703}{55.944}
\emmoveto{71.703}{55.934}
\emlineto{71.892}{55.843}
\emmoveto{71.892}{55.833}
\emlineto{72.080}{55.741}
\emmoveto{72.080}{55.731}
\emlineto{72.268}{55.639}
\emmoveto{72.268}{55.629}
\emlineto{72.455}{55.537}
\emmoveto{72.455}{55.527}
\emlineto{72.642}{55.436}
\emmoveto{72.642}{55.426}
\emlineto{72.828}{55.333}
\emmoveto{72.828}{55.323}
\emlineto{73.014}{55.232}
\emmoveto{73.014}{55.222}
\emlineto{73.200}{55.130}
\emmoveto{73.200}{55.120}
\emlineto{73.385}{55.028}
\emmoveto{73.385}{55.018}
\emlineto{73.569}{54.926}
\emmoveto{73.569}{54.916}
\emlineto{73.753}{54.824}
\emmoveto{73.753}{54.814}
\emlineto{73.937}{54.721}
\emmoveto{73.937}{54.711}
\emlineto{74.120}{54.619}
\emmoveto{74.120}{54.609}
\emlineto{74.303}{54.517}
\emmoveto{74.303}{54.507}
\emlineto{74.485}{54.415}
\emmoveto{74.485}{54.405}
\emlineto{74.667}{54.313}
\emmoveto{74.667}{54.303}
\emlineto{74.849}{54.211}
\emmoveto{74.849}{54.201}
\emlineto{75.030}{54.109}
\emmoveto{75.030}{54.099}
\emlineto{75.210}{54.006}
\emmoveto{75.210}{53.996}
\emlineto{75.390}{53.904}
\emmoveto{75.390}{53.894}
\emlineto{75.570}{53.802}
\emmoveto{75.570}{53.792}
\emlineto{75.749}{53.699}
\emmoveto{75.749}{53.689}
\emlineto{75.928}{53.597}
\emmoveto{75.928}{53.587}
\emlineto{76.106}{53.495}
\emmoveto{76.106}{53.485}
\emlineto{76.284}{53.392}
\emmoveto{76.284}{53.382}
\emlineto{76.461}{53.290}
\emmoveto{76.461}{53.280}
\emlineto{76.638}{53.187}
\emmoveto{76.638}{53.177}
\emlineto{76.814}{53.085}
\emmoveto{76.814}{53.075}
\emlineto{76.990}{52.982}
\emmoveto{76.990}{52.972}
\emlineto{77.166}{52.880}
\emmoveto{77.166}{52.870}
\emlineto{77.341}{52.777}
\emmoveto{77.341}{52.767}
\emlineto{77.516}{52.675}
\emmoveto{77.516}{52.665}
\emlineto{77.690}{52.572}
\emmoveto{77.690}{52.562}
\emlineto{77.863}{52.469}
\emmoveto{77.863}{52.459}
\emlineto{78.037}{52.367}
\emmoveto{78.037}{52.357}
\emlineto{78.210}{52.264}
\emmoveto{78.210}{52.254}
\emlineto{78.382}{52.161}
\emmoveto{78.382}{52.151}
\emlineto{78.554}{52.058}
\emmoveto{78.554}{52.048}
\emlineto{78.725}{51.956}
\emmoveto{78.725}{51.946}
\emlineto{78.896}{51.853}
\emmoveto{78.896}{51.843}
\emlineto{79.067}{51.750}
\emmoveto{79.067}{51.740}
\emlineto{79.237}{51.647}
\emmoveto{79.237}{51.637}
\emlineto{79.407}{51.544}
\emmoveto{79.407}{51.534}
\emlineto{79.576}{51.441}
\emmoveto{79.576}{51.431}
\emlineto{79.745}{51.339}
\emmoveto{79.745}{51.329}
\emlineto{79.913}{51.236}
\emmoveto{79.913}{51.226}
\emlineto{80.081}{51.133}
\emmoveto{80.081}{51.123}
\emlineto{80.248}{51.030}
\emmoveto{80.248}{51.020}
\emlineto{80.415}{50.927}
\emmoveto{80.415}{50.917}
\emlineto{80.582}{50.823}
\emmoveto{80.582}{50.813}
\emlineto{80.748}{50.720}
\emmoveto{80.748}{50.710}
\emlineto{80.913}{50.617}
\emmoveto{80.913}{50.607}
\emlineto{81.078}{50.514}
\emmoveto{81.078}{50.504}
\emlineto{81.243}{50.411}
\emmoveto{81.243}{50.401}
\emlineto{81.407}{50.308}
\emmoveto{81.407}{50.298}
\emlineto{81.571}{50.204}
\emmoveto{81.571}{50.194}
\emlineto{81.734}{50.101}
\emmoveto{81.734}{50.091}
\emlineto{81.897}{49.998}
\emmoveto{81.897}{49.988}
\emlineto{82.059}{49.895}
\emmoveto{82.059}{49.885}
\emlineto{82.221}{49.791}
\emmoveto{82.221}{49.781}
\emlineto{82.383}{49.688}
\emmoveto{82.383}{49.678}
\emlineto{82.544}{49.585}
\emmoveto{82.544}{49.575}
\emlineto{82.704}{49.481}
\emmoveto{82.704}{49.471}
\emlineto{82.864}{49.378}
\emmoveto{82.864}{49.368}
\emlineto{83.024}{49.275}
\emmoveto{83.024}{49.265}
\emlineto{83.183}{49.171}
\emmoveto{83.183}{49.161}
\emlineto{83.342}{49.068}
\emmoveto{83.342}{49.058}
\emlineto{83.500}{48.964}
\emmoveto{83.500}{48.954}
\emlineto{83.658}{48.861}
\emmoveto{83.658}{48.851}
\emlineto{83.815}{48.757}
\emmoveto{83.815}{48.747}
\emlineto{83.972}{48.654}
\emmoveto{83.972}{48.644}
\emlineto{84.129}{48.550}
\emmoveto{84.129}{48.540}
\emlineto{84.285}{48.446}
\emmoveto{84.285}{48.436}
\emlineto{84.440}{48.343}
\emmoveto{84.440}{48.333}
\emlineto{84.595}{48.239}
\emmoveto{84.595}{48.229}
\emlineto{84.750}{48.135}
\emmoveto{84.750}{48.125}
\emlineto{84.904}{48.032}
\emmoveto{84.904}{48.022}
\emlineto{85.058}{47.928}
\emmoveto{85.058}{47.918}
\emlineto{85.211}{47.824}
\emmoveto{85.211}{47.814}
\emlineto{85.364}{47.720}
\emmoveto{85.364}{47.710}
\emlineto{85.516}{47.616}
\emmoveto{85.516}{47.606}
\emlineto{85.668}{47.513}
\emmoveto{85.668}{47.503}
\emlineto{85.819}{47.409}
\emmoveto{85.819}{47.399}
\emlineto{85.970}{47.305}
\emmoveto{85.970}{47.295}
\emlineto{86.121}{47.201}
\emmoveto{86.121}{47.191}
\emlineto{86.271}{47.097}
\emmoveto{86.271}{47.087}
\emlineto{86.421}{46.993}
\emmoveto{86.421}{46.983}
\emlineto{86.570}{46.889}
\emmoveto{86.570}{46.879}
\emlineto{86.718}{46.785}
\emmoveto{86.718}{46.775}
\emlineto{86.867}{46.681}
\emmoveto{86.867}{46.671}
\emlineto{87.014}{46.577}
\emmoveto{87.014}{46.567}
\emlineto{87.162}{46.473}
\emmoveto{87.162}{46.463}
\emlineto{87.308}{46.369}
\emmoveto{87.308}{46.359}
\emlineto{87.455}{46.265}
\emmoveto{87.455}{46.255}
\emlineto{87.601}{46.161}
\emmoveto{87.601}{46.151}
\emlineto{87.746}{46.057}
\emmoveto{87.746}{46.047}
\emlineto{87.891}{45.953}
\emmoveto{87.891}{45.943}
\emlineto{88.036}{45.848}
\emmoveto{88.036}{45.838}
\emlineto{88.180}{45.744}
\emmoveto{88.180}{45.734}
\emlineto{88.323}{45.640}
\emmoveto{88.323}{45.630}
\emlineto{88.466}{45.536}
\emmoveto{88.466}{45.526}
\emlineto{88.609}{45.432}
\emmoveto{88.609}{45.422}
\emlineto{88.751}{45.327}
\emmoveto{88.751}{45.317}
\emlineto{88.893}{45.223}
\emmoveto{88.893}{45.213}
\emlineto{89.035}{45.119}
\emmoveto{89.035}{45.109}
\emlineto{89.175}{45.014}
\emmoveto{89.175}{45.004}
\emlineto{89.316}{44.910}
\emmoveto{89.316}{44.900}
\emlineto{89.456}{44.806}
\emmoveto{89.456}{44.796}
\emlineto{89.595}{44.701}
\emmoveto{89.595}{44.691}
\emlineto{89.734}{44.597}
\emmoveto{89.734}{44.587}
\emlineto{89.873}{44.492}
\emmoveto{89.873}{44.482}
\emlineto{90.011}{44.388}
\emmoveto{90.011}{44.378}
\emlineto{90.148}{44.283}
\emmoveto{90.148}{44.273}
\emlineto{90.286}{44.179}
\emmoveto{90.286}{44.169}
\emlineto{90.422}{44.074}
\emmoveto{90.422}{44.064}
\emlineto{90.559}{43.970}
\emmoveto{90.559}{43.960}
\emlineto{90.694}{43.865}
\emmoveto{90.694}{43.855}
\emlineto{90.830}{43.760}
\emmoveto{90.830}{43.750}
\emlineto{90.965}{43.656}
\emmoveto{90.965}{43.646}
\emlineto{91.099}{43.551}
\emmoveto{91.099}{43.541}
\emlineto{91.233}{43.446}
\emmoveto{91.233}{43.436}
\emlineto{91.366}{43.342}
\emmoveto{91.366}{43.332}
\emlineto{91.499}{43.237}
\emmoveto{91.499}{43.227}
\emlineto{91.632}{43.132}
\emmoveto{91.632}{43.122}
\emlineto{91.764}{43.027}
\emmoveto{91.764}{43.017}
\emlineto{91.896}{42.923}
\emmoveto{91.896}{42.913}
\emlineto{92.027}{42.818}
\emmoveto{92.027}{42.808}
\emlineto{92.158}{42.713}
\emmoveto{92.158}{42.703}
\emlineto{92.288}{42.608}
\emmoveto{92.288}{42.598}
\emlineto{92.418}{42.503}
\emmoveto{92.418}{42.493}
\emlineto{92.547}{42.399}
\emmoveto{92.547}{42.389}
\emlineto{92.676}{42.294}
\emmoveto{92.676}{42.284}
\emlineto{92.804}{42.189}
\emmoveto{92.804}{42.179}
\emlineto{92.932}{42.084}
\emmoveto{92.932}{42.074}
\emlineto{93.060}{41.979}
\emmoveto{93.060}{41.969}
\emlineto{93.187}{41.874}
\emmoveto{93.187}{41.864}
\emlineto{93.313}{41.769}
\emmoveto{93.313}{41.759}
\emlineto{93.439}{41.664}
\emmoveto{93.439}{41.654}
\emlineto{93.565}{41.559}
\emmoveto{93.565}{41.549}
\emlineto{93.690}{41.454}
\emmoveto{93.690}{41.444}
\emlineto{93.815}{41.349}
\emmoveto{93.815}{41.339}
\emlineto{93.939}{41.244}
\emmoveto{93.939}{41.234}
\emlineto{94.063}{41.139}
\emmoveto{94.063}{41.129}
\emlineto{94.186}{41.033}
\emmoveto{94.186}{41.023}
\emlineto{94.309}{40.928}
\emmoveto{94.309}{40.918}
\emlineto{94.431}{40.823}
\emmoveto{94.431}{40.813}
\emlineto{94.553}{40.718}
\emmoveto{94.553}{40.708}
\emlineto{94.674}{40.613}
\emmoveto{94.674}{40.603}
\emlineto{94.795}{40.507}
\emmoveto{94.795}{40.497}
\emlineto{94.916}{40.402}
\emmoveto{94.916}{40.392}
\emlineto{95.036}{40.297}
\emmoveto{95.036}{40.287}
\emlineto{95.156}{40.192}
\emmoveto{95.156}{40.182}
\emlineto{95.275}{40.087}
\emmoveto{95.275}{40.077}
\emlineto{95.393}{39.981}
\emmoveto{95.393}{39.971}
\emlineto{95.512}{39.876}
\emmoveto{95.512}{39.866}
\emlineto{95.629}{39.770}
\emmoveto{95.629}{39.760}
\emlineto{95.747}{39.665}
\emmoveto{95.747}{39.655}
\emlineto{95.863}{39.560}
\emmoveto{95.863}{39.550}
\emlineto{95.980}{39.454}
\emmoveto{95.980}{39.444}
\emlineto{96.096}{39.349}
\emmoveto{96.096}{39.339}
\emlineto{96.211}{39.243}
\emmoveto{96.211}{39.233}
\emlineto{96.326}{39.138}
\emmoveto{96.326}{39.128}
\emlineto{96.440}{39.033}
\emmoveto{96.440}{39.023}
\emlineto{96.554}{38.927}
\emmoveto{96.554}{38.917}
\emlineto{96.668}{38.822}
\emmoveto{96.668}{38.812}
\emlineto{96.781}{38.716}
\emmoveto{96.781}{38.706}
\emlineto{96.894}{38.611}
\emmoveto{96.894}{38.601}
\emlineto{97.006}{38.505}
\emmoveto{97.006}{38.495}
\emlineto{97.117}{38.399}
\emmoveto{97.117}{38.389}
\emlineto{97.229}{38.294}
\emmoveto{97.229}{38.284}
\emlineto{97.339}{38.188}
\emmoveto{97.339}{38.178}
\emlineto{97.450}{38.083}
\emmoveto{97.450}{38.073}
\emlineto{97.559}{37.977}
\emmoveto{97.559}{37.967}
\emlineto{97.669}{37.871}
\emmoveto{97.669}{37.861}
\emlineto{97.778}{37.766}
\emmoveto{97.778}{37.756}
\emlineto{97.886}{37.660}
\emmoveto{97.886}{37.650}
\emlineto{97.994}{37.554}
\emmoveto{97.994}{37.544}
\emlineto{98.102}{37.448}
\emmoveto{98.102}{37.438}
\emlineto{98.209}{37.343}
\emmoveto{98.209}{37.333}
\emlineto{98.315}{37.237}
\emmoveto{98.315}{37.227}
\emlineto{98.421}{37.131}
\emmoveto{98.421}{37.121}
\emlineto{98.527}{37.025}
\emmoveto{98.527}{37.015}
\emlineto{98.632}{36.920}
\emmoveto{98.632}{36.910}
\emlineto{98.737}{36.814}
\emmoveto{98.737}{36.804}
\emlineto{98.841}{36.708}
\emmoveto{98.841}{36.698}
\emlineto{98.945}{36.602}
\emmoveto{98.945}{36.592}
\emlineto{99.048}{36.496}
\emmoveto{99.048}{36.486}
\emlineto{99.151}{36.390}
\emmoveto{99.151}{36.380}
\emlineto{99.253}{36.284}
\emmoveto{99.253}{36.274}
\emlineto{99.355}{36.178}
\emmoveto{99.355}{36.168}
\emlineto{99.456}{36.073}
\emmoveto{99.456}{36.063}
\emlineto{99.557}{35.967}
\emmoveto{99.557}{35.957}
\emlineto{99.658}{35.861}
\emmoveto{99.658}{35.851}
\emlineto{99.758}{35.755}
\emmoveto{99.758}{35.745}
\emlineto{99.857}{35.649}
\emmoveto{99.857}{35.639}
\emlineto{99.957}{35.543}
\emmoveto{99.957}{35.533}
\emlineto{100.055}{35.436}
\emmoveto{100.055}{35.426}
\emlineto{100.153}{35.330}
\emmoveto{100.153}{35.320}
\emlineto{100.251}{35.224}
\emmoveto{100.251}{35.214}
\emlineto{100.348}{35.118}
\emmoveto{100.348}{35.108}
\emlineto{100.445}{35.012}
\emmoveto{100.445}{35.002}
\emlineto{100.541}{34.906}
\emmoveto{100.541}{34.896}
\emlineto{100.637}{34.800}
\emmoveto{100.637}{34.790}
\emlineto{100.732}{34.694}
\emmoveto{100.732}{34.684}
\emlineto{100.827}{34.588}
\emmoveto{100.827}{34.578}
\emlineto{100.922}{34.481}
\emmoveto{100.922}{34.471}
\emlineto{101.016}{34.375}
\emmoveto{101.016}{34.365}
\emlineto{101.109}{34.269}
\emmoveto{101.109}{34.259}
\emlineto{101.202}{34.163}
\emmoveto{101.202}{34.153}
\emlineto{101.295}{34.057}
\emmoveto{101.295}{34.047}
\emlineto{101.387}{33.950}
\emmoveto{101.387}{33.940}
\emlineto{101.478}{33.844}
\emmoveto{101.478}{33.834}
\emlineto{101.569}{33.738}
\emmoveto{101.569}{33.728}
\emlineto{101.660}{33.632}
\emmoveto{101.660}{33.622}
\emlineto{101.750}{33.525}
\emmoveto{101.750}{33.515}
\emlineto{101.840}{33.419}
\emmoveto{101.840}{33.409}
\emlineto{101.929}{33.313}
\emmoveto{101.929}{33.303}
\emlineto{102.018}{33.206}
\emmoveto{102.018}{33.196}
\emlineto{102.106}{33.100}
\emmoveto{102.106}{33.090}
\emlineto{102.194}{32.994}
\emmoveto{102.194}{32.984}
\emlineto{102.281}{32.887}
\emmoveto{102.281}{32.877}
\emlineto{102.368}{32.781}
\emmoveto{102.455}{32.665}
\emlineto{102.626}{32.462}
\emmoveto{102.711}{32.345}
\emlineto{102.880}{32.142}
\emmoveto{102.964}{32.026}
\emlineto{103.129}{31.823}
\emmoveto{103.212}{31.706}
\emlineto{103.375}{31.503}
\emmoveto{103.455}{31.387}
\emlineto{103.616}{31.183}
\emmoveto{103.695}{31.067}
\emlineto{103.852}{30.864}
\emmoveto{103.930}{30.747}
\emlineto{104.085}{30.544}
\emmoveto{104.161}{30.427}
\emlineto{104.313}{30.224}
\emmoveto{104.388}{30.107}
\emlineto{104.537}{29.903}
\emmoveto{104.611}{29.787}
\emlineto{104.757}{29.583}
\emmoveto{104.829}{29.466}
\emlineto{104.973}{29.263}
\emmoveto{105.043}{29.146}
\emlineto{105.184}{28.942}
\emmoveto{105.253}{28.826}
\emlineto{105.391}{28.622}
\emmoveto{105.459}{28.505}
\emlineto{105.594}{28.301}
\emmoveto{105.660}{28.184}
\emlineto{105.792}{27.980}
\emmoveto{105.858}{27.863}
\emlineto{105.987}{27.659}
\emmoveto{106.051}{27.542}
\emlineto{106.177}{27.339}
\emmoveto{106.239}{27.222}
\emlineto{106.363}{27.017}
\emmoveto{106.424}{26.900}
\emlineto{106.544}{26.696}
\emmoveto{106.604}{26.579}
\emlineto{106.722}{26.375}
\emmoveto{106.780}{26.258}
\emlineto{106.895}{26.054}
\emmoveto{106.952}{25.937}
\emlineto{107.064}{25.732}
\emmoveto{107.119}{25.615}
\emlineto{107.228}{25.411}
\emmoveto{107.282}{25.294}
\emlineto{107.389}{25.089}
\emmoveto{107.441}{24.972}
\emlineto{107.545}{24.768}
\emmoveto{107.596}{24.651}
\emlineto{107.697}{24.446}
\emmoveto{107.746}{24.329}
\emlineto{107.844}{24.124}
\emmoveto{107.892}{24.007}
\emlineto{107.988}{23.803}
\emmoveto{108.034}{23.685}
\emlineto{108.127}{23.481}
\emmoveto{108.172}{23.363}
\emlineto{108.261}{23.159}
\emmoveto{108.305}{23.041}
\emlineto{108.392}{22.837}
\emmoveto{108.435}{22.719}
\emlineto{108.518}{22.515}
\emmoveto{108.559}{22.397}
\emlineto{108.640}{22.193}
\emmoveto{108.680}{22.075}
\emlineto{108.758}{21.870}
\emmoveto{108.796}{21.753}
\emlineto{108.872}{21.548}
\emmoveto{108.908}{21.431}
\emlineto{108.981}{21.226}
\emmoveto{109.016}{21.108}
\emlineto{109.086}{20.903}
\emmoveto{109.120}{20.786}
\emlineto{109.187}{20.581}
\emmoveto{109.219}{20.464}
\emlineto{109.283}{20.259}
\emmoveto{109.314}{20.141}
\emlineto{109.375}{19.936}
\emmoveto{109.405}{19.819}
\emlineto{109.463}{19.614}
\emmoveto{109.491}{19.496}
\emlineto{109.547}{19.291}
\emmoveto{109.574}{19.174}
\emlineto{109.626}{18.968}
\emmoveto{109.652}{18.851}
\emlineto{109.701}{18.646}
\emmoveto{109.725}{18.528}
\emlineto{109.772}{18.323}
\emmoveto{109.795}{18.206}
\emlineto{109.839}{18.000}
\emmoveto{109.860}{17.883}
\emlineto{109.901}{17.678}
\emmoveto{109.921}{17.560}
\emlineto{109.959}{17.355}
\emmoveto{109.977}{17.237}
\emlineto{110.013}{17.032}
\emmoveto{110.030}{16.915}
\emlineto{110.062}{16.709}
\emmoveto{110.078}{16.592}
\emlineto{110.107}{16.386}
\emmoveto{110.121}{16.269}
\emlineto{110.148}{16.064}
\emmoveto{110.161}{15.946}
\emlineto{110.185}{15.741}
\emmoveto{110.196}{15.623}
\emlineto{110.217}{15.418}
\emmoveto{110.227}{15.300}
\emlineto{110.245}{15.095}
\emmoveto{110.254}{14.977}
\emlineto{110.269}{14.772}
\emmoveto{110.276}{14.654}
\emlineto{110.289}{14.449}
\emmoveto{110.294}{14.331}
\emlineto{110.304}{14.126}
\emmoveto{110.308}{14.009}
\emlineto{110.315}{13.803}
\emmoveto{110.318}{13.685}
\emlineto{110.322}{13.480}
\emmoveto{110.323}{13.362}
\emlineto{110.324}{13.157}
\emmoveto{110.324}{13.040}
\emlineto{110.323}{12.834}
\emmoveto{110.321}{12.717}
\emlineto{110.317}{12.511}
\emmoveto{110.314}{12.394}
\emlineto{110.306}{12.188}
\emmoveto{110.302}{12.071}
\emlineto{110.292}{11.865}
\emmoveto{110.286}{11.748}
\emlineto{110.273}{11.542}
\emmoveto{110.266}{11.425}
\emlineto{110.250}{11.220}
\emmoveto{110.241}{11.102}
\emlineto{110.222}{10.897}
\emmoveto{110.212}{10.779}
\emlineto{110.190}{10.574}
\emmoveto{110.179}{10.456}
\emlineto{110.155}{10.251}
\emshow{72.180}{17.700}{Classical particle}
\emshow{1.000}{10.000}{-1.40e-2}
\emshow{1.000}{17.000}{1.74e-2}
\emshow{1.000}{24.000}{4.88e-2}
\emshow{1.000}{31.000}{8.02e-2}
\emshow{1.000}{38.000}{1.12e-1}
\emshow{1.000}{45.000}{1.43e-1}
\emshow{1.000}{52.000}{1.74e-1}
\emshow{1.000}{59.000}{2.06e-1}
\emshow{1.000}{66.000}{2.37e-1}
\emshow{1.000}{73.000}{2.69e-1}
\emshow{1.000}{80.000}{3.00e-1}
\emshow{12.000}{5.000}{-1.00e-4}
\emshow{23.800}{5.000}{1.20e-1}
\emshow{35.600}{5.000}{2.40e-1}
\emshow{47.400}{5.000}{3.60e-1}
\emshow{59.200}{5.000}{4.80e-1}
\emshow{71.000}{5.000}{6.00e-1}
\emshow{82.800}{5.000}{7.20e-1}
\emshow{94.600}{5.000}{8.40e-1}
\emshow{106.400}{5.000}{9.60e-1}
\emshow{118.200}{5.000}{1.08e0}
\emshow{130.000}{5.000}{1.20e0}
\centerline{\bf{Fig.A.3}}
\eject

 
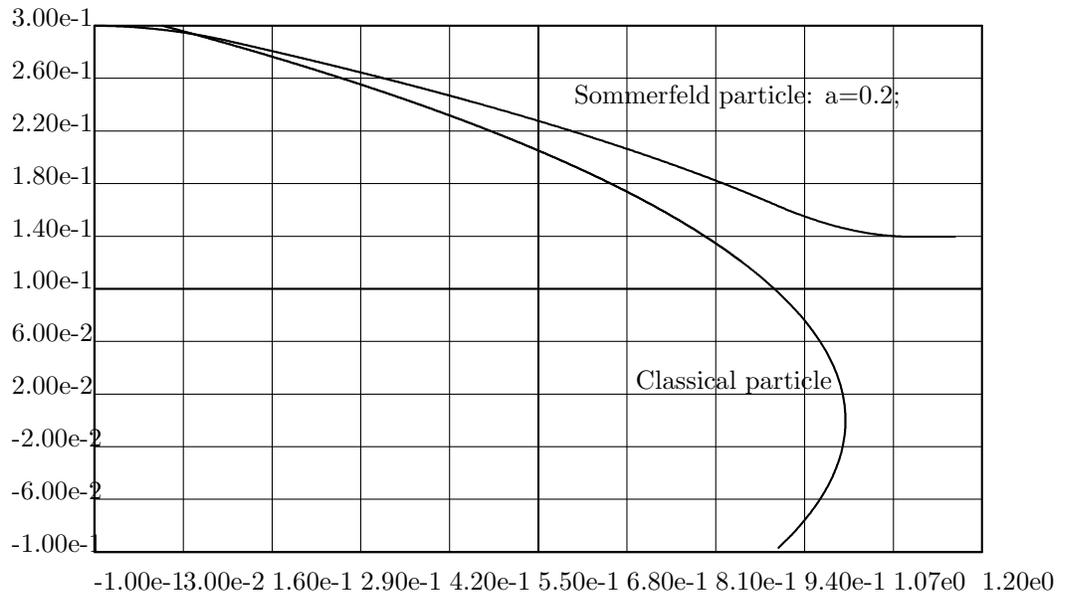
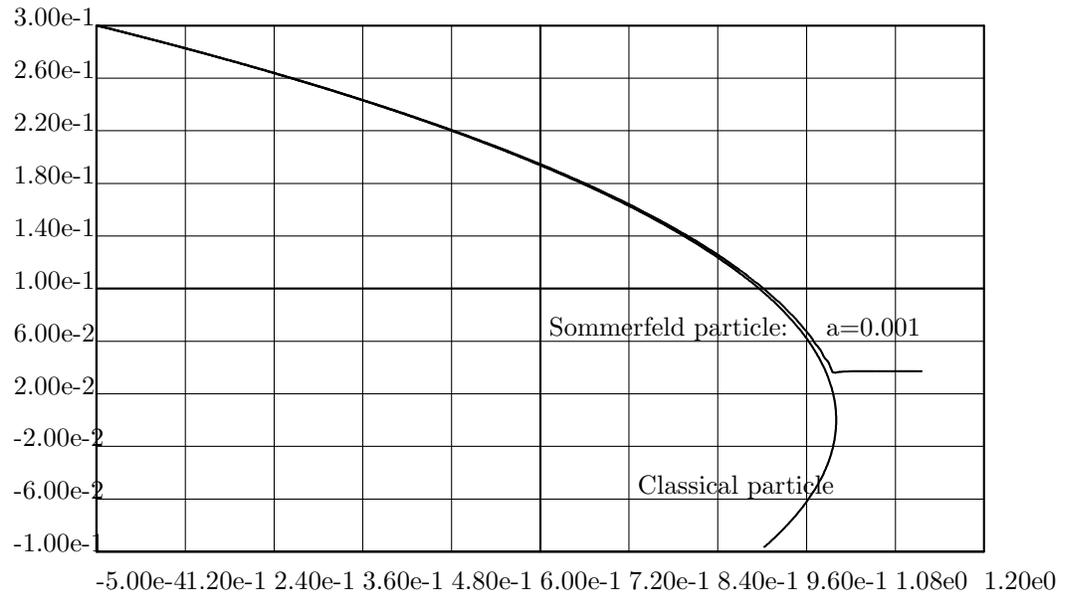
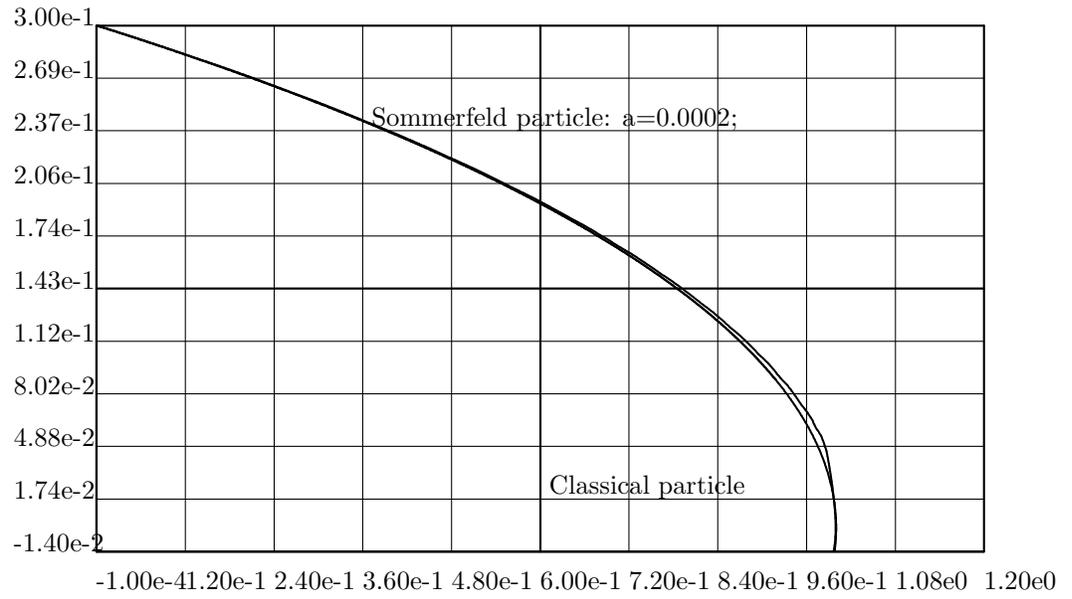
\end{document}